\documentclass[useAMS,usenatbib]{mn2e}
\usepackage{natbib}
\usepackage{graphicx} 
 \usepackage{amssymb}
 
\DeclareMathVersion{bold}

\newcommand{\iindex}{\emph{index }}
\newcommand{\sscale}{\emph{scale }}
\newcommand{\xx}{\vec{x}}

\title[Biparametric Adaptive Filter]{Biparametric Adaptive Filter: detection of compact sources in complex microwave backgrounds} 

\author[M. L\'opez-Caniego and P. Vielva]{M. L\'opez-Caniego and P. Vielva \\
Instituto de F\'isica de Cantabria (CSIC-Universidad de Cantabria), Avda. de Los Castros S/N, 39005, Santander, Spain}

\begin{document}
\maketitle
\begin{abstract}
  In this article we consider the detection of compact sources in maps
  of the Cosmic Microwave Background radiation (CMB) following the
  philosophy behind the Mexican Hat Wavelet Family (MHW$n$) of linear
  filters. We present a new analytical filter, the Biparametric
  Adaptive Filter (BAF), that is able to adapt itself to the
  statistical properties of the background as well as to the profile of
  the compact sources, maximizing the amplification and improving the
  detection process. We have tested the performance of this filter
  using realistic simulations of the microwave sky between 30 and 857
  GHz as observed by the Planck satellite, where complex backgrounds
  can be found. We demonstrate that doing a local analysis on flat
  patches allows one to find a combination of the optimal \sscale of
  the filter $R$ and the \iindex of the filter $g$ that will produce a
  global maximum in the amplification, enhancing the signal-to-noise
  ratio (SNR) of the detected sources in the filtered map and
  improving the total number of detections above a threshold. We
  conclude that the new filter is able to improve the overall
  performance of the MHW2, increasing the SNR of the detections and,
  therefore, the number of detections above a $5\sigma$ threshold. The
  improvement of the new filter in terms of SNR is particularly
  important in the vicinity of the galactic plane and in the presence
  of strong galactic emission. Finally, we compare the sources detected by each method and find that the new
  filter is able to detect more new sources than the MHW2 at all frequencies 
  and in clean regions of the sky. The BAF is also less affected by spurious detections, associated to compact structures in the vicinity of the galactic plane.

\end{abstract}


\newpage

\section{Introduction} \label{sec:intro} \footnotetext{E-mail:
  caniego@ifca.unican.es} The emission of extragalactic point sources
at microwave frequencies is known to be one of the most critical
contaminants of the cosmic microwave background (CMB) anisotropies
\citep{dezotti05,toffolatti05}. This foreground emission is a strong
source of bias in the estimation of the CMB temperature and
polarization angular power spectra. Even at the frequency range of 60 ---
90 GHz (which covers an optimal window for observing the CMB), this effect is 
important. In particular, point sources contaminate the power spectra for multipoles $\ell \geqslant 800$
\citep{tegmark97,tucci05}, which is translated into a bias when
determining the cosmological parameters from these quantities. In
particular, the scalar spectral index $n_s$ and the optical depth
$\tau$ are two of the most biased
parameters\footnote{J. A. Rubi\~no-Mart{\'\i}n and R. B. Barreiro,
  private communication}. In addition, the foreground emission of
point sources also introduces a high level of non-Gaussianity in the
CMB anisotropies \citep[e.g.,][]{argueso03}. This non-Gaussian signal
is a very important confusion noise when someone is probing
non-standard models of structure formation, which actually predict a
certain degree of non-Gaussianity on the CMB anisotropies, typically
at a similar or lower level than the Gaussian deviation caused by the
point sources \citep[e.g.,][]{komatsu03,curto09}.  In addition to the
\emph{role} of the extragalactic point sources as CMB contaminants,
characterising their properties at the microwave frequencies is a very
important field \textit{per se}. For example, studying the number of
objects per flux interval provides useful information to
understand the history of galaxy evolution
\citep[e.g.,][]{gonzaleznuevo08,tucci11}.

A large number of works have been presented in the literature to
mitigate the impact of extragalactic point sources on the science that
can be extracted from the analysis of the CMB anisotropies. Two global
approaches are usually followed, being, in fact, complementary. On the
one hand, detection algorithms are applied to the CMB images for
removing or masking the emission due to the brightest point
sources. On the other hand, statistical modelling of the background of
the residual point sources is adopted. This modelling is followed in
the determination of the cosmological parameters as well as when
exploring the compatibility of the CMB signal with a given
non-Gaussian scenario.  This work lies in the former category: the
detection of the brightest point sources. The literature on this field
is quite large, not only in relation to the development of specific
tools for the CMB problem, but also to the adaptation of techniques
originally developed in radio \citep[e.g.,][]{hogbom74} and optical
\citep[e.g.,][]{bertin96} astronomy. We refer to \cite{herranz10} for
a complete and recent tutorial on the detection of compact sources in
CMB maps.

Some of the most popular tools to perform the detection and flux
estimation of point sources are wavelets. Wavelets represent the
simplest case of point source detection, since it is just based on a
thresholding criterion, without any prior knowledge on the statistical
properties or the background \citep[as the Matched filter,
e.g.,][]{tegmark98,argueso09,lopezcaniego09} or, even more, on the
statistical properties of the background and the signal \citep[as the
Bayesian methods, e.g.,][]{lopezcaniego05,carvalho09,argueso11}. For CMB experiments
with a PSF well defined by a Gaussian function, the Mexican Hat
Wavelet (MHW, built as function proportional to the Laplacian of a
Gaussian kernel) has proved to be a very good tool
\citep[e.g.,][]{cayon00}. In addition, the wavelet scale $R$ can be
optimized to provide a larger number of detections, depending on the
statistical properties of the image \citep{vielva01}. Further
improvement can be achieved by applying subsequently the Laplacian
operator to the Gaussian kernel. This produces a series of wavelets
known as the Mexican Hat Wavelet Family \citep[MHW$n$,][ where MHW1 is
the standard MHW]{gonzaleznuevo06}. The degree of derivation $n$ is a
reflection of the statistical properties of the background. In particular, 
low values of $n$ define a filter that has most of the power at low frequencies 
(or equivalently, in the context of the CMB, at large scales). As the value of $n$ increases, 
the power of the filter moves towards smaller scales. Therefore, if an image is dominated by large scale structures in the background, then one should expect a better point source detection efficiency for a higher value of $n$ than for a smaller one. 

In \cite{lopezcaniego06}, using simulations based on the Planck Reference Sky Model available at
that time, it was shown that, on average, the MHW2 provides a larger number of point source detections than the MHW and the MF, and a better flux estimation  (see \cite{lopezcaniego06} for further details of the simulations, and \cite{tauber10} for  a description of the Planck mission). As a matter of fact, the MHW2 has been successfully applied to the WMAP maps, providing a larger number of detections than the Matched filter used by WMAP \citep{hinshaw07,wright09} as well as other methods \citep{chen08,chen09}.

In this work we generalize the idea behind the MHW$n$, by defining a
biparametric filter, $\psi_g\left(R\right)$ where the parameter
related to the background fluctuations $g$ is allowed to vary in a
continuous way. Therefore, the determination of the $g$ and $R$
parameters is done jointly, attending to the statistical properties of
the background.  The paper is organized as follows. In Section 2 the
biparametric filter is defined. We also discuss under which
circumstances the new filter defaults to other known filters. The
performance of the method is illustrated in Section 3 by analysing in
detail some simulated microwave images. A statistical analysis on how the biparametric filter behaves
as a function of the frequency and the galactic latitude is presented. Conclusions are given in Section 4.

\section{Methodology} \label{sec:method}

As mentioned above, considering the statistical properties of the
background in the vicinity of a compact object is a key step in the
design of a filter to be used for the detection of point sources. This
issue can be as simple as looking for the optimal \sscale that
provides the maximum amplification for wavelets or as sophisticated as
providing a complete likelihood and prior functions, for instance,
when applying Bayesian approaches.  In all-sky CMB experiments such as
WMAP \citep{bennet03} or Planck \citep{tauber10} one can deal with
very different backgrounds. For instance, the properties of the
background have a typical dependence on the Galactic latitude: the
major emission corresponds to the Galactic components (synchrotron,
free-free, dust) near the Galactic plane, whereas the CMB and the
background of extragalactic sources are, typically, more important at
intermediate and high latitudes.  In addition, depending on the
observational frequency, radio emissions (galactic and extragalactic)
are more important in the lower part of the microwave frequency range
(i.e., from 10 to 100 GHz), whereas thermal dust and the cosmic
infrared background are the major contributors to highest frequencies
(i.e., from 300 to 1000 GHz). However, at intermediate frequencies the
CMB is, overall, the most important background.  Each one of these
emissions has its own specificities and, therefore, it is suboptimal
to use the same filter at all frequencies and positions in the sky.

In this paper we explore the detection of objects in different types
of backgrounds commonly found in CMB experiments. These objects are,
in general, point-like sources that have been convolved with the beam
point spread function (PSF) of the instrument used for the
observation. They can be described as a signal $s(\xx)=A\tau(\xx)$,
where $A$ is the intrinsic flux density of the object and $\tau(\xx)$
is the PSF of the beam and $\xx$ is a unit vector in the sky. Although
the profile $\tau(\xx)$ can be described by any function, the
formalism adopted in this paper assumes that it is given by an
isotropic PSF, i.e., $\tau(\xx) \equiv \tau(|\xx|) = \tau(x)$. We
adopt this simplification because for most of the CMB experiments an
effective isotropic window function is usually defined, even when the
PSF has a certain degree of anisotropy. In any case, of course, the
filter proposed in this paper can be easily generalized for
non-circular beam profiles.

\subsection{The Biparametric Adaptive filter}
\label{subsec:baf}
Following the idea behind the MHW Family, in this paper we propose a
filter that has two free parameters that will allow us to better
incorporate into the filter the statistical properties of the
background. As for the case of wavelets, one of the parameters of the
filter is the \sscale $R$. The filter $\sscale$ is associated with a compression/expansion
of the typical scale of the PSF and provides the size of the filtering
kernel. In addition, we incorporate the \iindex of the filter 
$g$ that can be seen as a generalization of the role
played by the order of the Laplacian operator used in the definition
of the MHW Family~\citep[e.g.,][]{gonzaleznuevo06}\footnote{Notice
  that, whereas the order $n$ of the Laplacian operator is a natural
  number, the \iindex $g$ is defined as a real number $\geq0$}. It is
related to the filter location in Fourier space or, conversely, with
the filter oscillations in the real space. As it will be discussed in
the next subsection, this parameter is somehow associated with the
shape of the angular power spectrum of the background. The filter
definition is such that it is compensated \citep[e.g.][]{cayon00}, and, therefore, for the case adopted in this
paper of 2D Euclidean images, it behaves as a wavelet. The only case in which there is no compensation and, therefore, the filter is not a wavelet, is for the particular case in which the background is defined by a white noise field. In this particular situation, the filter defaults to a Gaussian kernel with $g=0$ (see next subsection).

Let us denote
the BAF (defined in terms of the two parameters $R$ and $g$) by $\Psi
(\xx; R, g, \vec{b})$. It denotes the value of the filter centred at
$\vec{b}$, in the position $\xx$.  Following the philosophy behind the 
wavelets, this filter can be seen as the scaling version of a
translated mother filter:
\begin{center}
\begin{equation} \label{eq:psi} 
\Psi(\xx; R, g, \vec{b}) \equiv \frac{1}{R^2} \bar{\psi}_g \left( 
\frac{|\xx - \vec{b}|}{R} \right) . 
\end{equation} 
\end{center}
This mother filter $\bar{\psi}_g$ defines the BAF and, in Fourier space, is given by:
\begin{equation}
\hat{\psi}_g \left(qR\right) =\frac{1}{\pi} \frac{1}{\Gamma \left( \frac{2+g}{2}\right)} \left(qR\right)^{g} \tau(qR).
\label{eq:filtro_general}
\end{equation}
In particular, and following the convention in \cite{sanz01}, convolving such a filter with a 2D image $f(\xx)$ we obtain the filter coefficients $\omega_g(R, \vec{b})$ defined as:
\begin{equation} \label{eq:wavcoef0} 
\omega_g \left(R, \vec{b} \right) = \int d\xx\,f(\xx)\Psi (\xx; R, g, \vec{b}). 
\end{equation} 
A very common situation in astronomy is to have a PSF defined by a Gaussian function $\tau(\xx) =\left( 1 / 2\pi{\sigma_b}^2 \right) e^{-\frac{1}{2}\left(\xx/\sigma_b\right)^2} $, where $\sigma_b$ is the Gaussian beam dispersion. For this particular case, the BAF is given by:
\begin{equation}
\hat{\psi}_g \left(qR\right) =\frac{1}{\pi} \frac{1}{\Gamma \left( \frac{2+g}{2}\right)} \left(qR\right)^{g}
e^{-\frac{1}{2}\left(qR\right)^2}.
\label{eq:filtro}
\end{equation}
and the filter coefficients at the position of a source with a profile $I_0\tau(qR)$, where $I_0$ is the amplitude of the source, is given by:
\begin{equation}
\omega_g\left( R \right)= \frac{I_0 2^{\frac{g+2}{2}} z^{g} } { \left(1+z^{2}\right)^{\frac{g+2}{2}} },
\label{eq:coeffwav}
\end{equation}
where $z \equiv R/\sigma_{b}$.
As it was introduced by~\cite{vielva01}, the filter parameters are determined by imposing a maximum
amplification $\mathcal{A}$ of the point source amplitude in the filter coefficients map:
\begin{equation}
\mathrm{{\mathcal{A}}} \equiv \frac{\omega_g \left(R \right)/\sigma_{\omega}}{I_0/\sigma},
\label{eq:ampli}
\end{equation}
where $\sigma$ is the dispersion of the image in the real space, $I_0$ is the amplitude of the source and $\sigma_{\omega}$ is the dispersion of the filter coefficients at the \sscale $R$ and \iindex $g$, that can be defined as:
\begin{equation} \label{eq:variance0} 
\sigma_{\omega}^2\left(R,g\right) \equiv 2 \pi \int_0^{\infty} dq \ q P\left(q\right) \hat{\psi}_g^2(Rq),
\end{equation} 
with $P\left(q\right)$ being the angular power spectrum of the analysed image.

\subsection{The BAF and other filtering kernels}
\label{subsec:baf_others}
The BAF introduced in equation~\ref{eq:filtro_general} defaults, under certain conditions, to other
well known filter kernels extensively used in literature. In this sense, the BAF can be considered
as a generalization of these filters.
The most obvious generalization is for the MHW$n$~\citep[e.g.,][$\hat{\psi}_n$]{gonzaleznuevo06}.  As previously mentioned,
when the profile of the point sources is described by a Gaussian function, the BAF is given by
equation~\ref{eq:filtro}. Both filters are related by: $\hat{\psi}_g = \left( 2^n / \pi \right) \hat{\psi}_n$, where $g = 2n$ and $n = 1, 2, 3, \ldots$.

Another interesting situation occurs when the background can be described by an angular power spectrum following a
power law $P(q)= C q^{-\gamma}$, for $q > 0$, and $C$ being a normalization constant. In this case, it is easy to show, from equation~\ref{eq:variance0},
that the variance of the filter coefficients is given by:
\begin{equation}
\sigma_{\omega}^{2}\left(R,g \right) = C\frac{\sigma_{b}^{\gamma -2}}{\pi} \frac{\Gamma \left( \frac{2g+2-\gamma}{2}\right) }{ \left[ \Gamma \left( \frac{2+g}{2}\right) \right]^2} z^{\gamma -2}.
\label{eq:variance}
\end{equation}
\noindent
From equation~\ref{eq:ampli}, one can demonstrate that the maximum amplification occurs when $R\equiv \sigma_b$ and $g \equiv \gamma$, i.e., the BAF defaults to the Matched filter.

\section{The Simulations} \label{sec:sims}

\begin{figure}
\begin{center}
\includegraphics[width=0.45\textwidth]{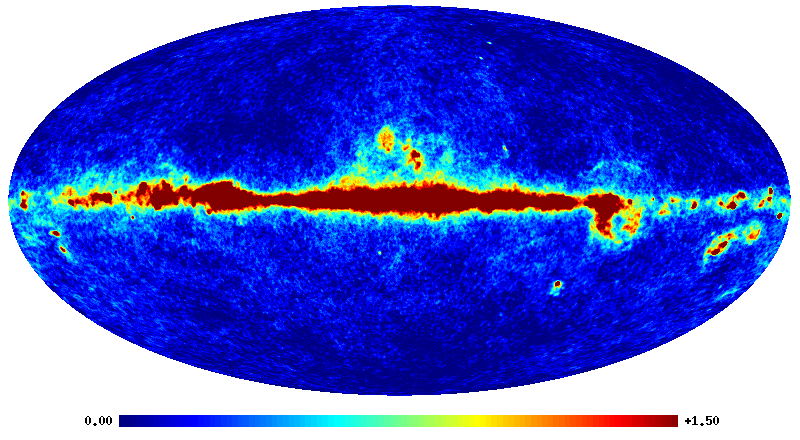}
\includegraphics[width=0.45\textwidth]{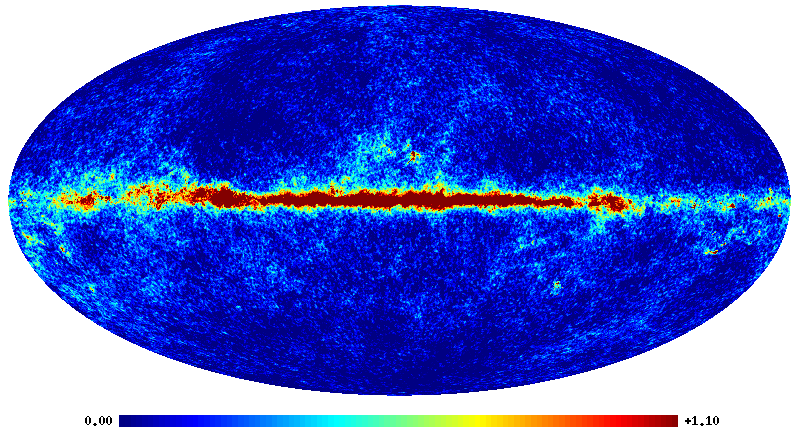}
\includegraphics[width=0.45\textwidth]{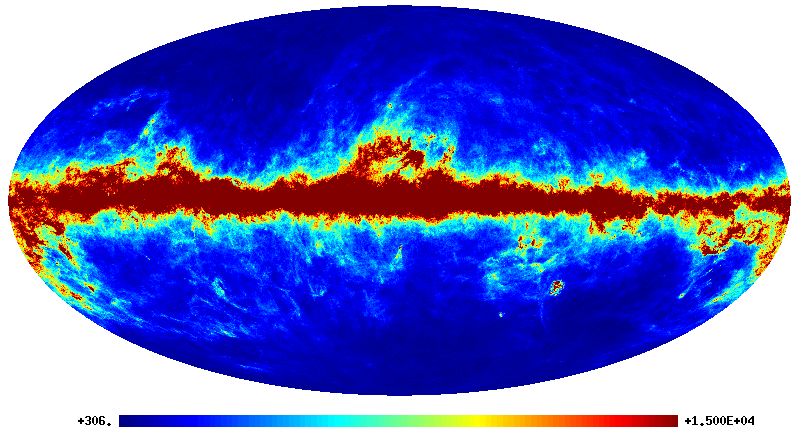}
\caption{From top to bottom, a combination of noise, CMB, galactic diffuse emission and compact source emission maps at 30, 143 and 857 GHz simulated with the pre-launch Planck Sky Model.}
\label{fig:plancksims}
\end{center}
\end{figure}

In order to test the performance of the new filter we will use
realistic simulations of the microwave sky at 30, 44, 70, 100, 143, 217, 353, 545 and
857 GHz. These simulations have been generated with the pre-launch Planck Sky
Model \citep{leach,delab11}, a software package developed within the
Planck Collaboration that allows one to make simulations at the
microwave frequencies of the CMB, Galactic diffuse emissions and
compact sources. The Galactic interstellar emission adopted in this
paper is described by a three component model of the interstellar
medium comprising of free-free, synchrotron and dust emissions plus
added small scale fluctuations to reproduce the non-Gaussian nature of
the interstellar emission. Free-free emission is based on the model
of \cite{Dickinson} assuming an electronic temperature of 7000 K,
where the spatial structure of the emission is estimated using a
H$\alpha$ template corrected for dust extinction. Synchrotron emission
is based on an extrapolation of the 408 MHz map of \cite{Haslam} from
which an estimate of the free-free emission was removed. The thermal
emission from interstellar dust is estimated using model 7 of
\cite{Finkbeiner}. Point sources are modelled with two main
categories: radio and infrared. Simulated radio sources are based on
the NVSS or SUMSS \citep{condon98,mauch03} and GB6 or PMN catalogues
\citep{griffith95,gregory96}. Measured fluxes at 1 and/or 4.85 GHz are
extrapolated to Planck frequencies assuming a distribution in flat and
steep populations. Infrared sources are based on the IRAS catalogue
\citep{beichman88}, and modelled as dusty galaxies. A detailed
description of each component can be found in \cite{leach}. These
simulations use the HEALPix pixelization scheme \citep{gorski05} with
NSIDE=1024 at 30, 44 and 70 GHz and NSIDE=2048 for the rest. In Figure
\ref{fig:plancksims} we show three of these simulations, at 30, 143
and 857 GHz.

\subsection{Patch analysis}
\begin{figure*}
\begin{center}
\includegraphics[width=0.32\textwidth]{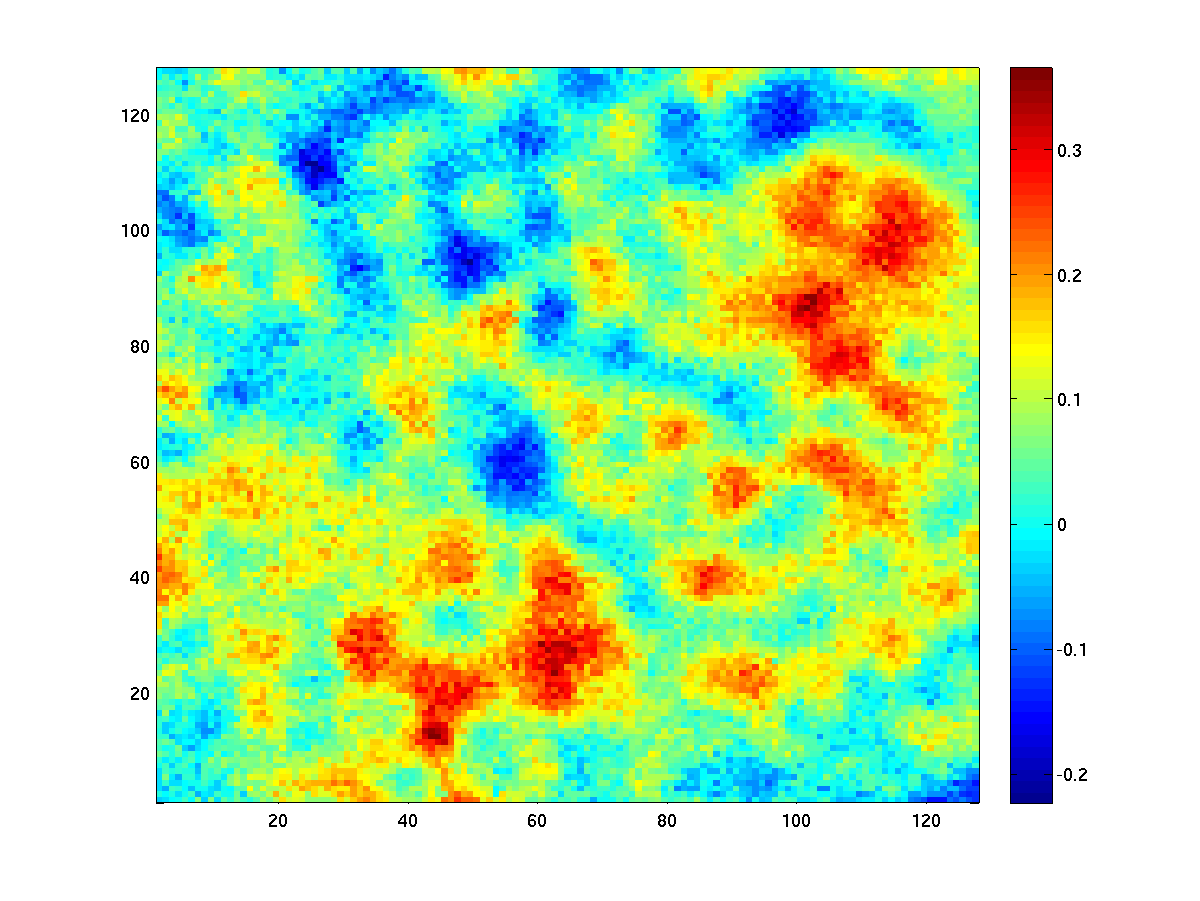}
\includegraphics[width=0.32\textwidth]{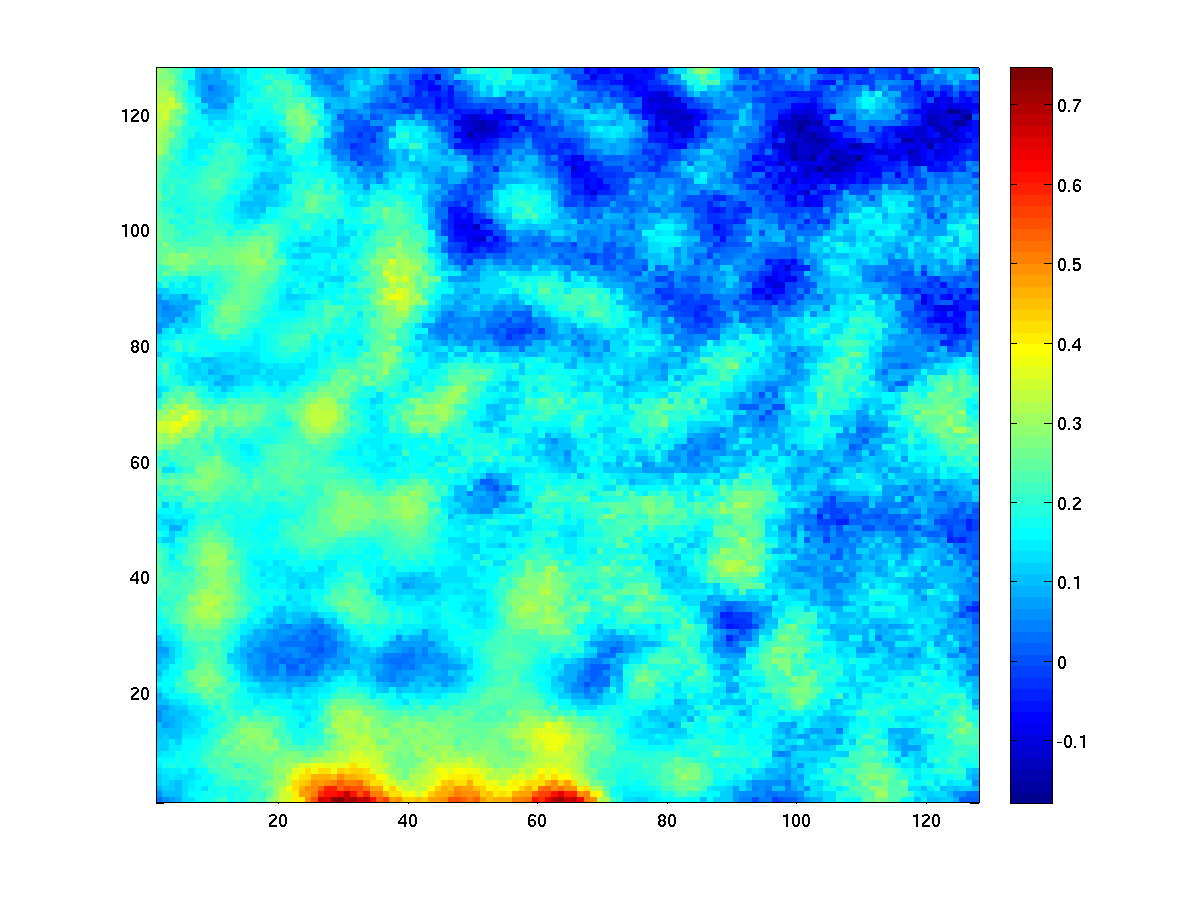}
\includegraphics[width=0.32\textwidth]{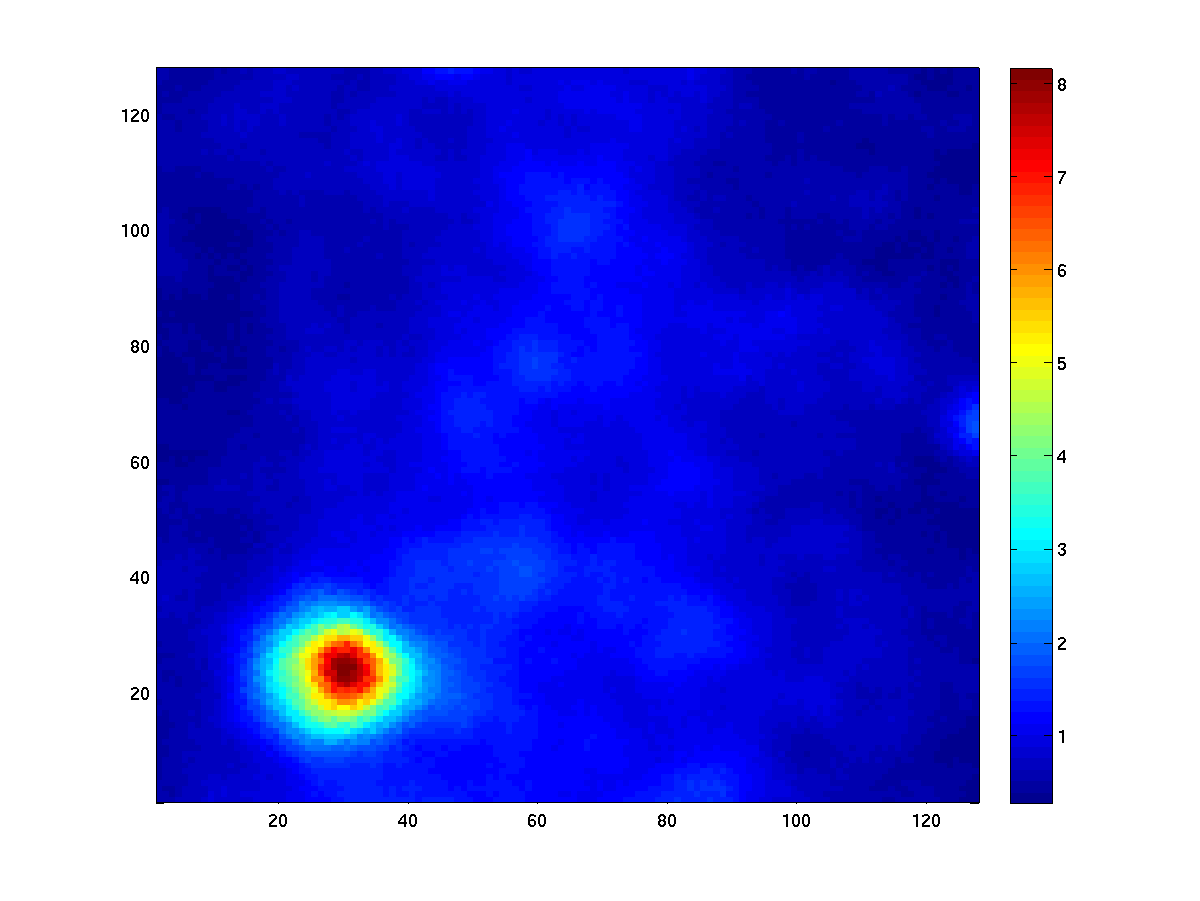}
\includegraphics[width=0.32\textwidth]{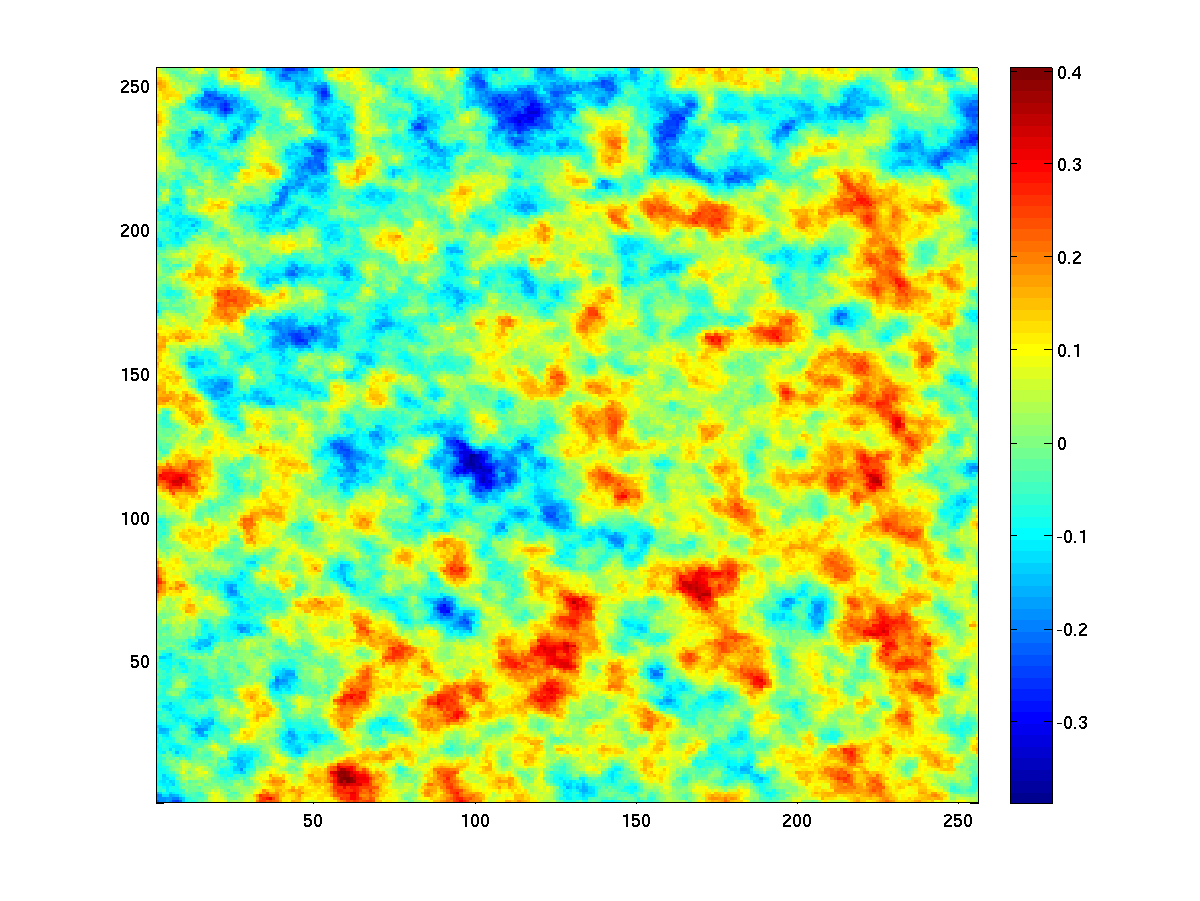}
\includegraphics[width=0.32\textwidth]{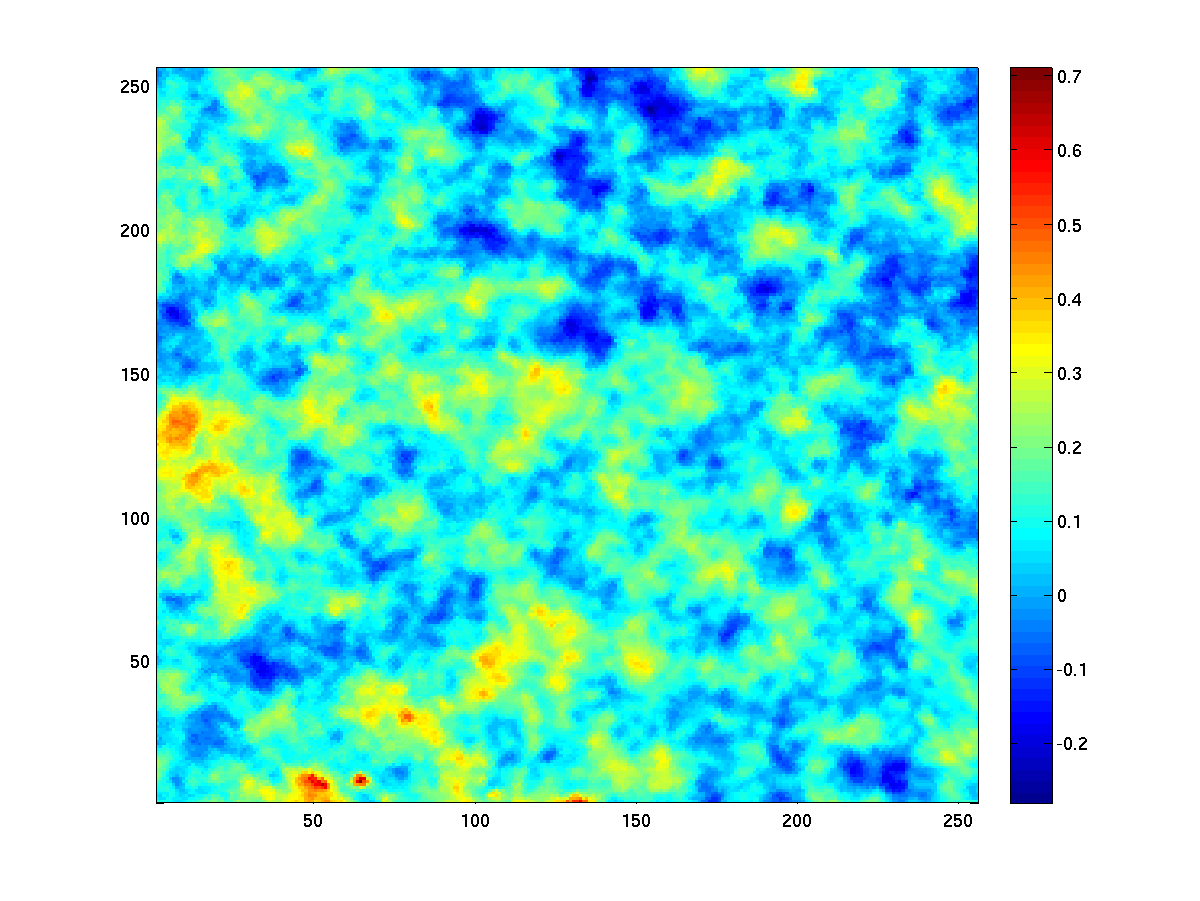}
\includegraphics[width=0.32\textwidth]{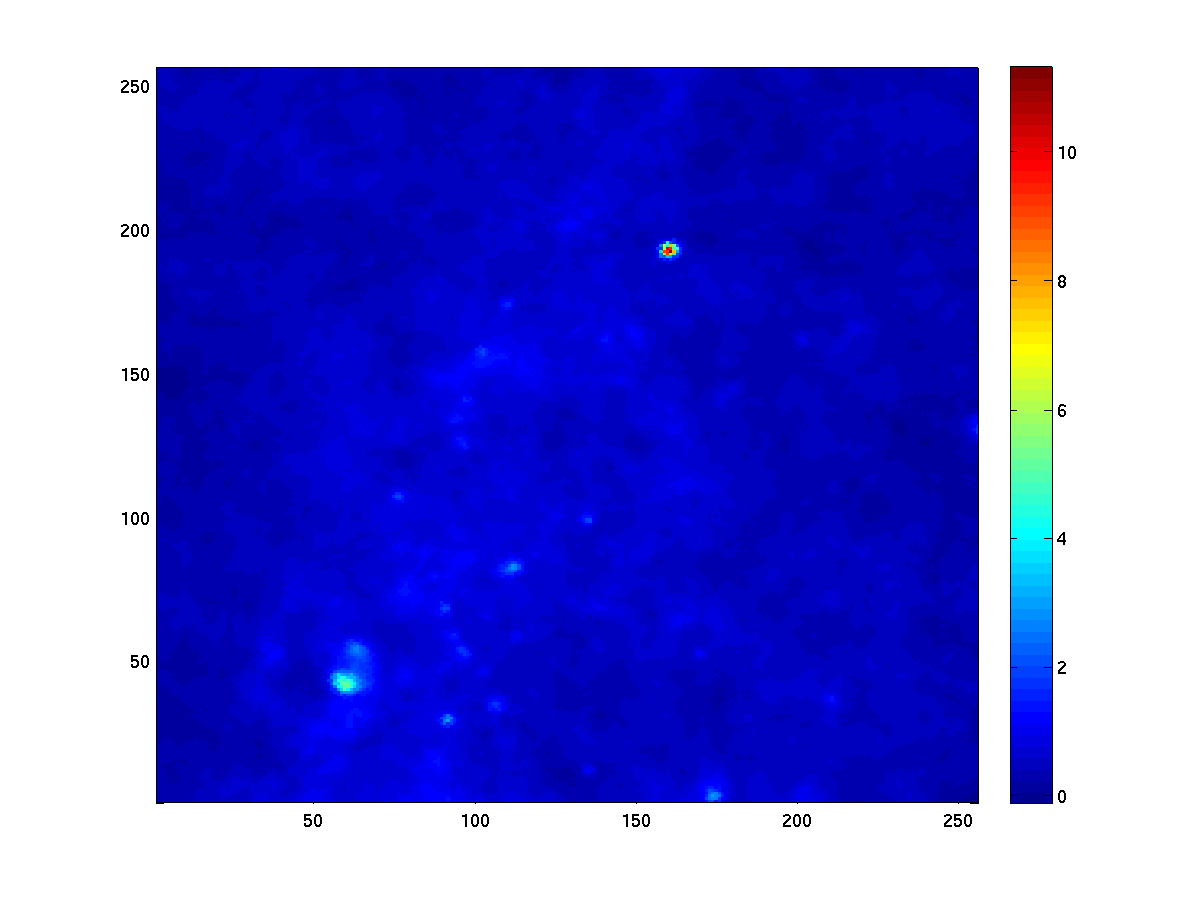}
\includegraphics[width=0.32\textwidth]{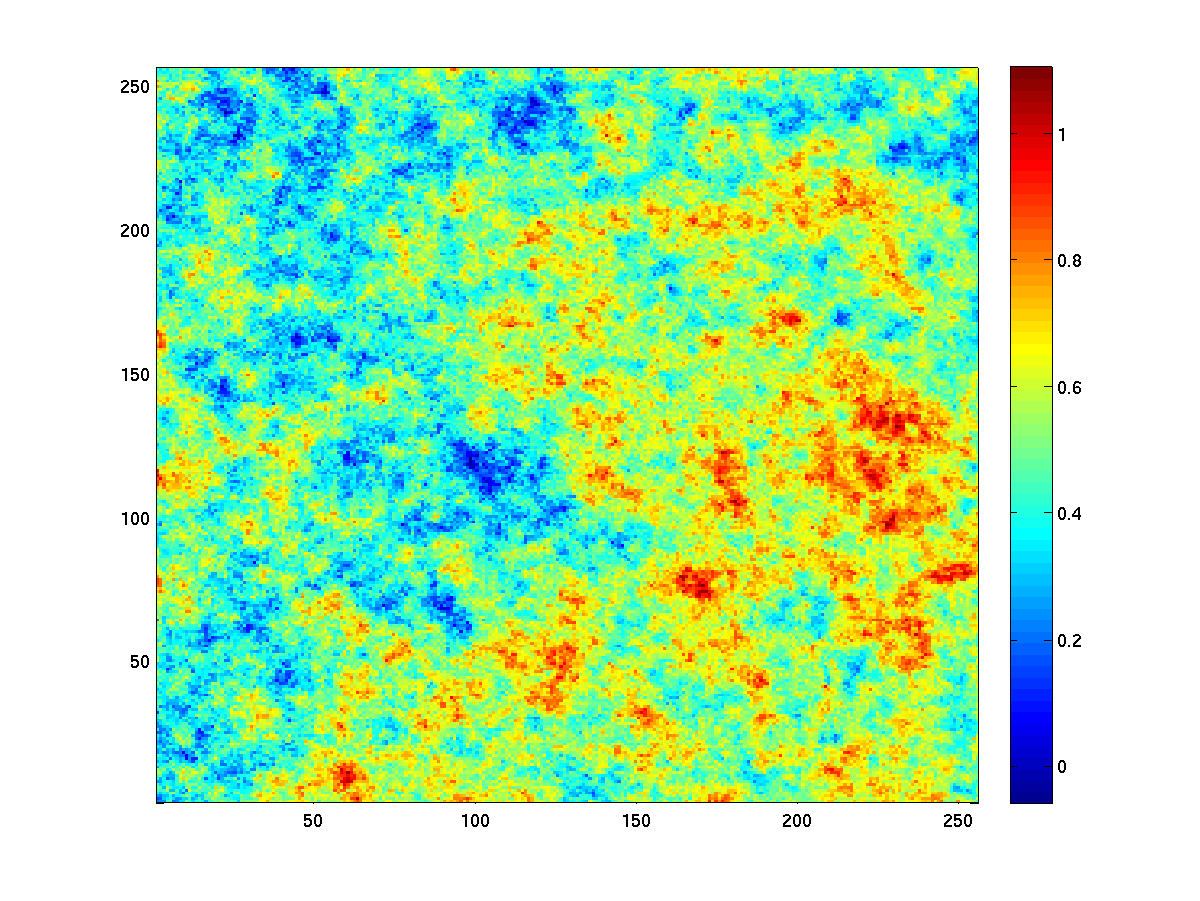}
\includegraphics[width=0.32\textwidth]{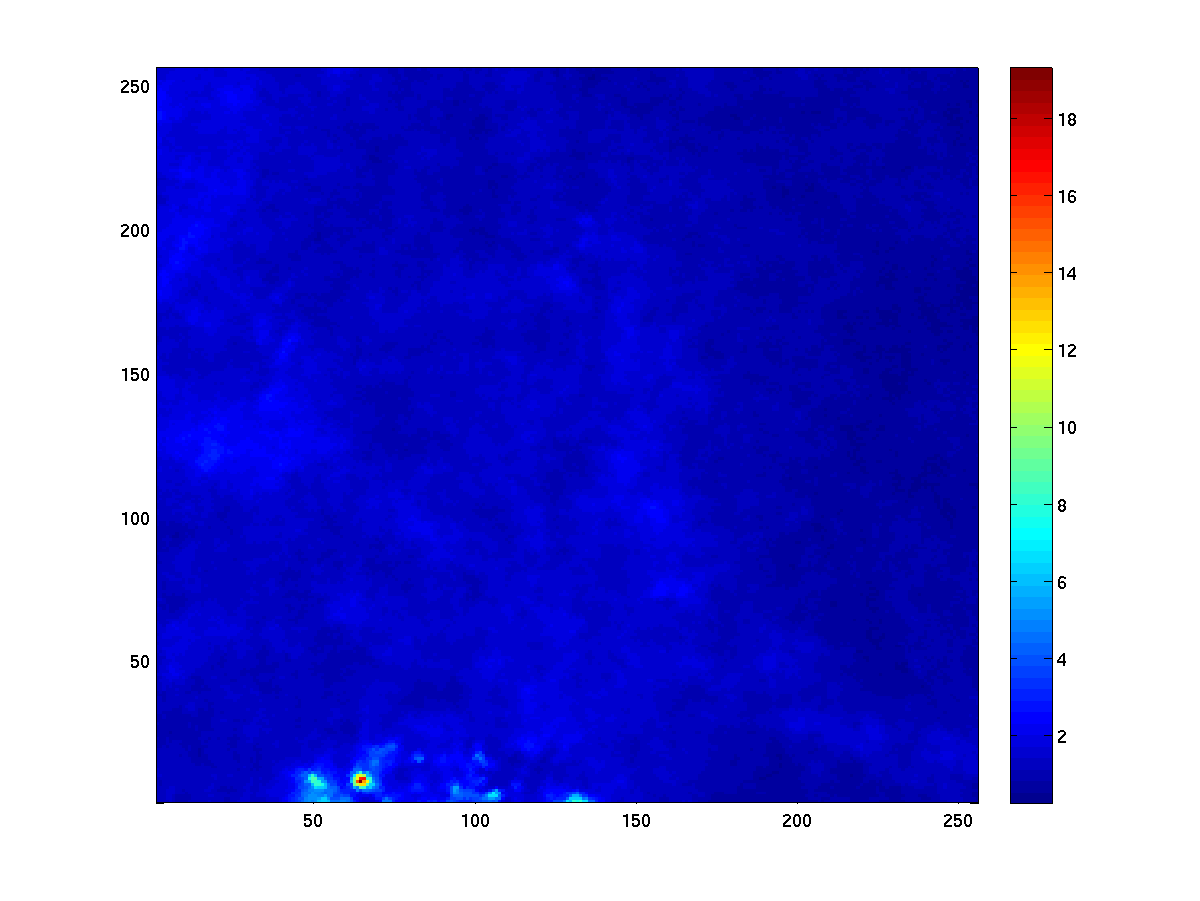}
\includegraphics[width=0.32\textwidth]{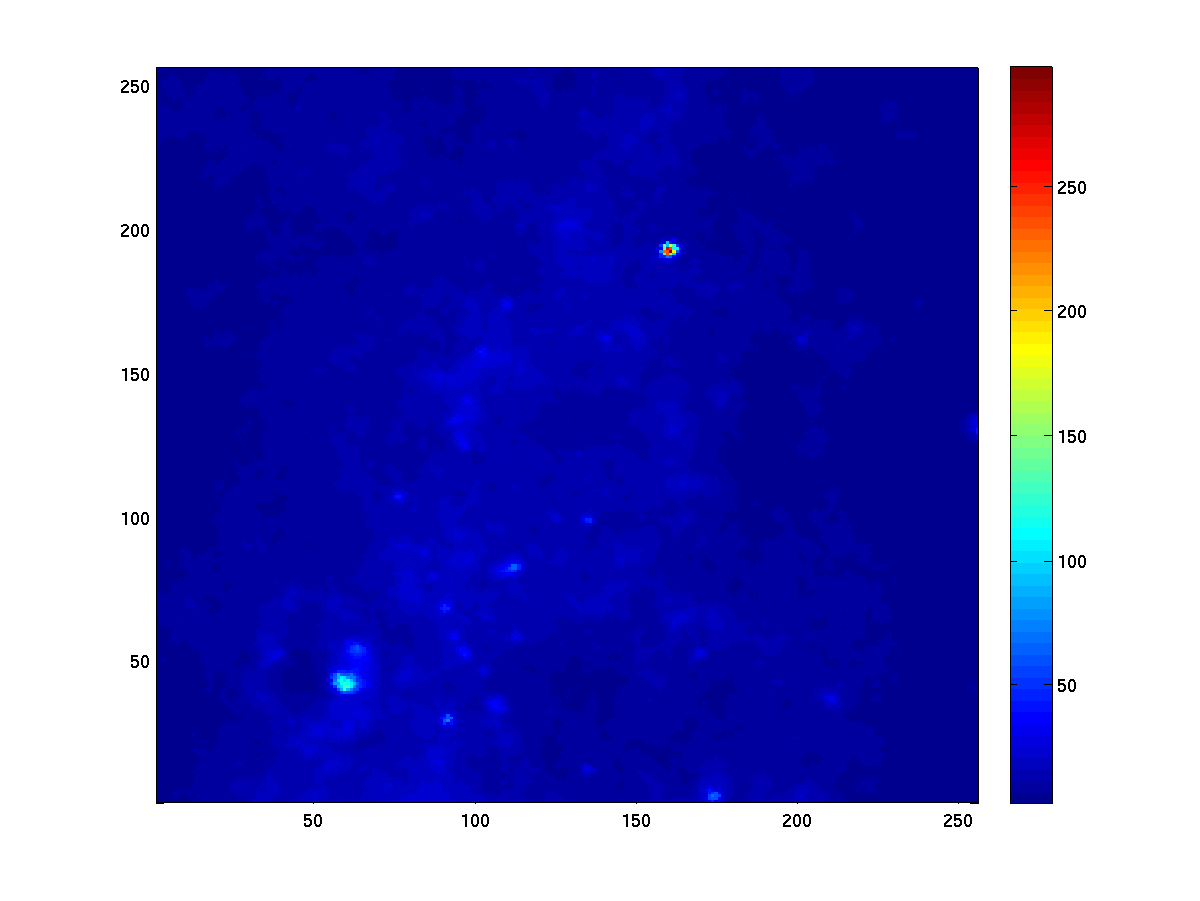}
\caption{Selected regions with increasing background complexities extracted from the 30, 143 and 857 GHz simulated maps. At 30 GHz these patches have
  $128\times128$ pixels in size, where the pixel size is 3.43 arcminutes. At
  143 and 857 GHz these patches have $256\times256$ pixels in size, where the
  pixel size is 1.71 arcminutes. The upper panels correspond to the 30
  GHz regions, the middle panels to the 143 GHz regions and the lower
  panels to the 857 GHz regions. The left, center and right panels
  correspond to regions 1, 2 and 3 respectively. The coordinates of
  the center of these three regions can be found in Table
  \ref{tab:tabla_res_parches}.}
\label{fig:regiones}
\end{center}
\end{figure*}

As it was mentioned above, the purpose of this work is to study the
performance of the new filter in different backgrounds. For this
reason, and to better illustrate the problem, we have studied three 
different regions of the sky at each of the following three
frequencies, 30, 143 and 857 GHz. For each region, we have projected
flat patches of $7.3\times7.3$ square degrees. At 30 GHz each patch has $128\times128$
pixels and a pixel size of 3.43 arcminutes. At 143 and 857 GHz, each
patch has $256\times256$ pixels and a pixel size of 1.71 arcminutes. Each
region has been selected by visual inspection to have increasing
background complexities. In general, the statistical properties of the
background in those patches with lower Galactic latitudes are more
complex than those in the higher latitudes. The coordinates of the
centres of the patches can be found in Table
\ref{tab:tabla_res_parches}. In Figure \ref{fig:regiones} one can see
the nine selected regions.

In order to determine the filter characteristics that best adapts to
the properties of each patch we look for the optimal \sscale $R$ and
\iindex $g$ of the filter that maximize the amplification of the
sources present in the patch. An increase in the amplification will
allow one to detect more sources above a certain threshold. In Table
\ref{tab:tabla_res_parches} one can see the values of $g$ obtained for
each patch and each frequency. Note that the optimal values of $g$ and
$R$ vary from one region of the sky to another. For the 30 and 143 GHz
cases, these variations are small even for different Galactic
latitudes, particularly for the 143 GHz case, but at 857 GHz, where
the Galactic dust and the far-infrared background dominate, the
variations are large. This simple analysis already shows the
importance of optimizing not only the \sscale but also the \iindex of
the filter.

\begin{figure*}
\begin{center}
\includegraphics[width=0.32\textwidth]{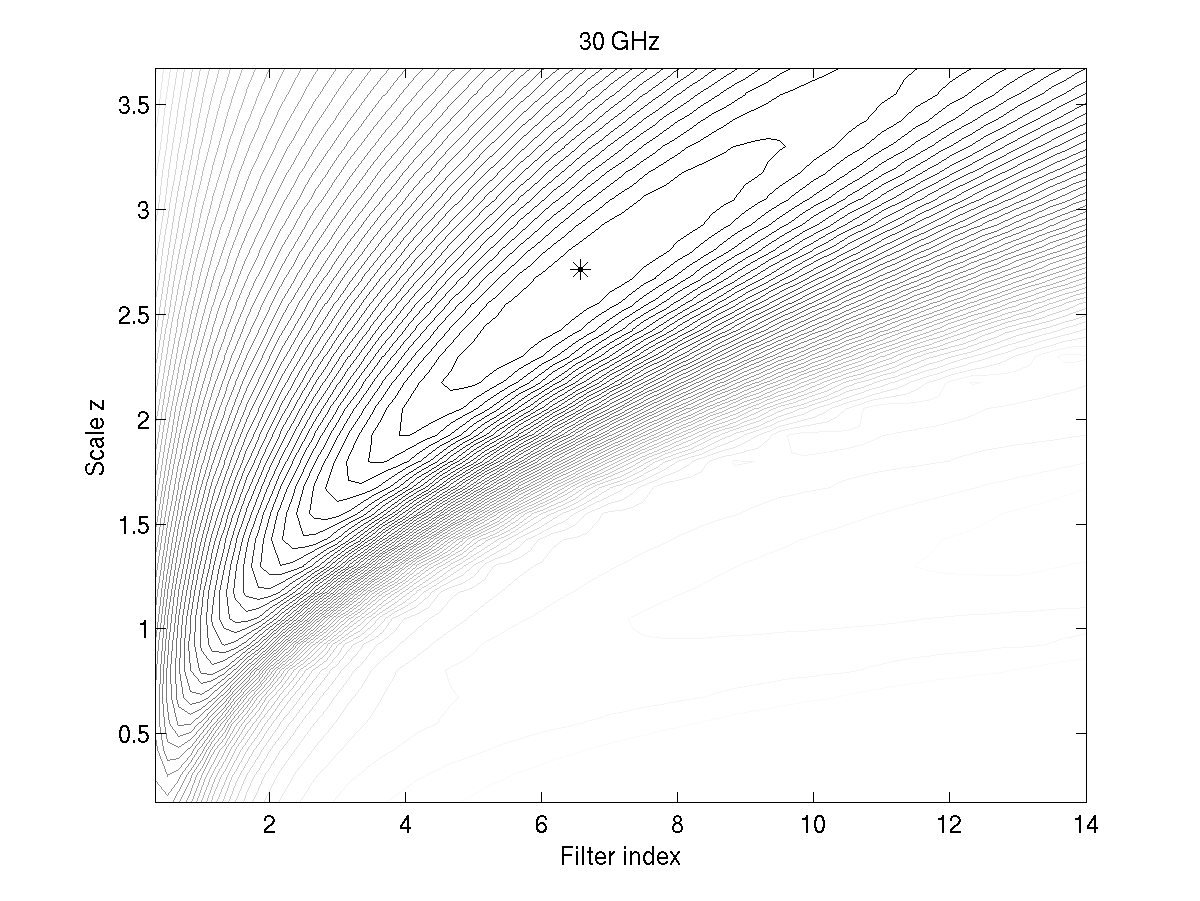}
\includegraphics[width=0.32\textwidth]{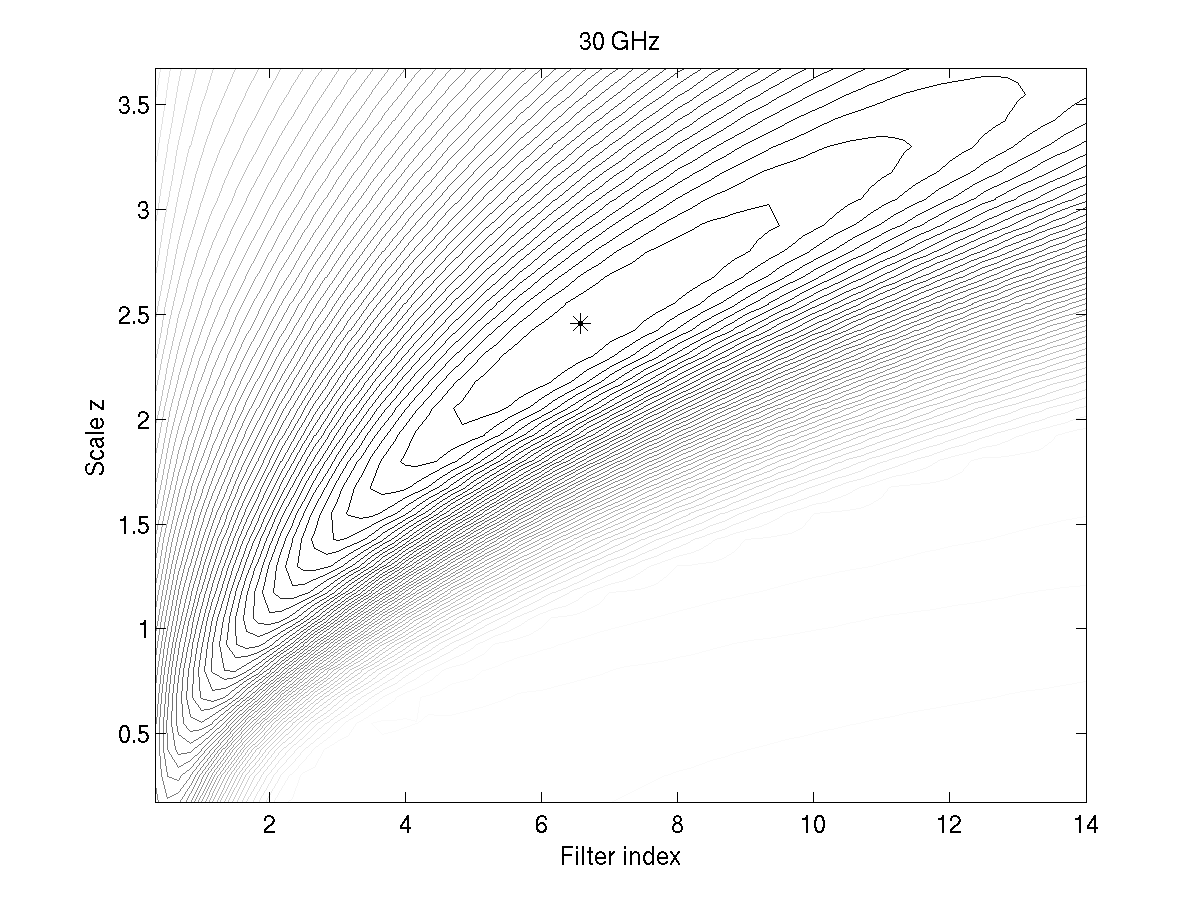}
\includegraphics[width=0.32\textwidth]{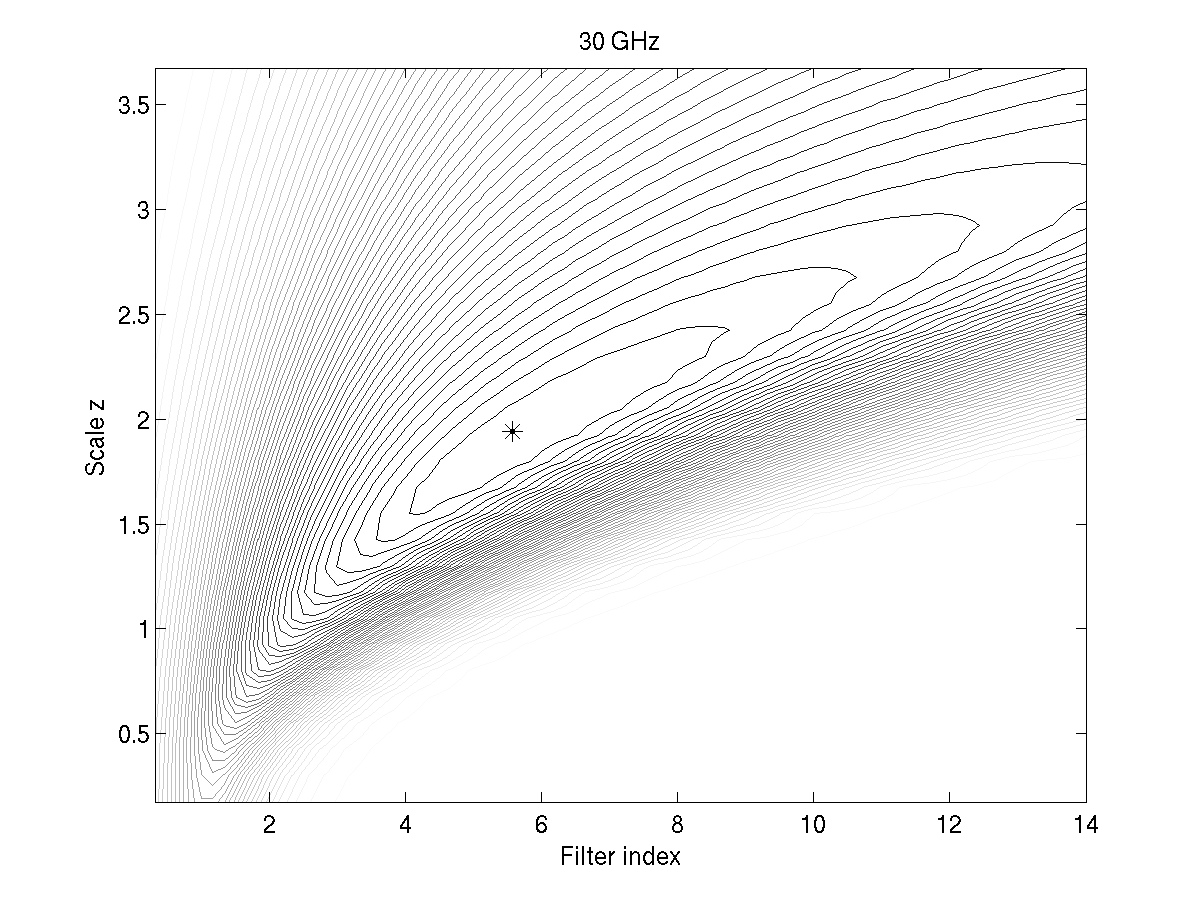}
\includegraphics[width=0.32\textwidth]{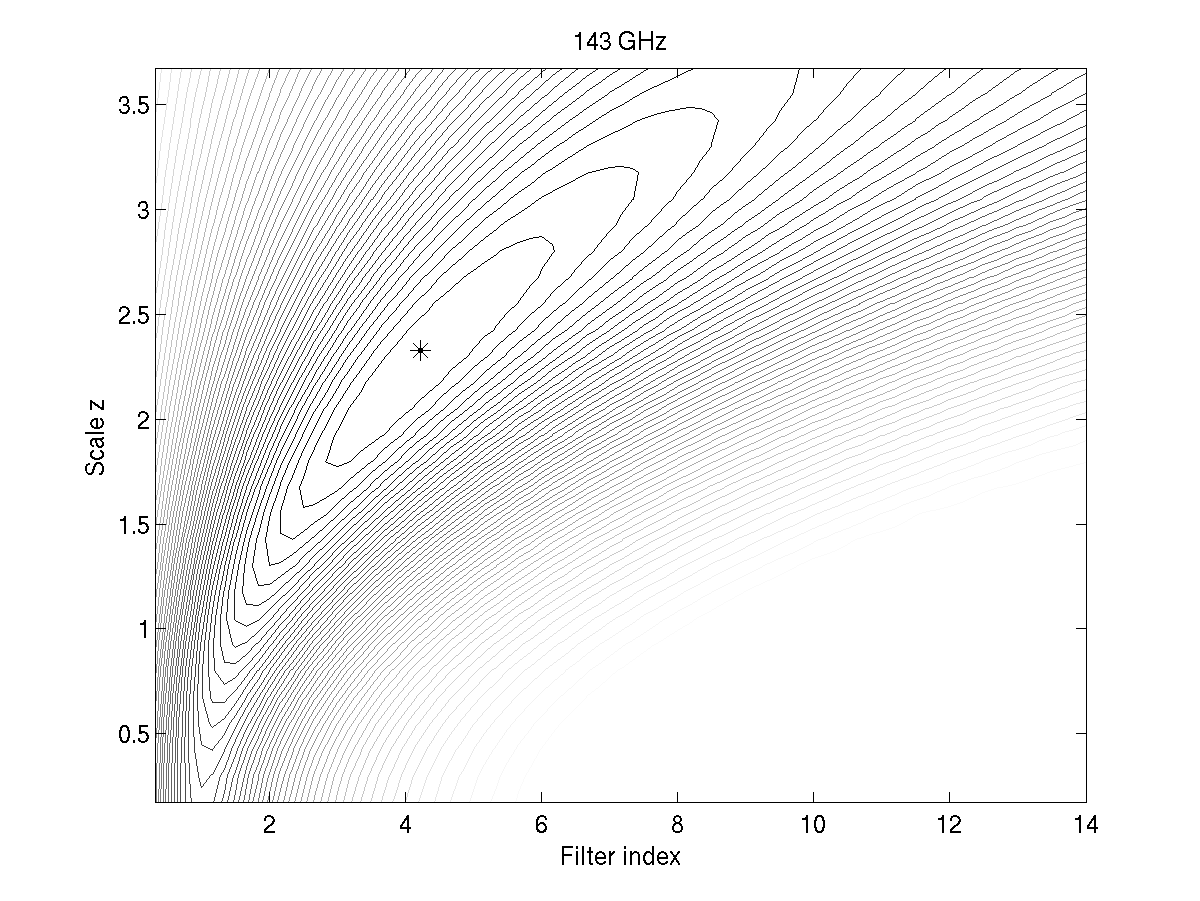}
\includegraphics[width=0.32\textwidth]{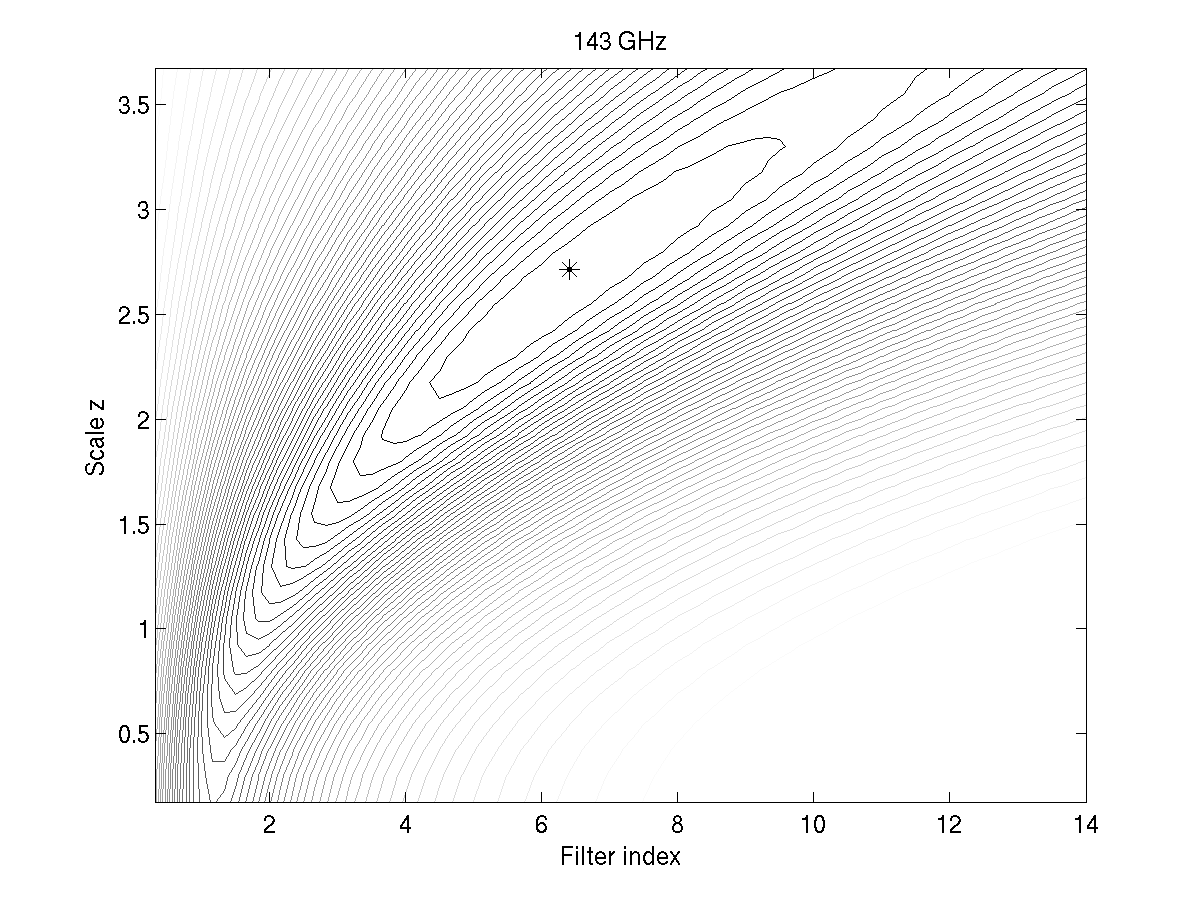}
\includegraphics[width=0.32\textwidth]{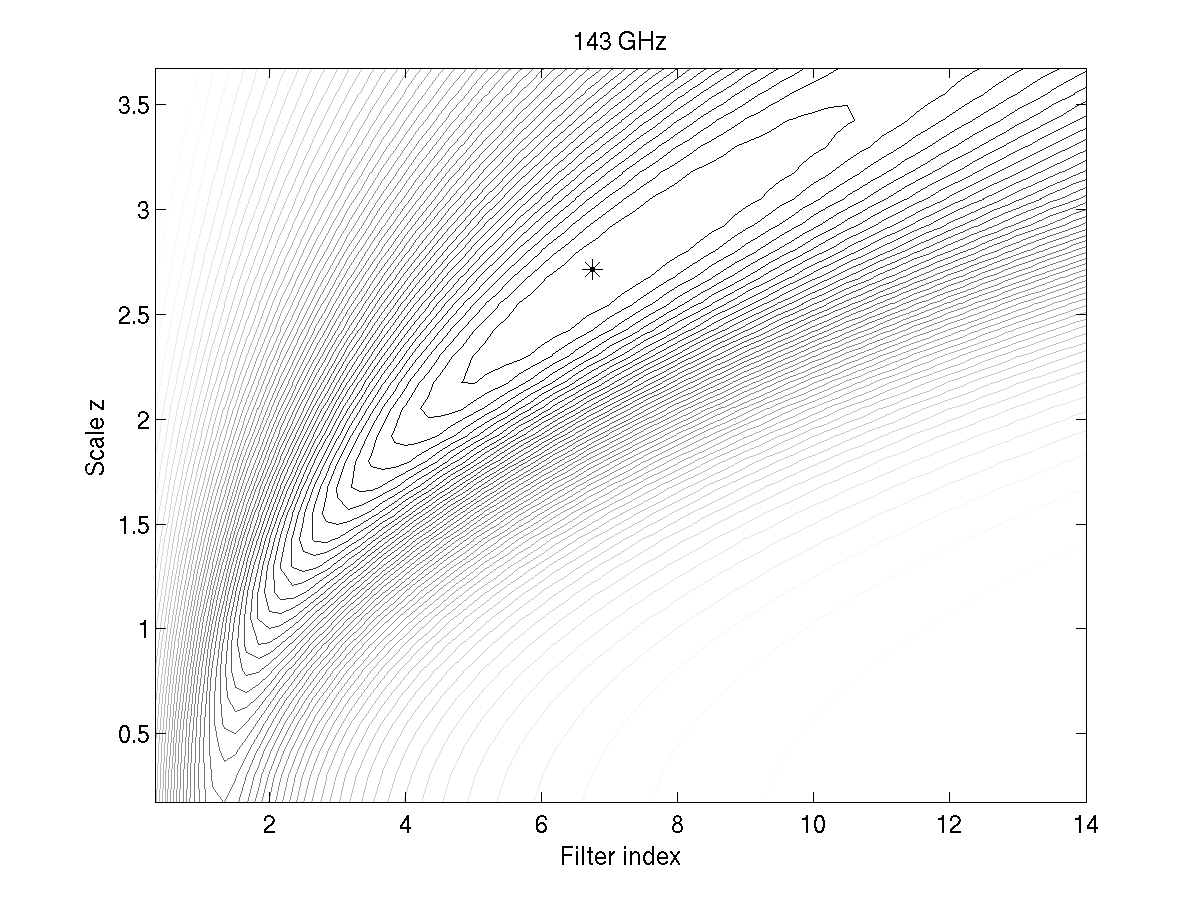}
\includegraphics[width=0.32\textwidth]{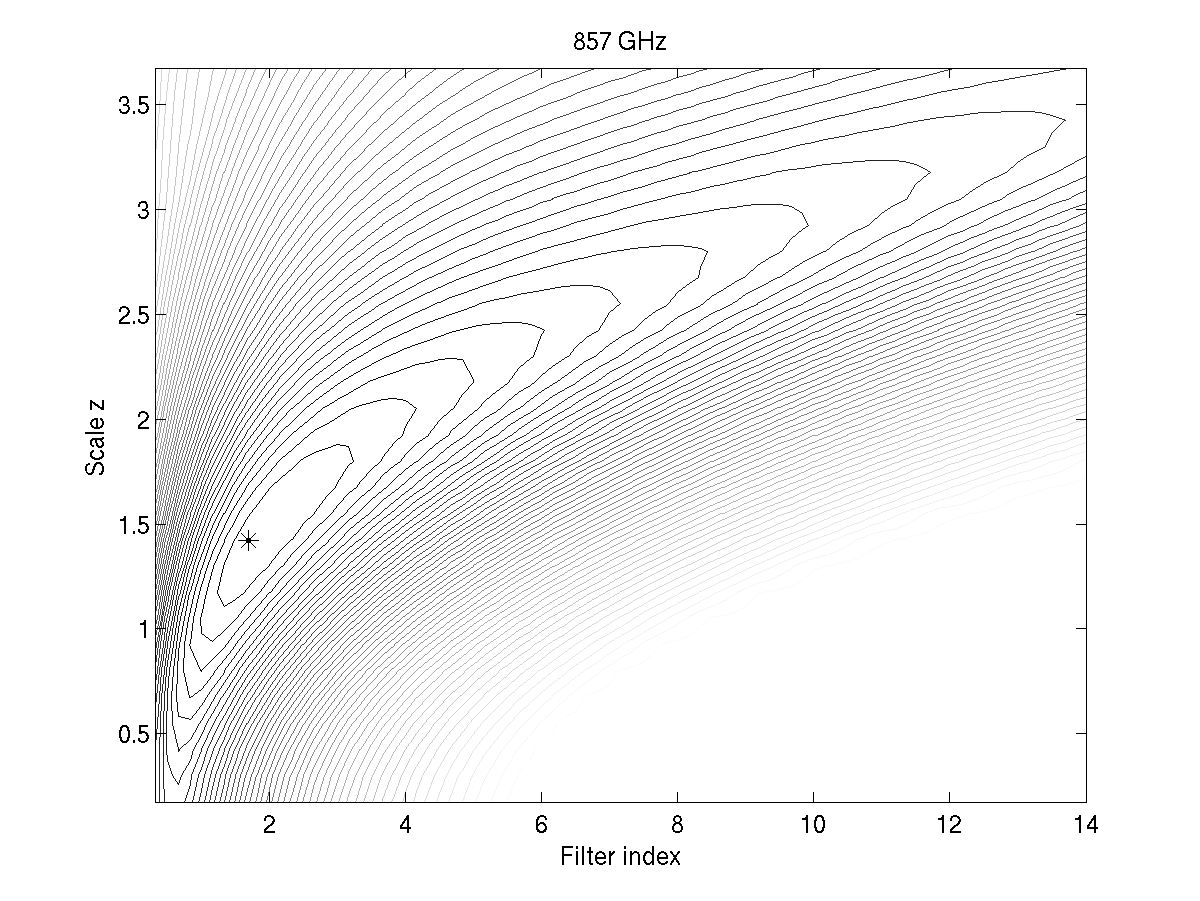}
\includegraphics[width=0.32\textwidth]{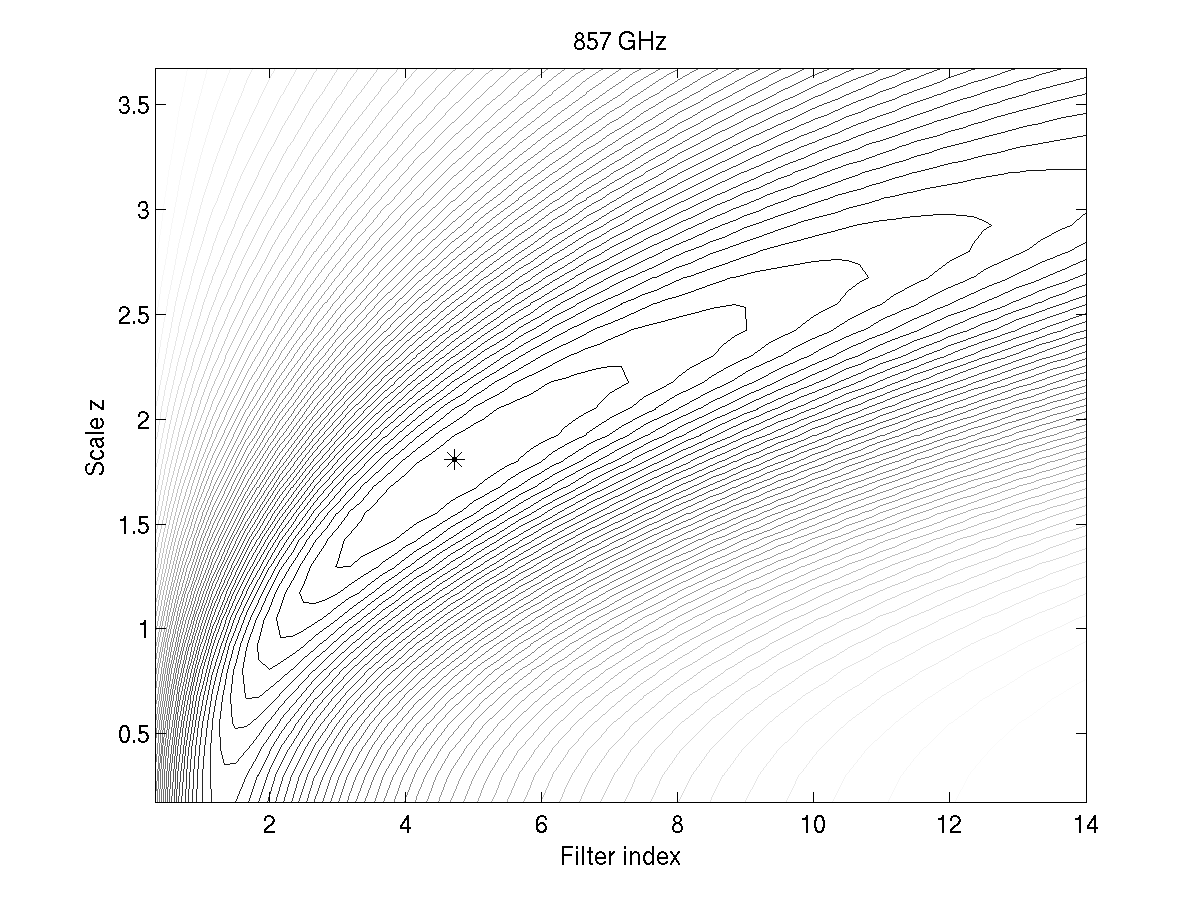}
\includegraphics[width=0.32\textwidth]{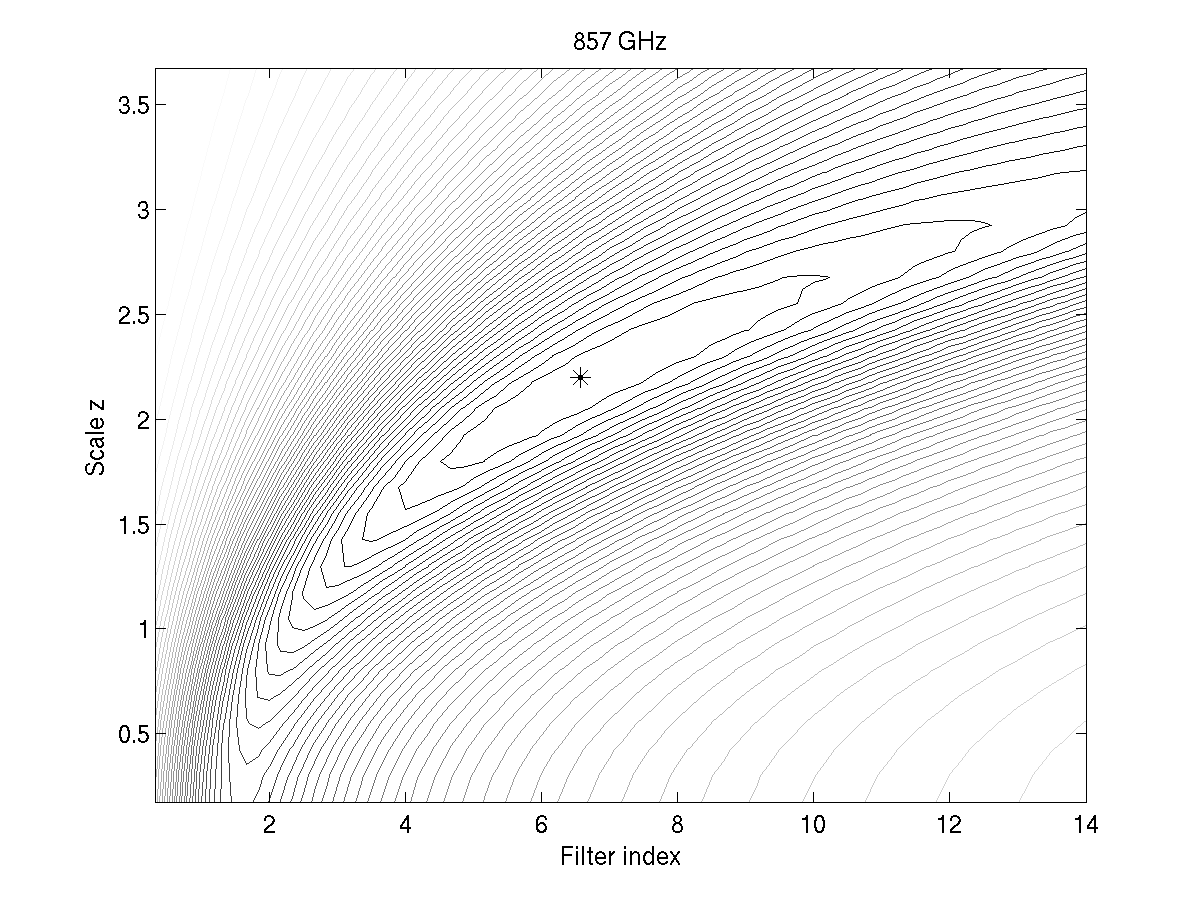}
\caption{Point source amplification as a function of the \iindex $g$ vs. the \sscale $z=R/\sigma_b$ of the filter for a selection of regions. The upper, middle and lower panels correspond to the 30, 143 and 857 GHz maps respectively. These panels correspond, from left to right, to the regions 1, 2 and 3. Every contour line represents 1/25th of the maximum amplification.}
\label{fig:amplif_g_R}
\end{center}
\end{figure*}

\begin{figure*}
\begin{center}
\includegraphics[width=0.32\textwidth]{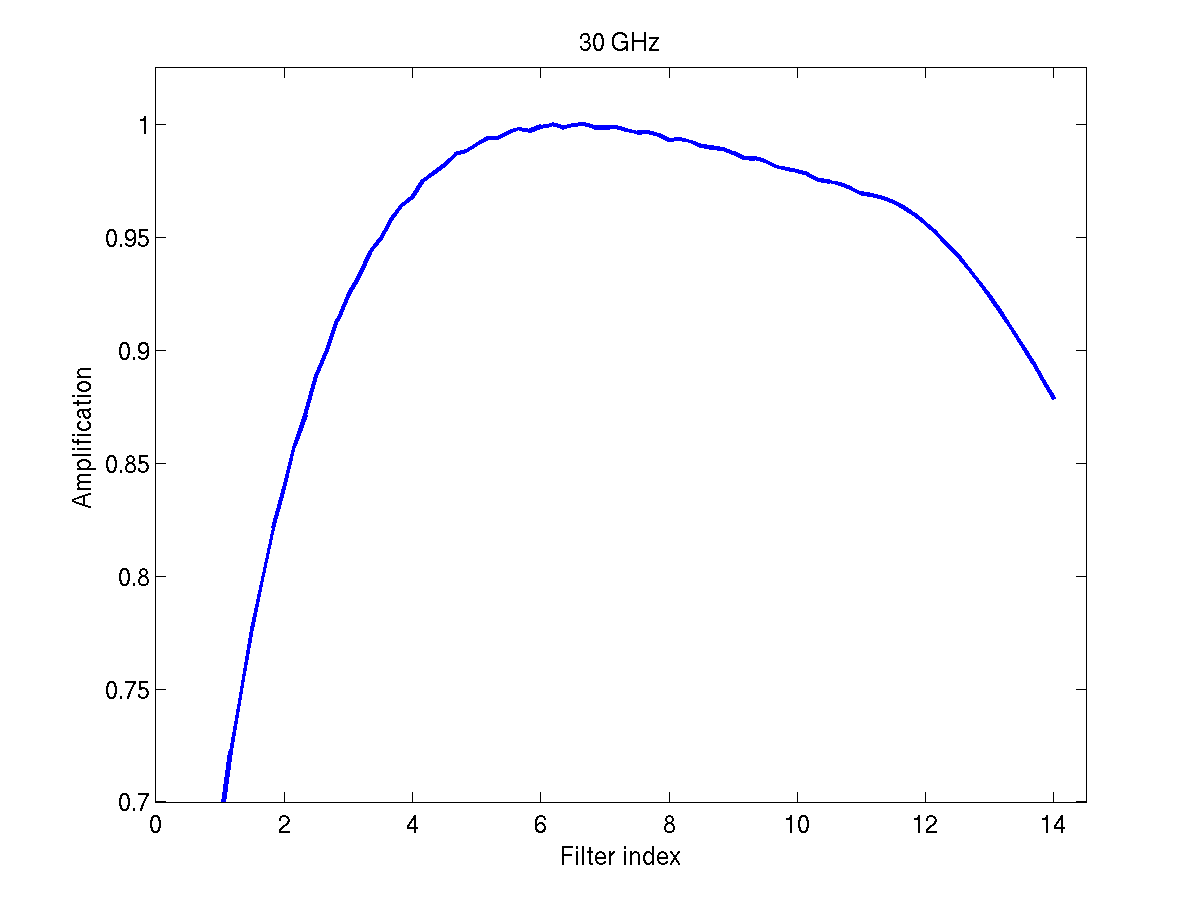}
\includegraphics[width=0.32\textwidth]{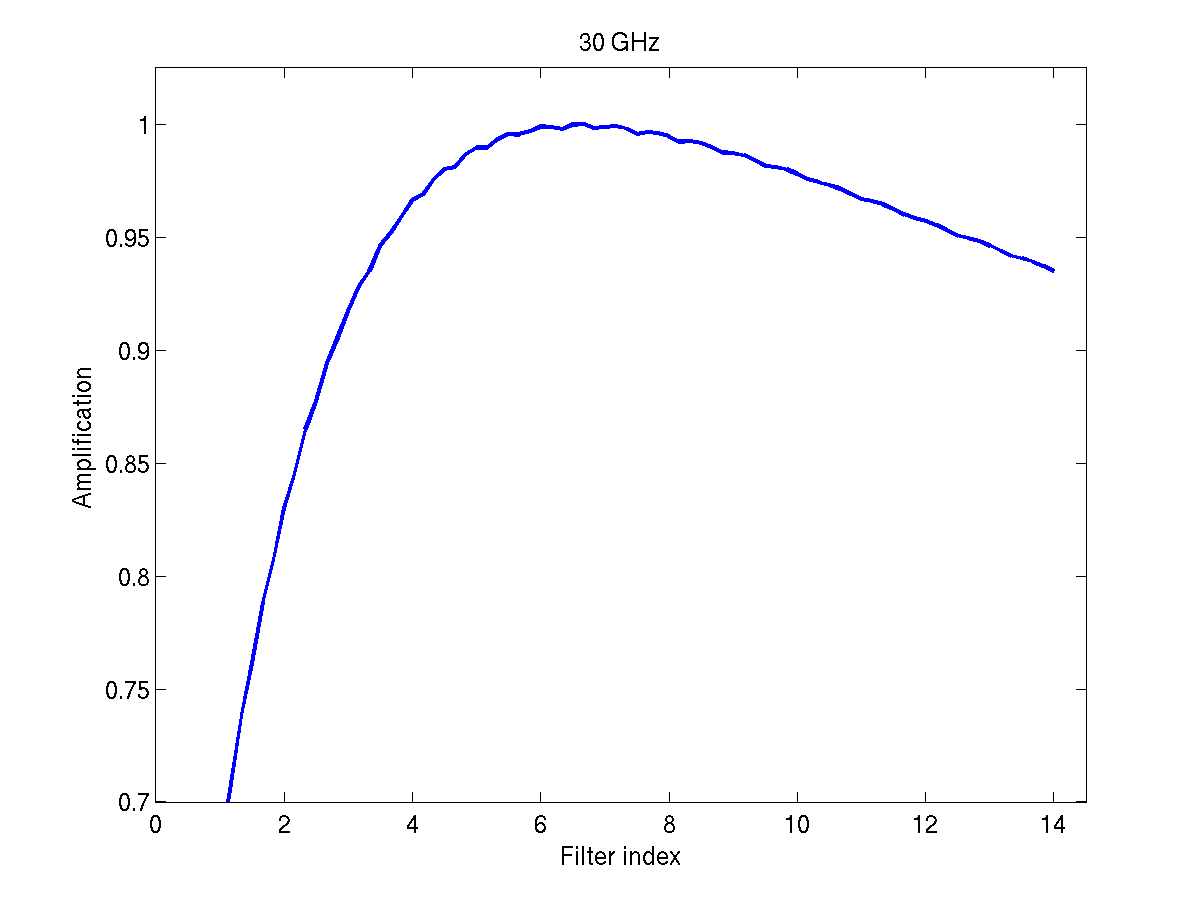}
\includegraphics[width=0.32\textwidth]{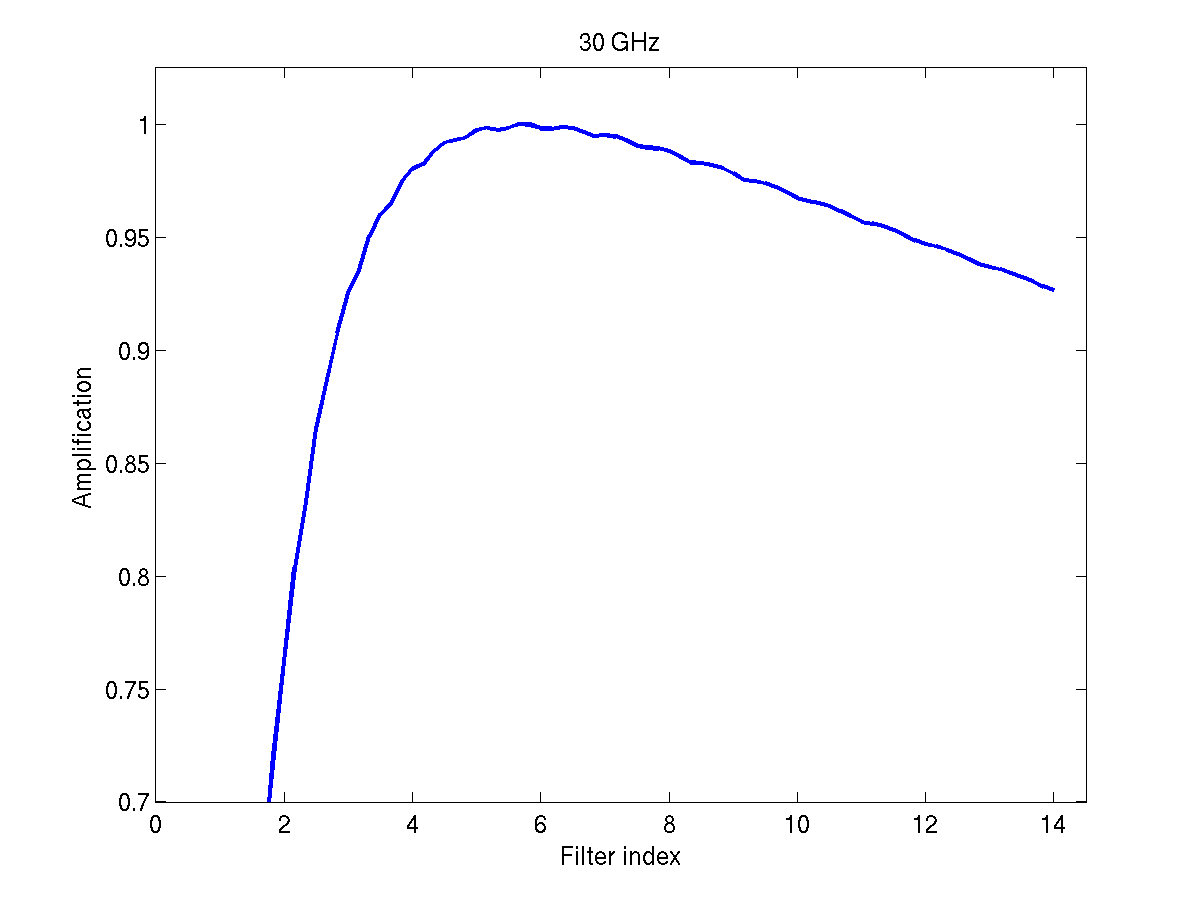}
\includegraphics[width=0.32\textwidth]{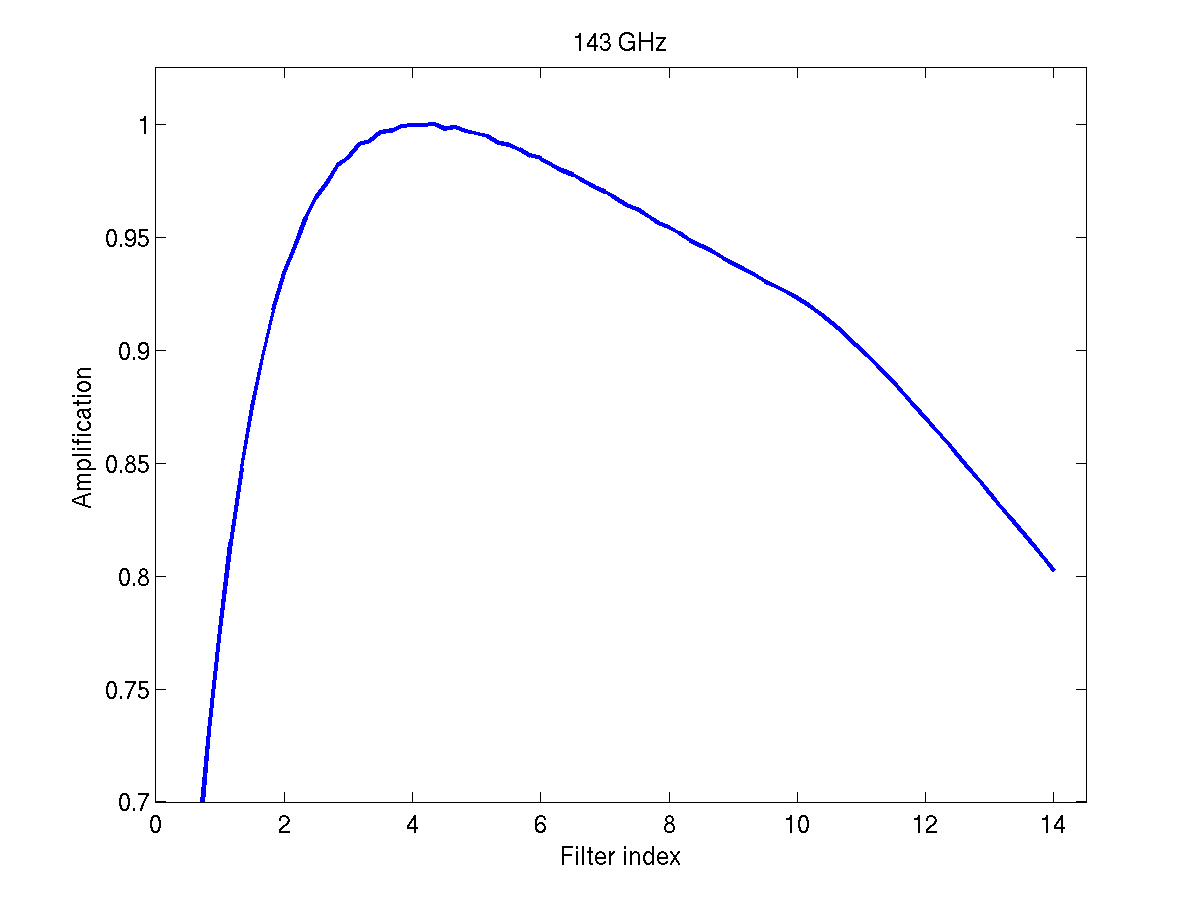}
\includegraphics[width=0.32\textwidth]{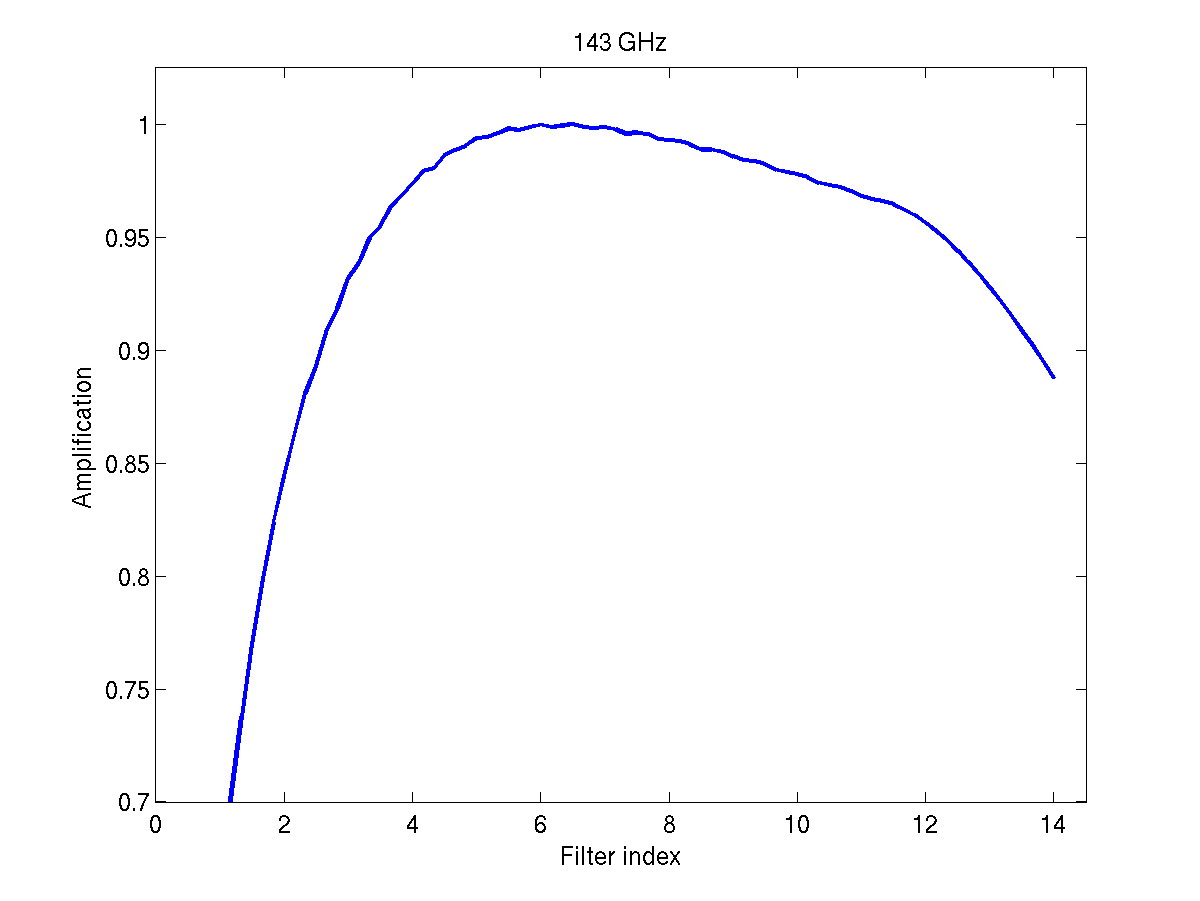}
\includegraphics[width=0.32\textwidth]{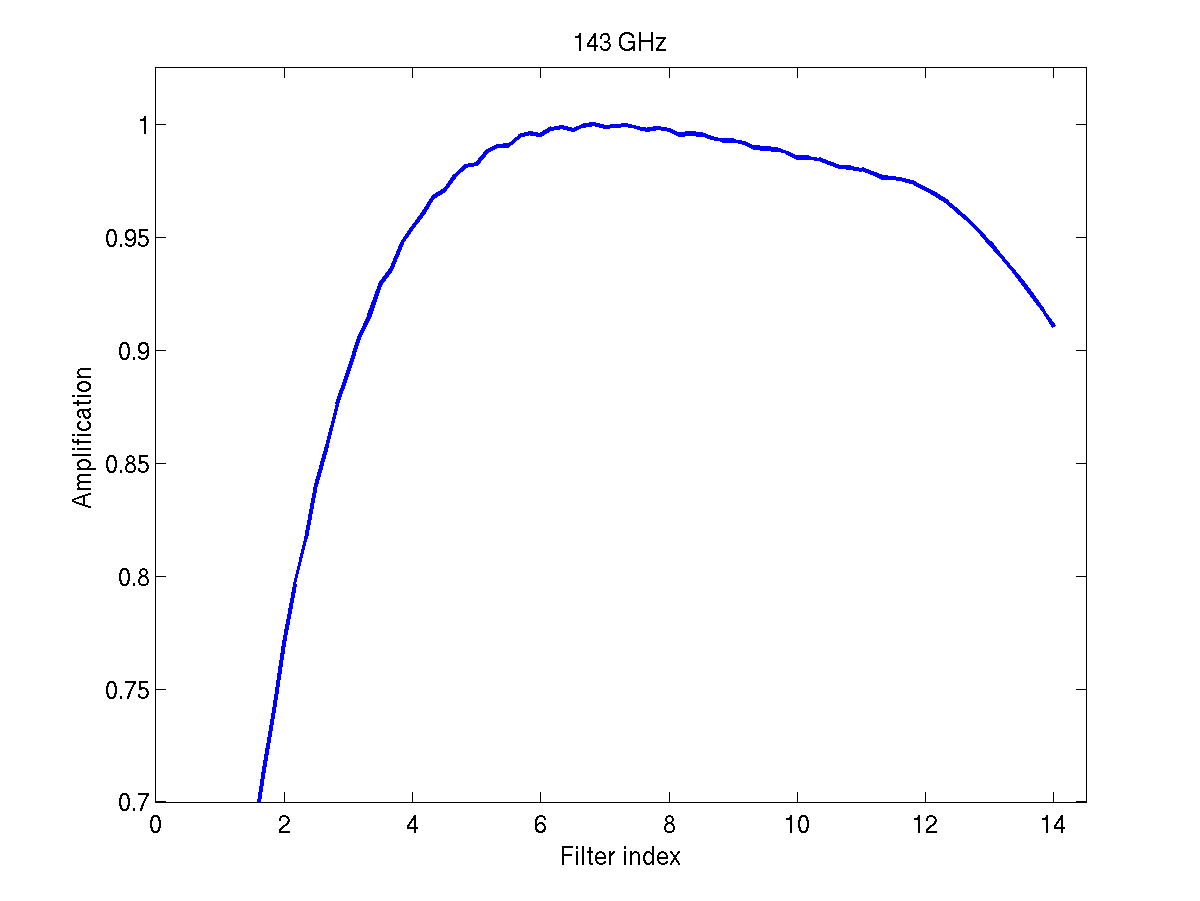}
\includegraphics[width=0.32\textwidth]{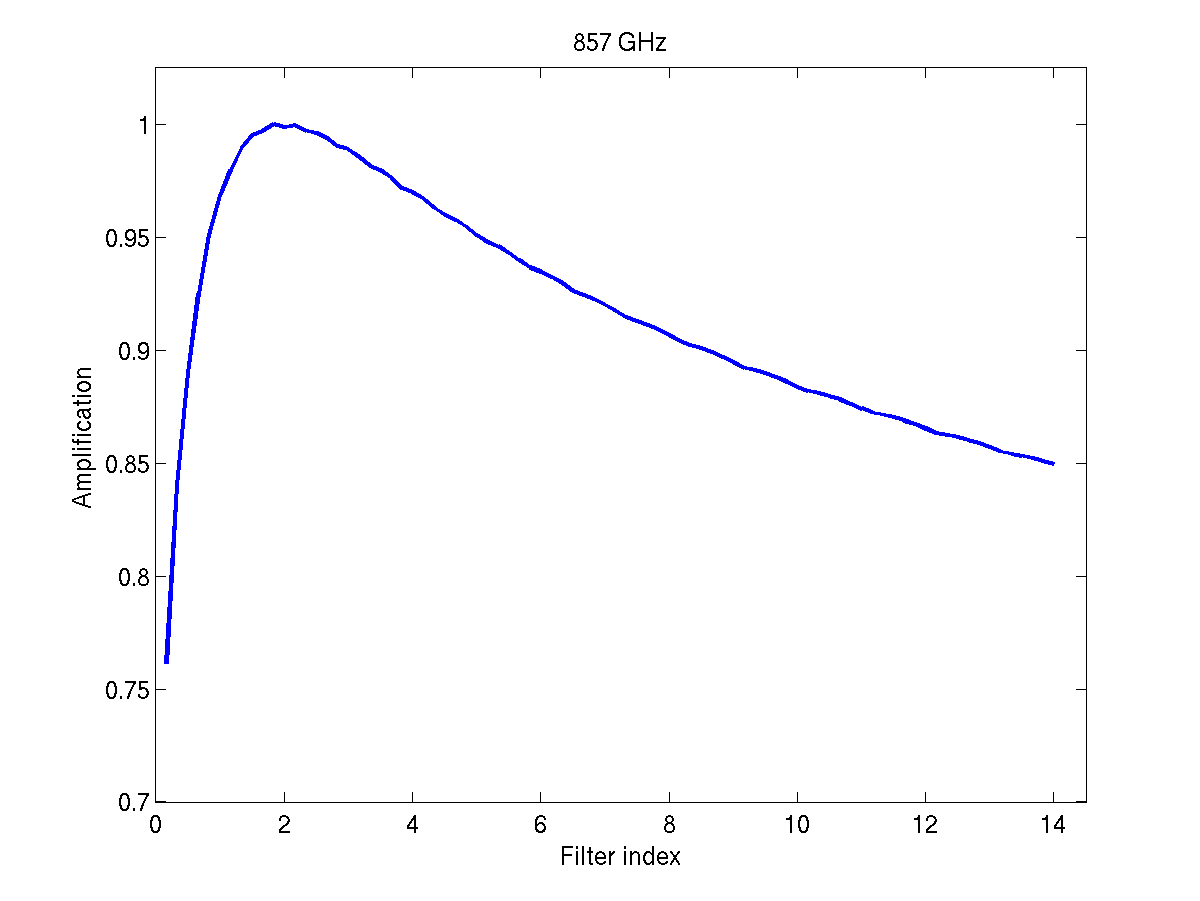}
\includegraphics[width=0.32\textwidth]{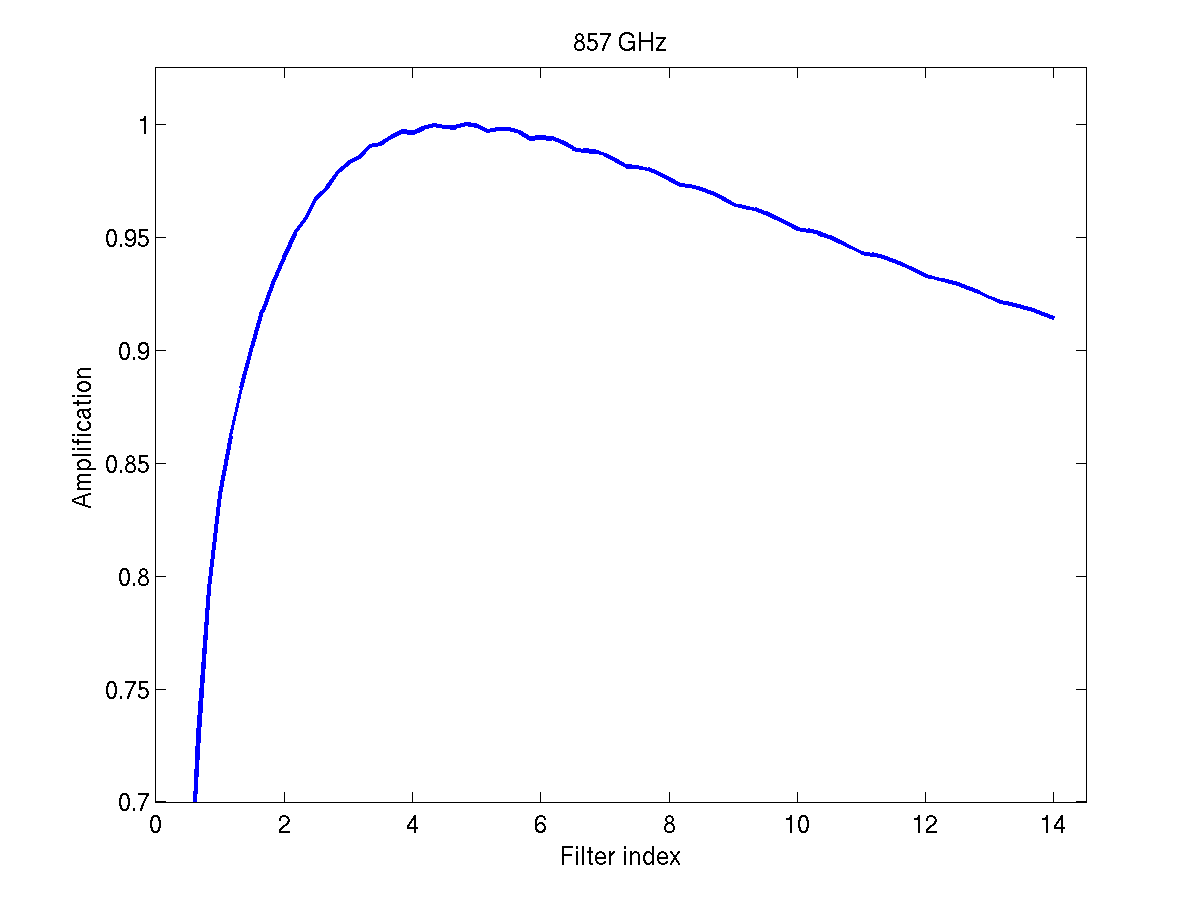}
\includegraphics[width=0.32\textwidth]{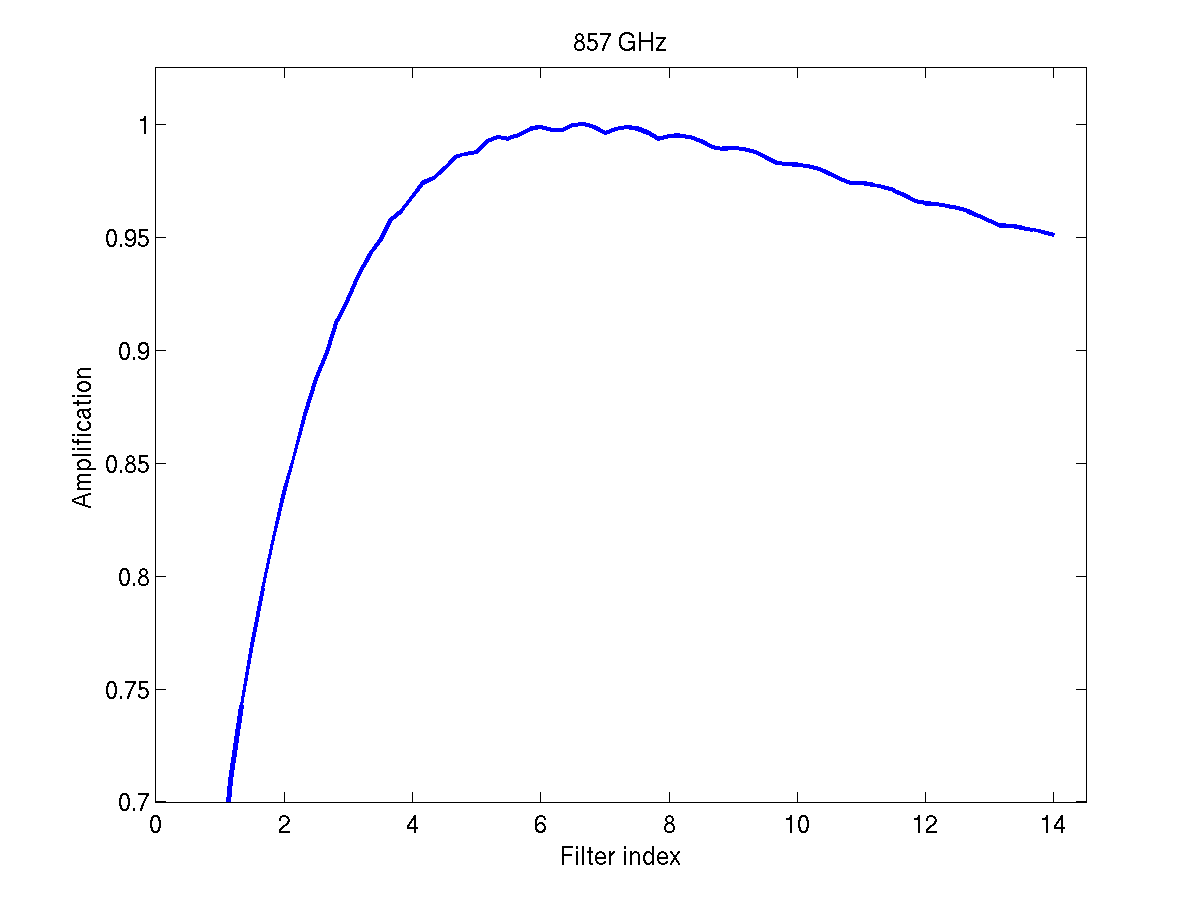}
\caption{Point source amplification as a function of the as a function of the \iindex $g$, conditioned on the optimal \sscale $R$. The upper, middle and lower panels correspond to  the 30, 143 and 857 GHz maps respectively. These panels correspond, from left to right, to the regions 1, 2 and 3. Note that the amplification has been normalized to the maximal amplification in each case.}
\label{fig:amplif}
\end{center}
\end{figure*}

To further illustrate this exercise, in Figure \ref{fig:amplif_g_R}
we show, for the selected regions, the amplification as a function of
the \sscale $R$ and \iindex $g$ of the filter. Since the behaviour of
the amplification as a function of the optimal \sscale $R$ has been
already studied in previous works
\citep[e.g.][]{vielva01,lopezcaniego06}, we will concentrate on the
properties of the \iindex $g$. In this respect, in Figure
\ref{fig:amplif} we show the results of the same analysis but
representing the amplification as function of the \iindex $g$,
conditioned on the optimal \sscale $R$. As we mentioned above, for
each region there is a combination of $R$ and $g$ that produces a
single maximum in the amplification. In particular, if we look to the
upper left panel of Figure \ref{fig:amplif} corresponding to the
region 1 at 30 GHz, the maximum in the amplification is reached for
$g=6.66$. If we use, instead, the standard MHW at the optimal \sscale, $g=2$, the amplification is only $82\%$ of
that of the maximum. If we compare the MHW2 at the optimal \sscale,
$g=4$, the amplification is $\sim97\%$. If we look at one of the regions
for the 143 GHz case, for example the middle central panel of Figure
\ref{fig:amplif}, the maximum appears at $g=6.5$. In this case we can
see a similar behaviour for the MHW at the optimal \sscale as before,
the amplification is $\sim83\%$ of that of the maximum and for the MHW2 at
the optimal \sscale the amplification is $\sim97\%$. More
interesting is the case of 857 GHz, where the maximum in $g$ changes
significantly from one region of the sky to another. In this
frequency, the maximum of the lower left panel of Figure
\ref{fig:amplif} is at $g=1.83$ and the MHW and MHW2 at the optimal
\sscale produce an amplification of $\sim95\%$ and $\sim100\%$,
respectively. Then, the maximum in the lower central panel is at
$g=4.33$, and the MHW and MHW2 at the optimal \sscale produce an
amplification of $\sim94\%$ and $\sim100\%$, respectively. Finally, the maximum
in the lower right panel is found at $g=6.66$, and the MHW and MHW2 at
the optimal \sscale produce an amplification of $\sim82\%$ and $\sim97\%$,
respectively.
\begin{table}
\begin{center}
\begin{tabular}{|c|c|c|c|c|c|}
\hline Region & GLAT & GLON & $g_{30}$ & $g_{143}$ & $g_{857}$ \\ 
\hline 1 & +60 & 70 & 6.66 & 4.33 & 1.83 \\ 
\hline 2 & -36 & 290 & 6.66 & 6.50 & 4.83 \\ 
\hline 3 & +01 & 123 & 5.66 & 6.83 & 6.66 \\ 
\hline 
\end{tabular} 
\caption{This table shows the galactic coordinates of the centres of the three regions that we have selected visually in the simulated maps at 30, 143 and 857 GHz in order to have increasing background complexities. We also show the optimal value of the \iindex of the filter $g$ that we have found in each case.} 
\label{tab:tabla_res_parches}
\end{center}
\end{table}

\subsection{All-sky analysis}
\begin{figure*}
 \begin{center}
\includegraphics[width=0.32\textwidth]{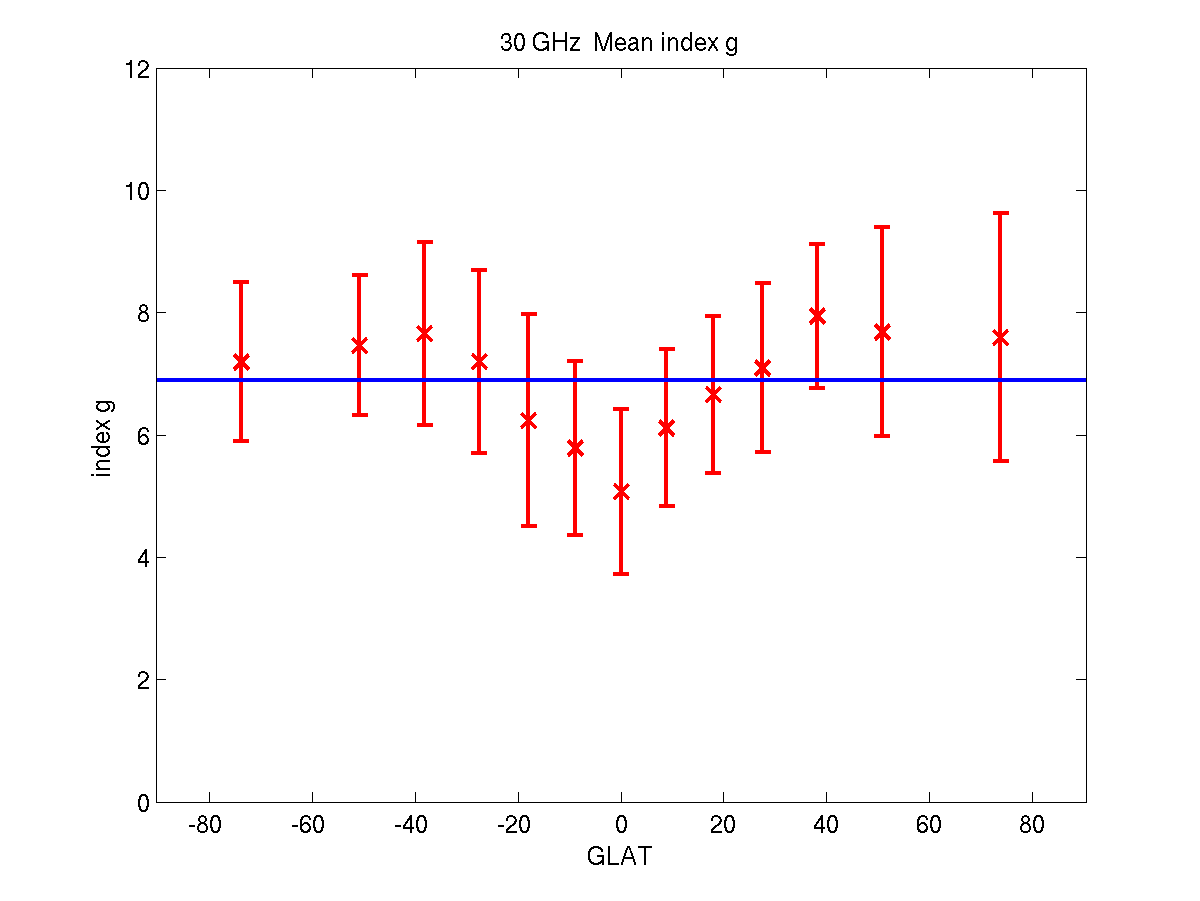}
\includegraphics[width=0.32\textwidth]{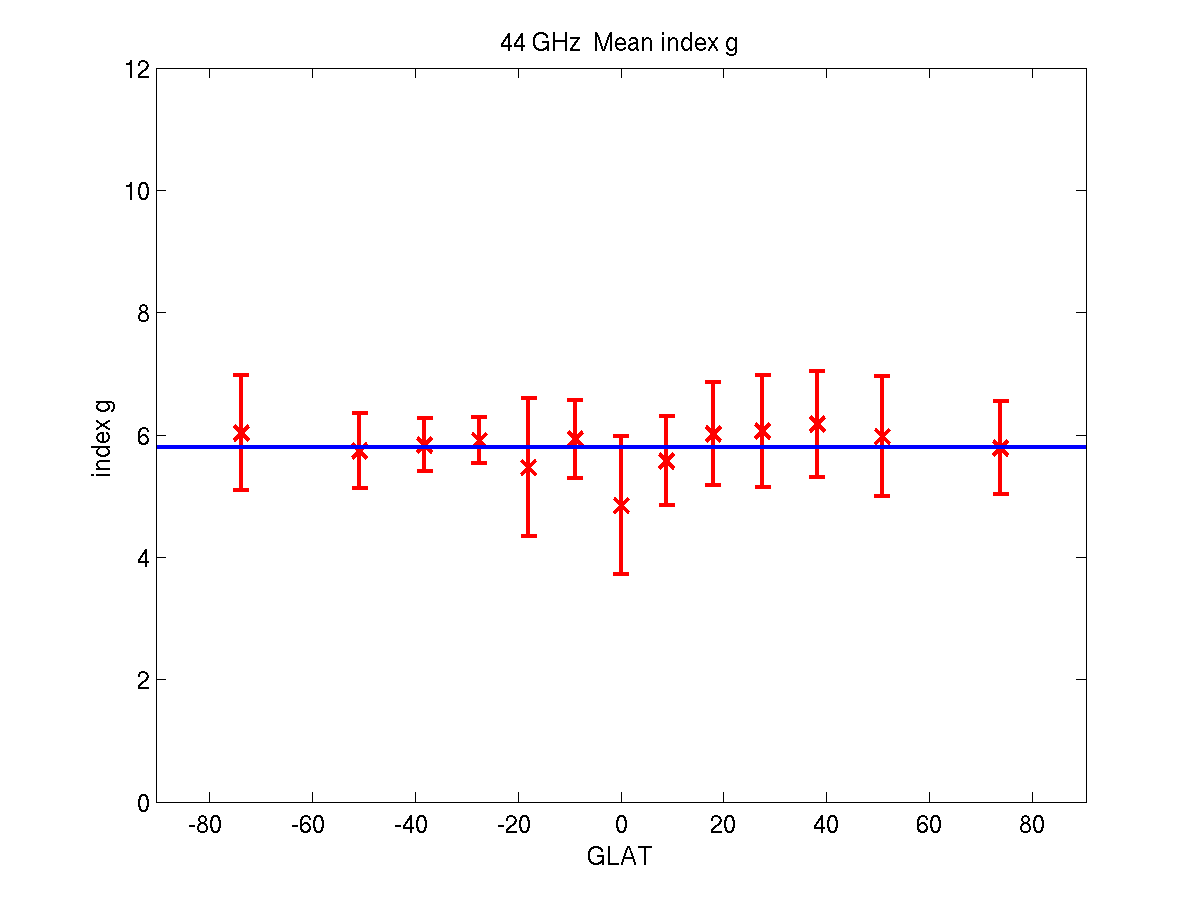}
\includegraphics[width=0.32\textwidth]{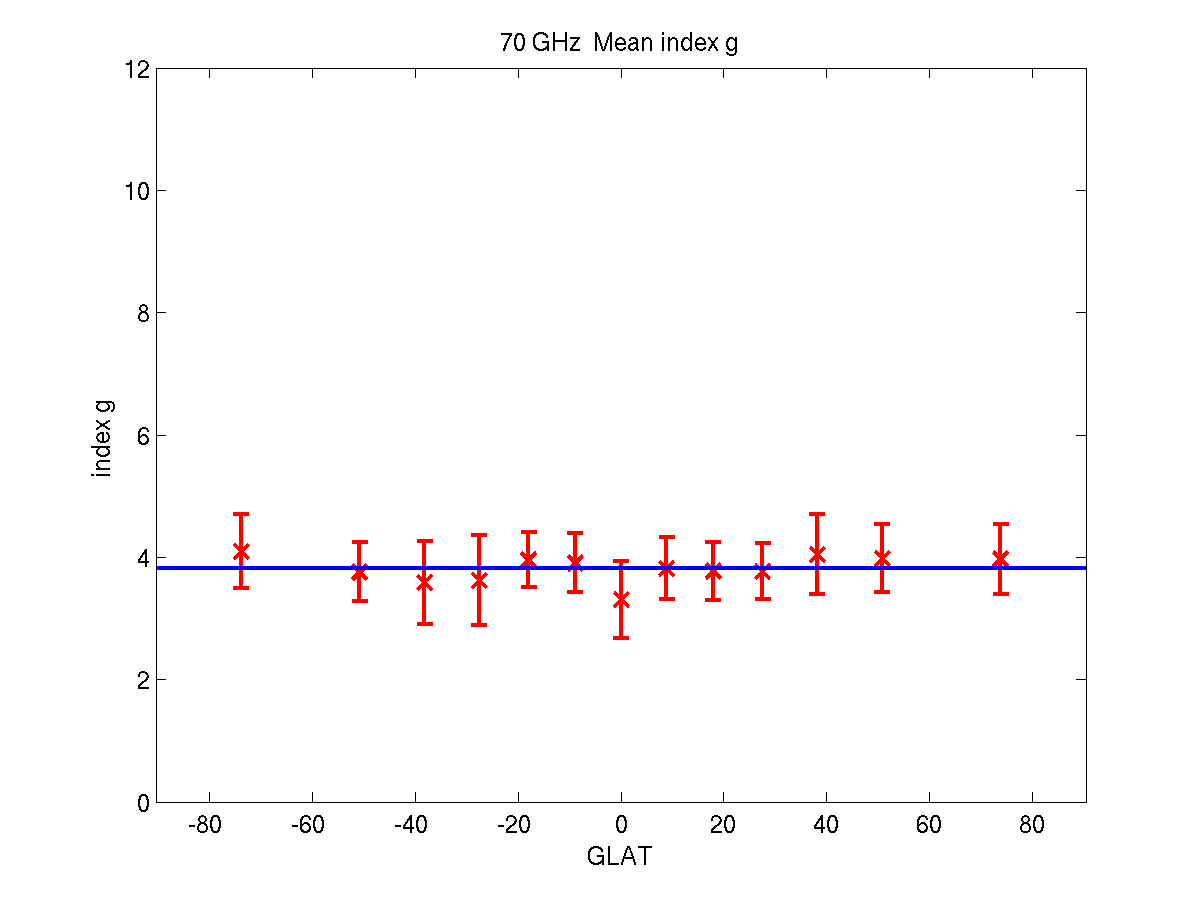}
\includegraphics[width=0.32\textwidth]{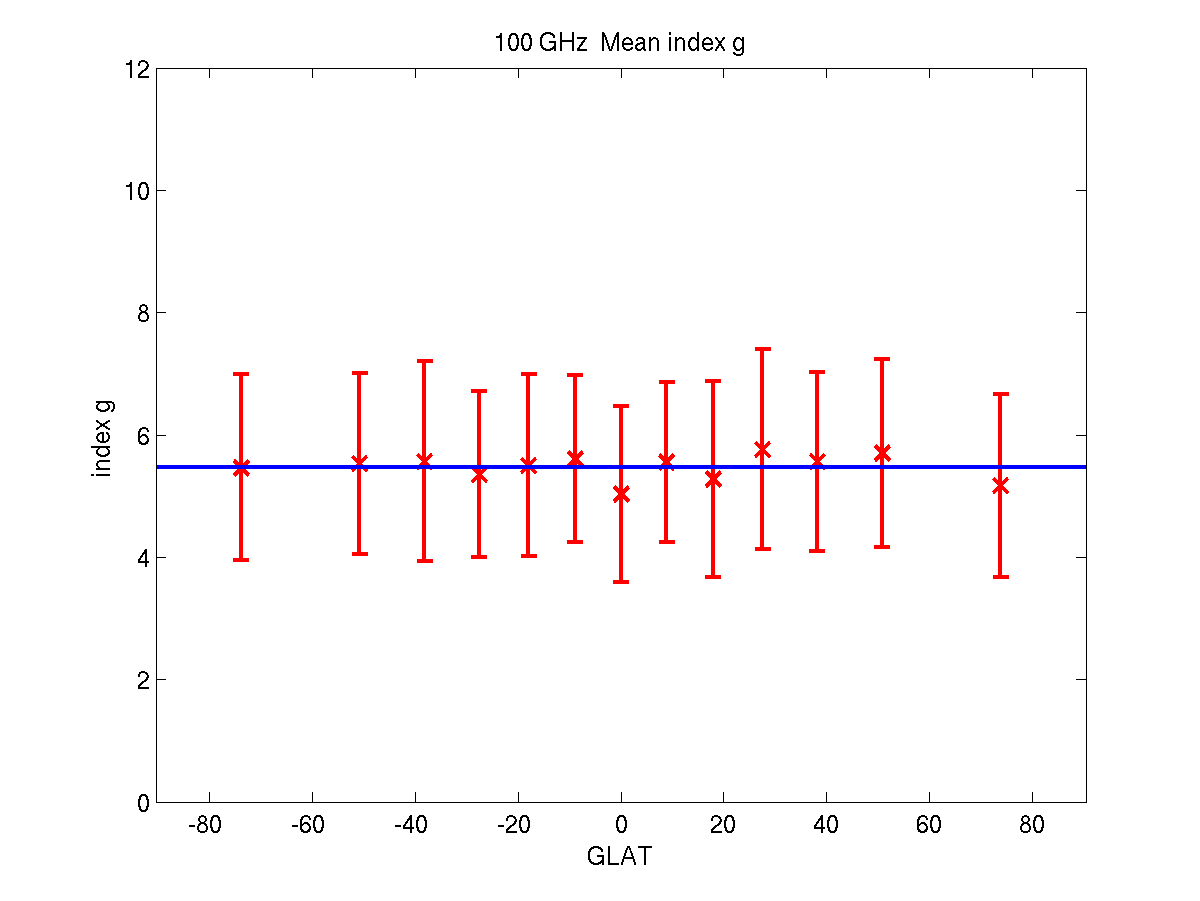}
\includegraphics[width=0.32\textwidth]{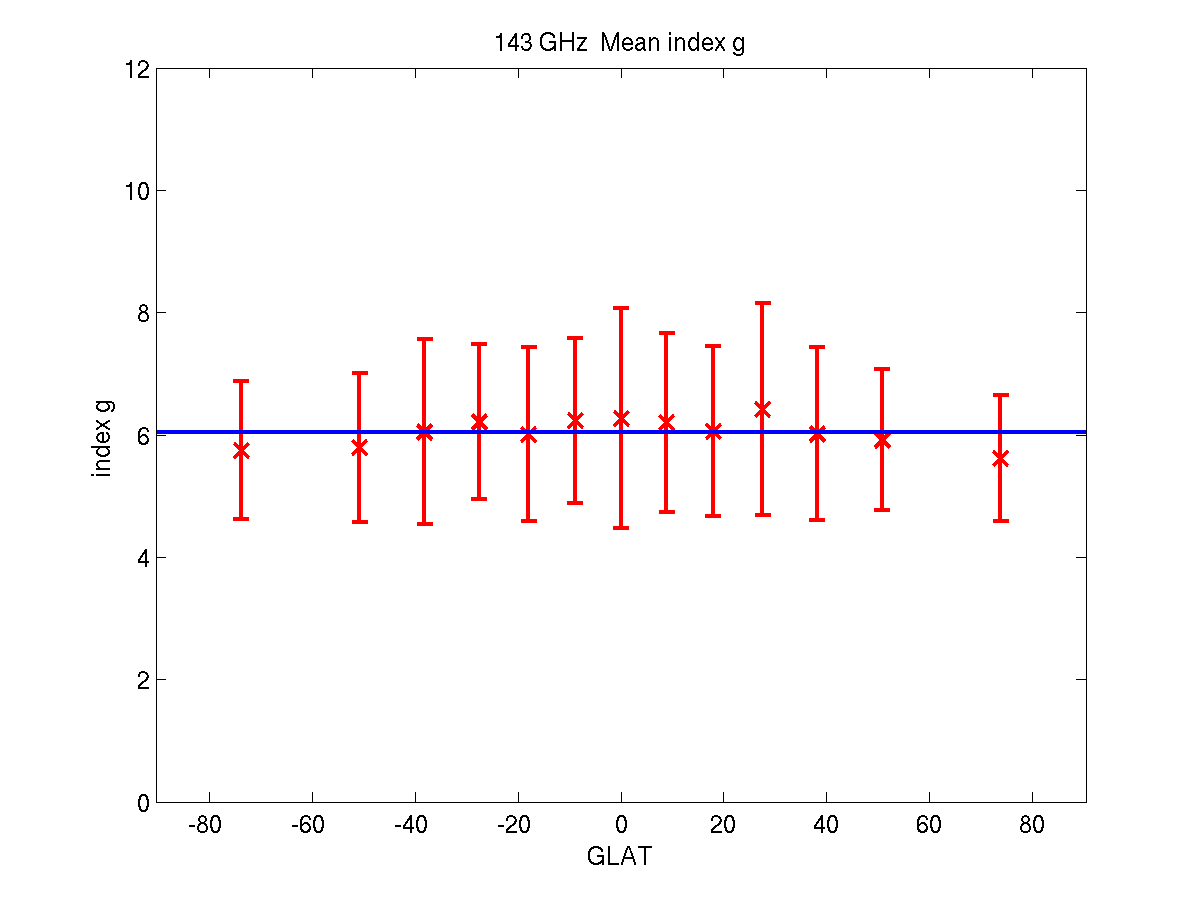}
\includegraphics[width=0.32\textwidth]{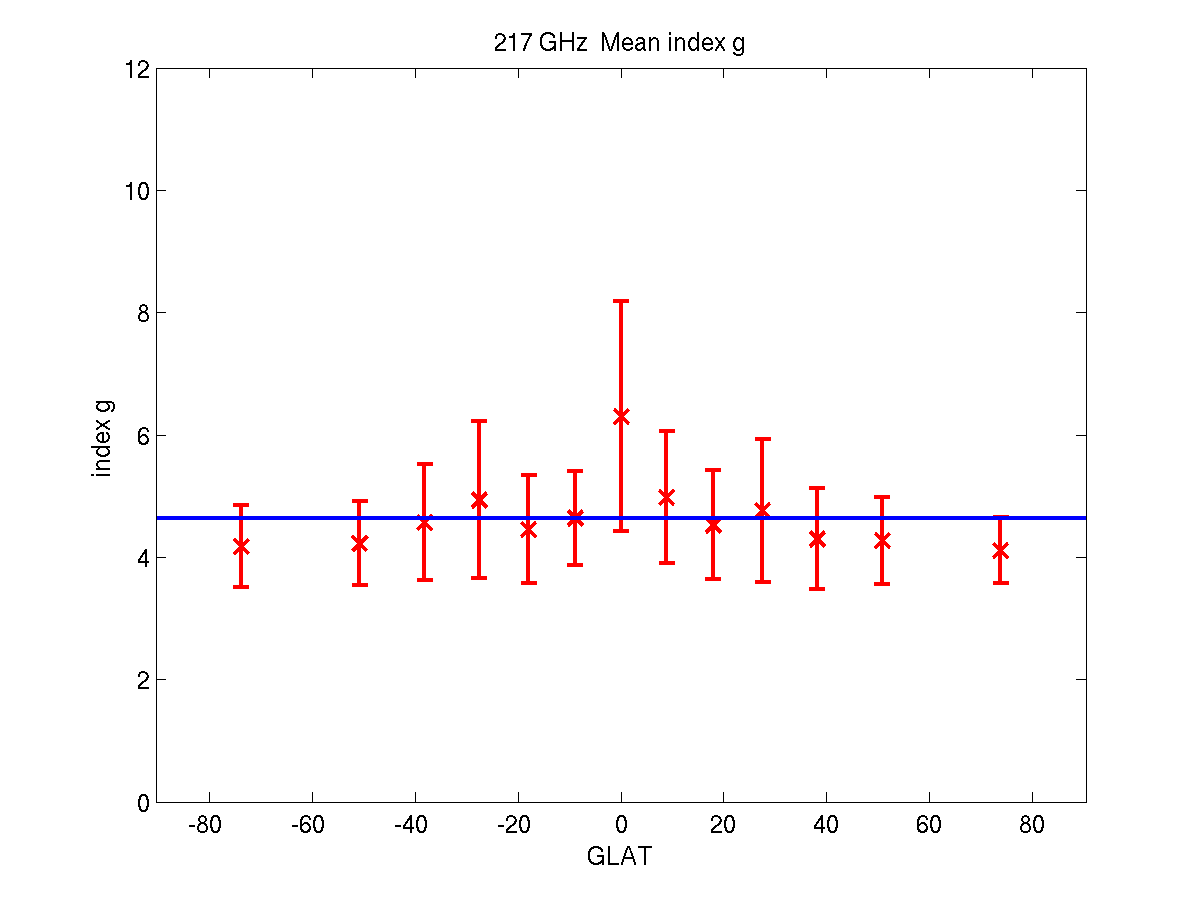}
\includegraphics[width=0.32\textwidth]{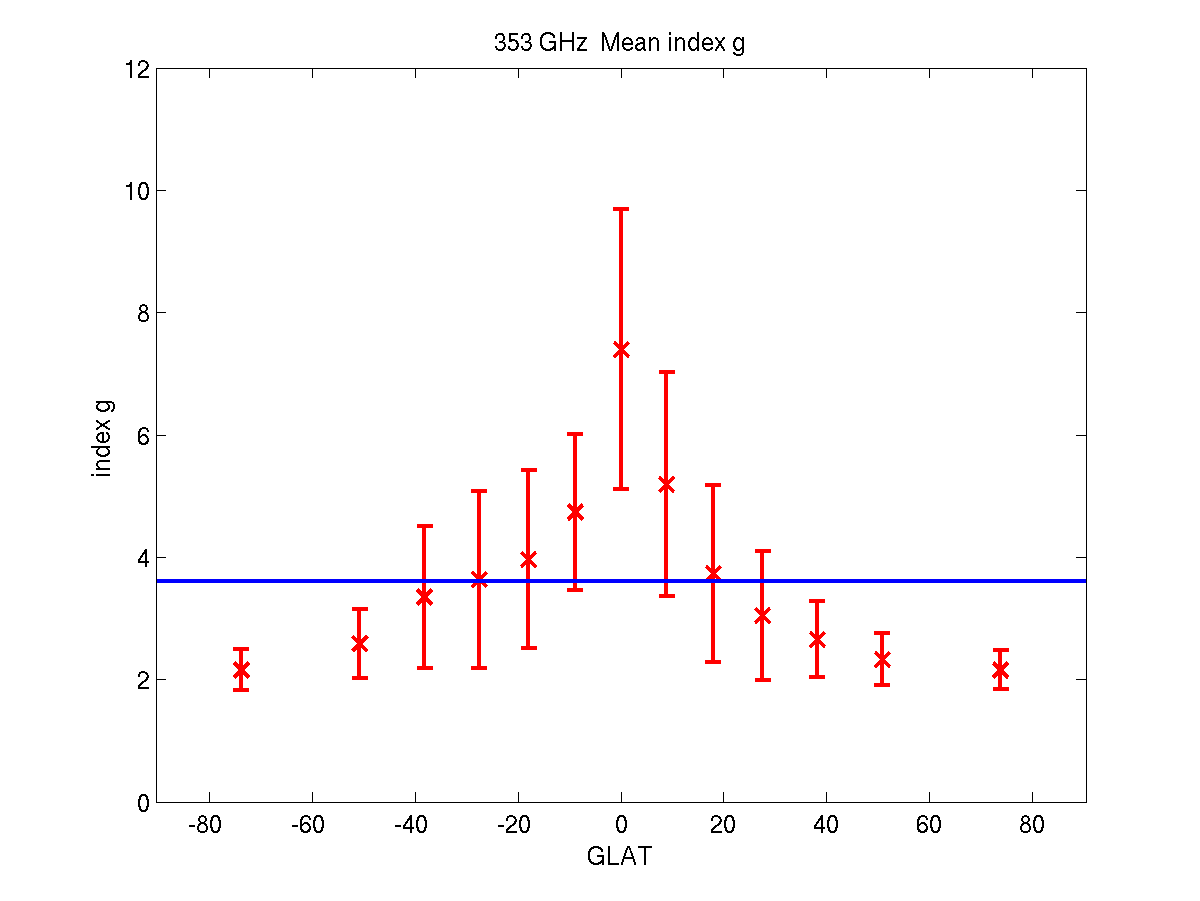}
\includegraphics[width=0.32\textwidth]{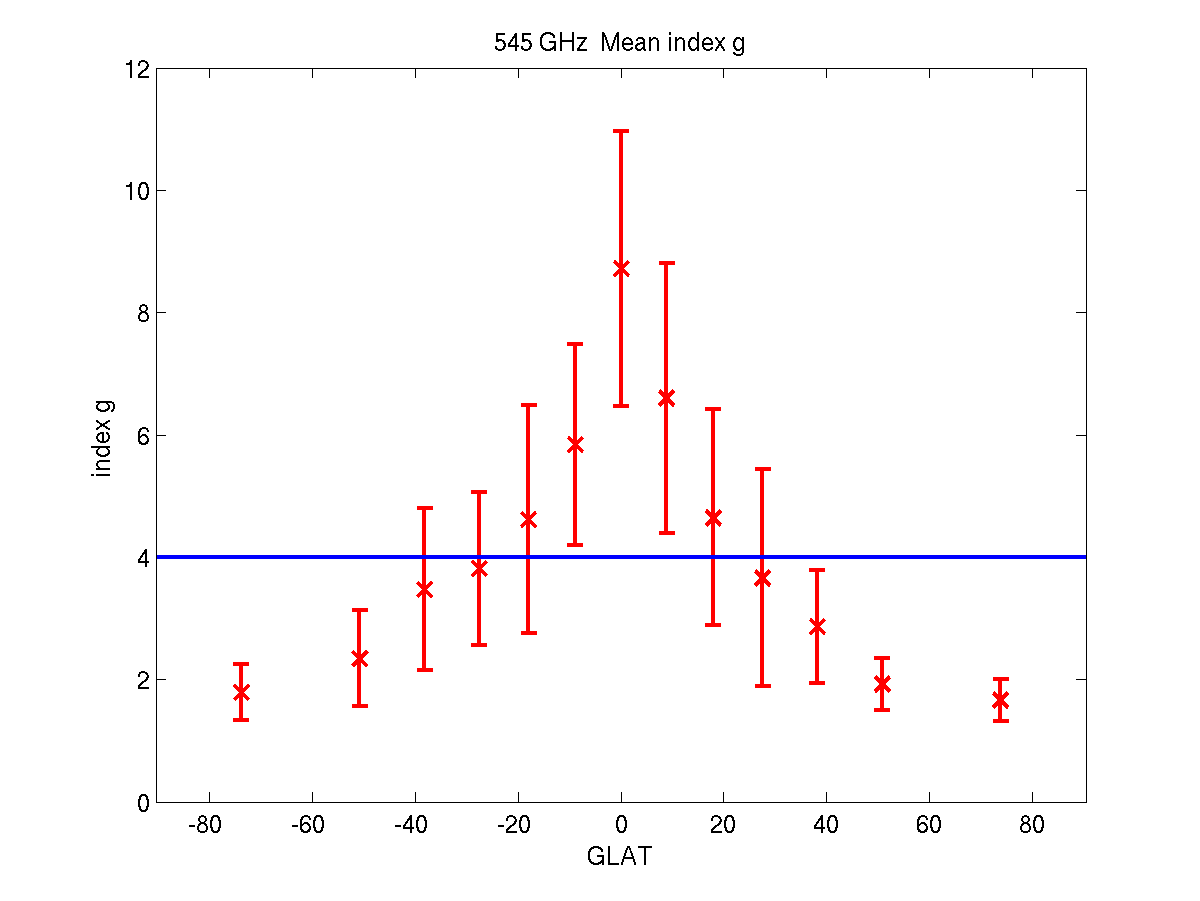}
\includegraphics[width=0.32\textwidth]{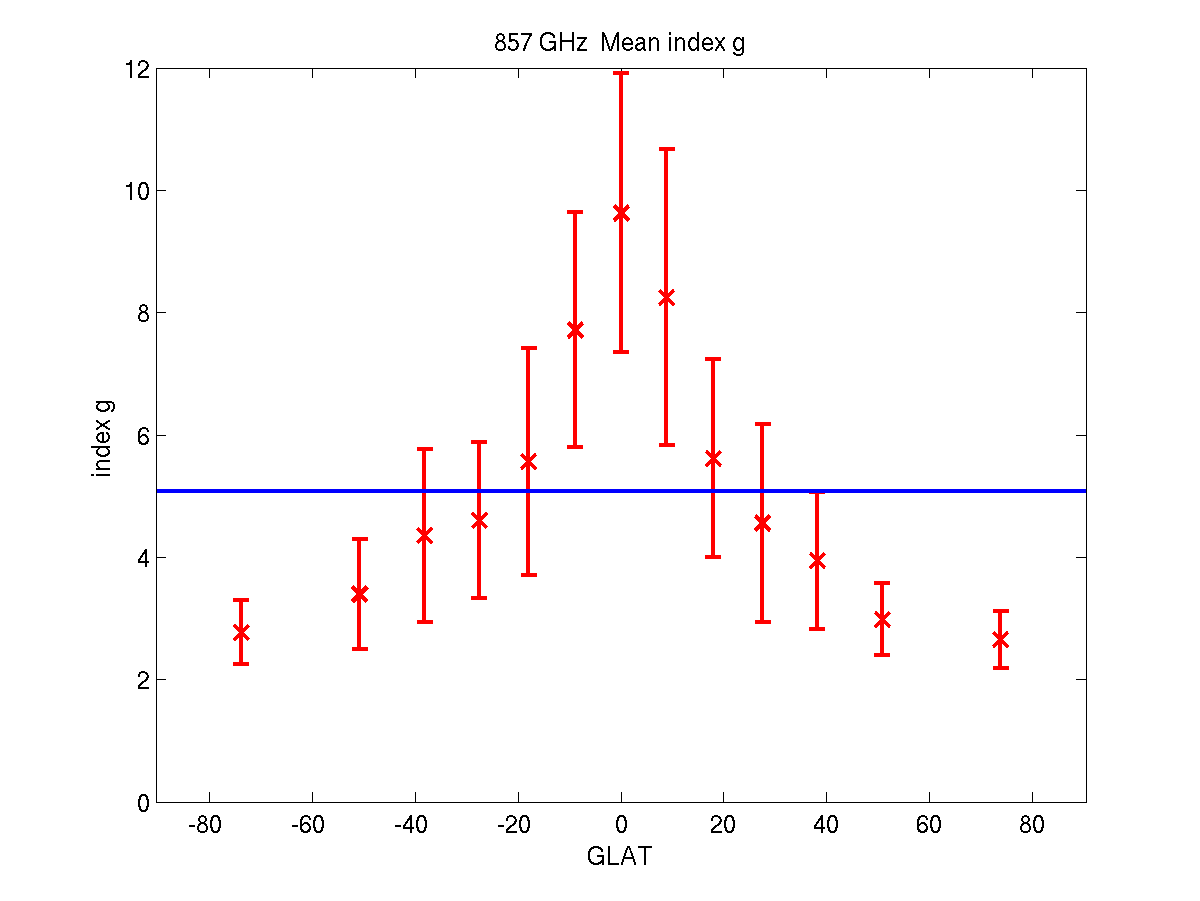}
 \caption{The mean filter \iindex $g$ per the galactic latitude bin and its dispersion are shown. The galactic latitude bins have been chosen to cover an equal area on the sky. In blue the mean \iindex $g$ across the sky is given.}
 \label{fig:mean_index_0}
 \end{center}
 \end{figure*}

In the previous section we have studied the performance of the BAF
optimizing the two parameters, the \sscale of the filter $R$ and the
\iindex of the filter $g$, that produce a maximum in the amplification
of the filtered map as compared with the unfiltered one. This study
was done looking at three visually selected regions in the sky for
each of the three considered frequencies. The results for this
analysis with only nine regions already show that choosing the right
\iindex $g$ and \sscale of the filter $R$ can increase the performance
of the filter significantly. In this section we want to improve the
statistics of the analysis by increasing the percentage of sky and
frequency coverage of the study. For this purpose, we use nine
full-sky simulations between 30 and 857 GHz, dividing each of the nine
simulations into 1344 flat patches that effectively overlap to cover
the $100\%$ of the maps. Then we apply our maximization techniques to
obtain the pair of optimal values $R$ and $g$ that define the filter,
taking into account the local statistics of the background in each 
particular patch. In addition, we make considerations in terms of
galactic latitudes since we know that the complexity of the background
increases when one gets closer to the galactic plane. In Figure
\ref{fig:mean_index_0} one can see the results of this analysis with
respect to the \iindex $g$ for the nine frequency bands. In this figure we show
the average filter \iindex and its dispersion per Galactic latitude
bin, where the bins have been chosen to cover the same area in the
sky.

For the 30 GHz case, the mean of the average filter \iindex is
$\sim7 $, except for the galactic region between [-20,20] where
the synchrotron radiation dominates and the \iindex $g$ decreases a
bit (see upper left panel of Figure \ref{fig:mean_index_0}). Then, the
behaviour of the filter \iindex between 44 and 143 GHz is fairly flat,
although it varies between 4-6. It is important to note that the
behaviour of $g$ starts to change again at 217 GHz and becomes more
and more steep when we get closer to 857 GHz, extending each time to
higher galactic latitudes. This is a clear correlation between the
contribution of the dust emission and the Far-Infrared background to
the background signal in the patches. It reflects how the filter is
adapting to background complexity. This effect can be clearly seen in
the lower panels of Figure \ref{fig:mean_index_0}.
 
 In order to do a qualitative comparison between the MHW2 and the BAF, we crossmatch the catalogs of objects detected with each technique at each frequency and use the common objects to build several sets of figures. 

\begin{figure*}
\begin{center}
\includegraphics[width=0.32\textwidth]{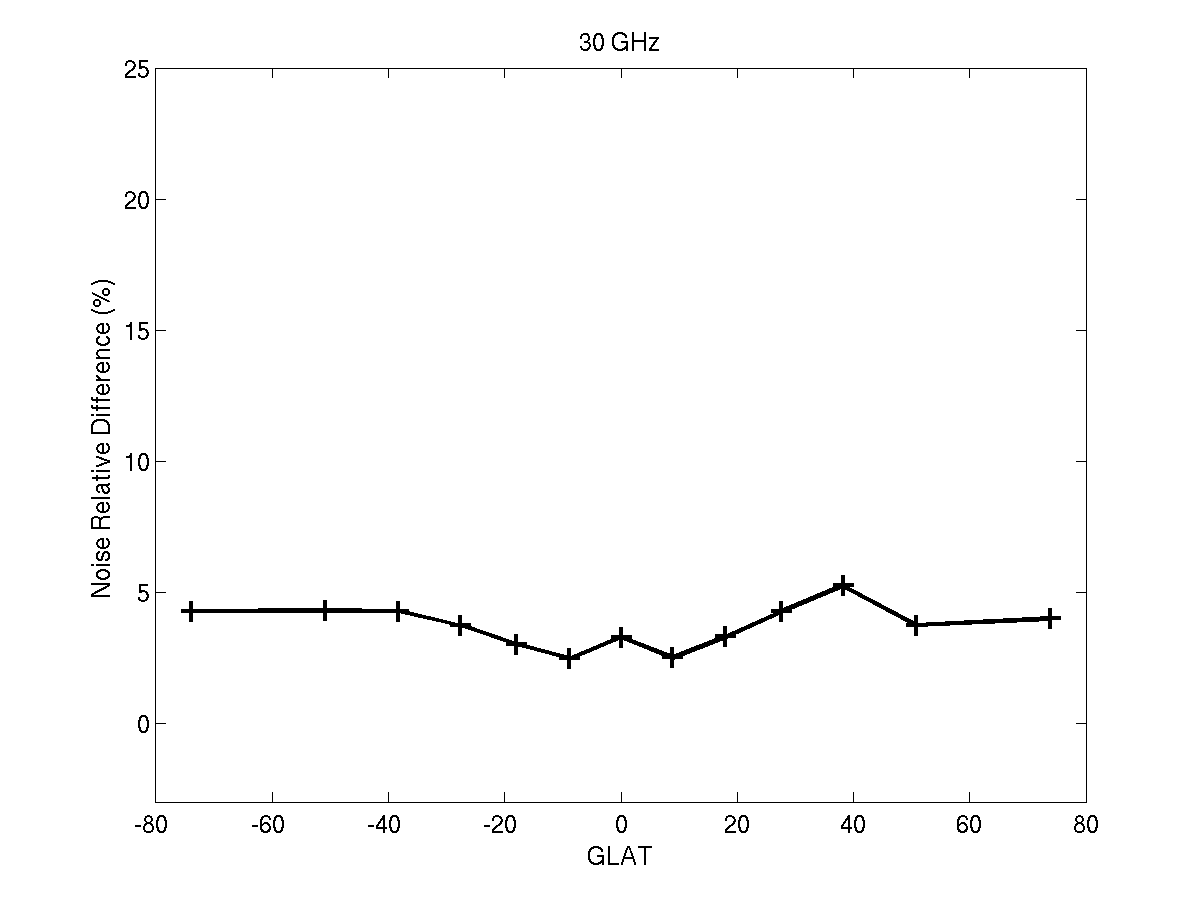}
\includegraphics[width=0.32\textwidth]{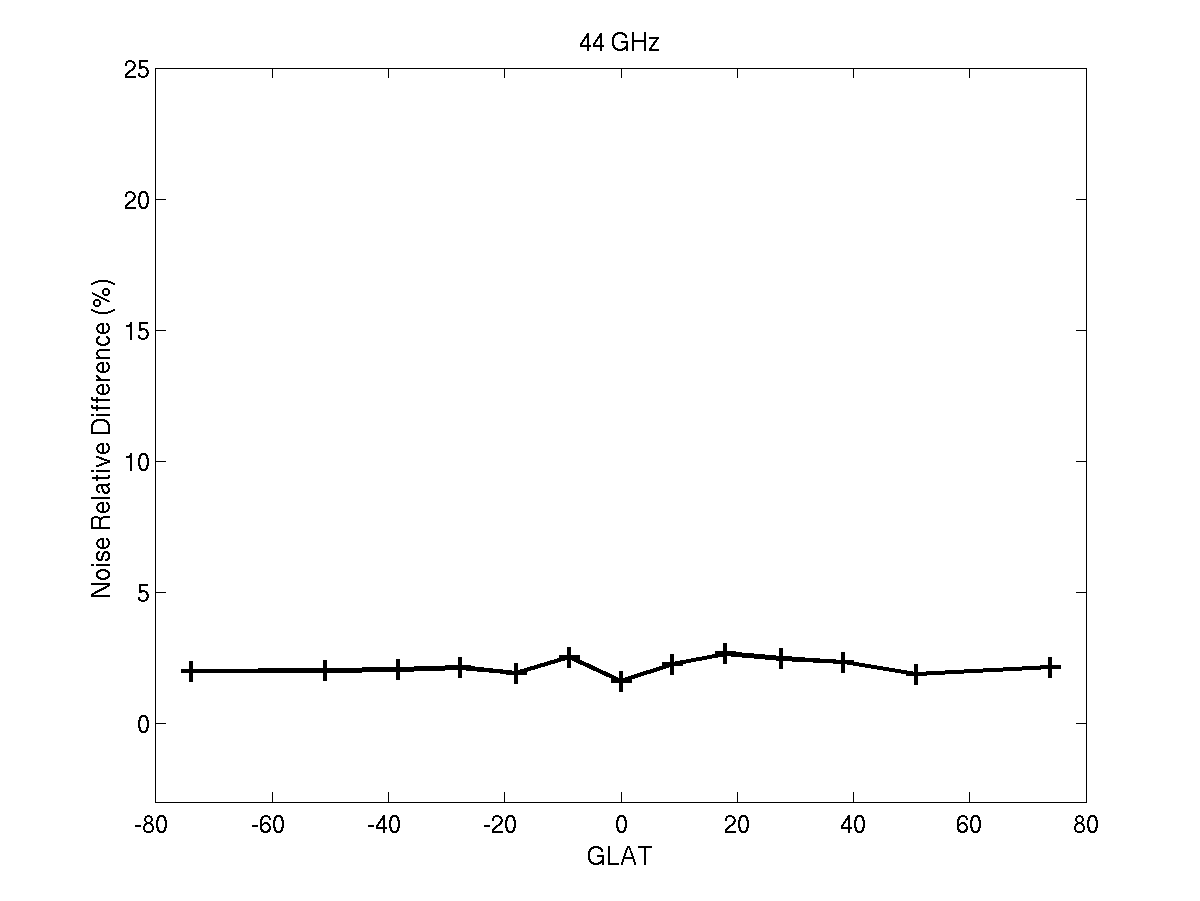}
\includegraphics[width=0.32\textwidth]{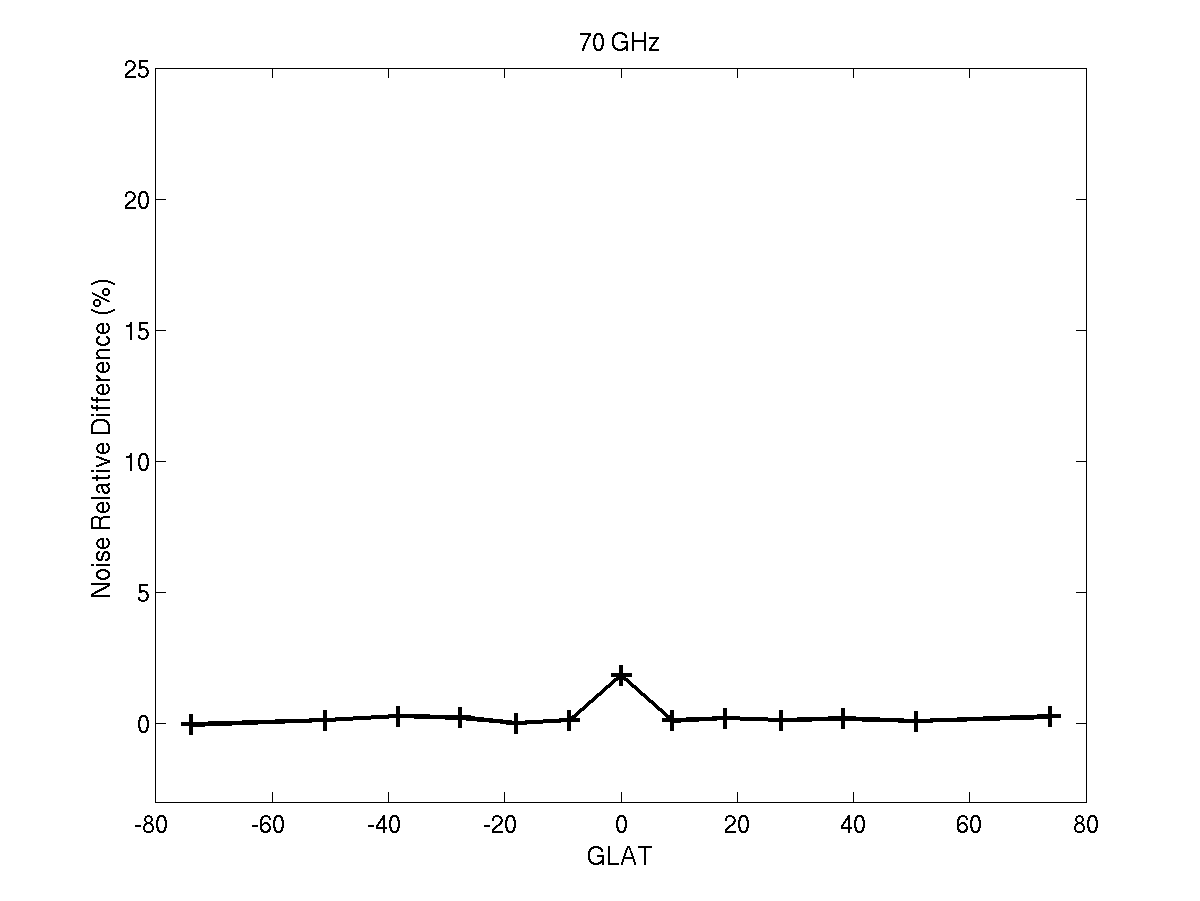}
\includegraphics[width=0.32\textwidth]{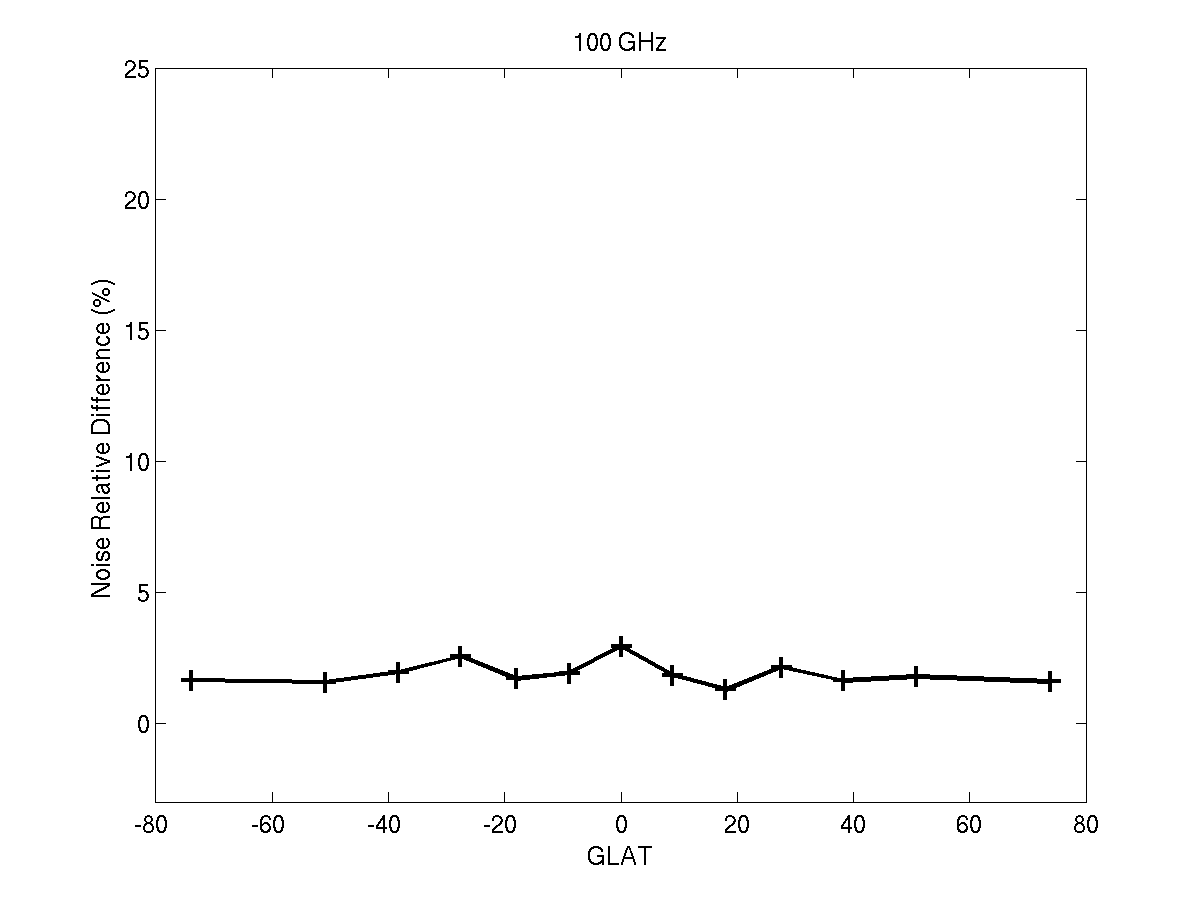}
\includegraphics[width=0.32\textwidth]{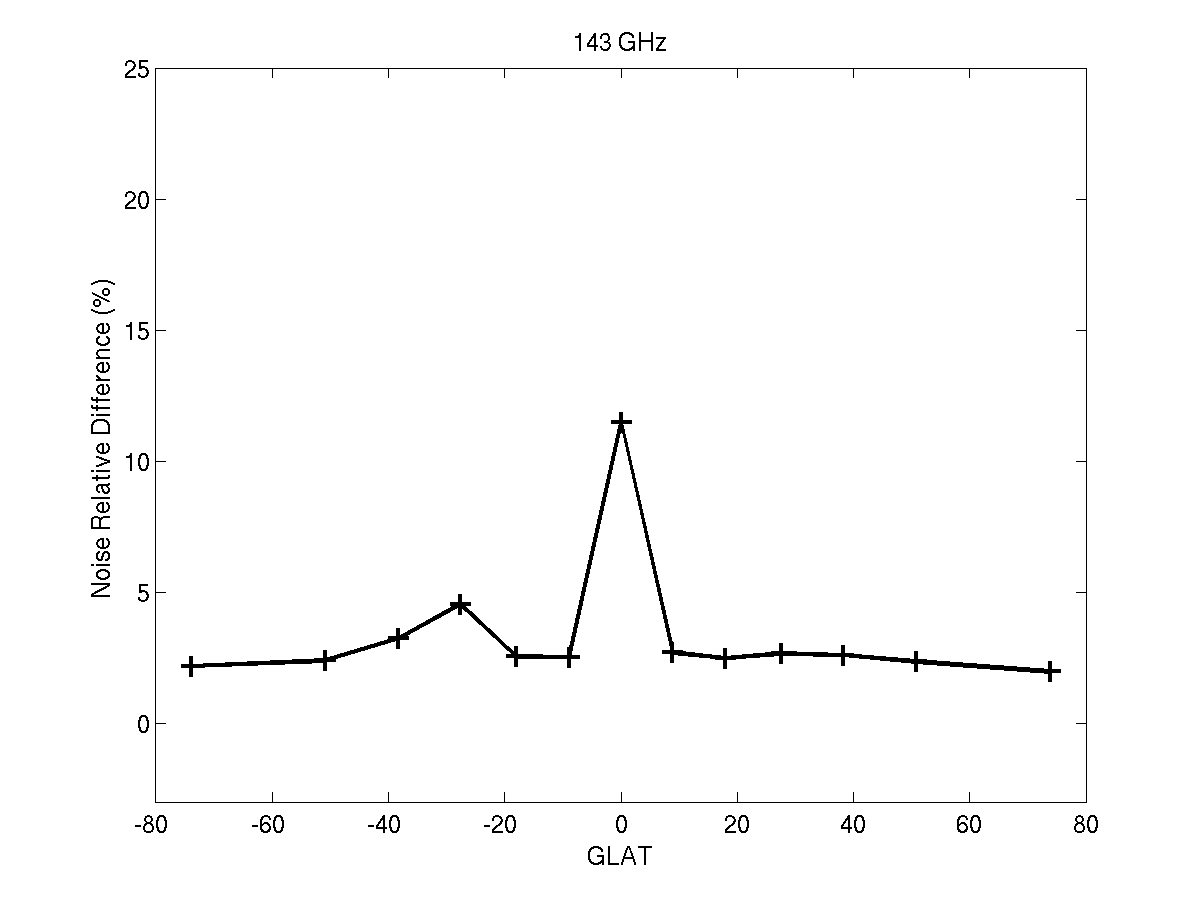}
\includegraphics[width=0.32\textwidth]{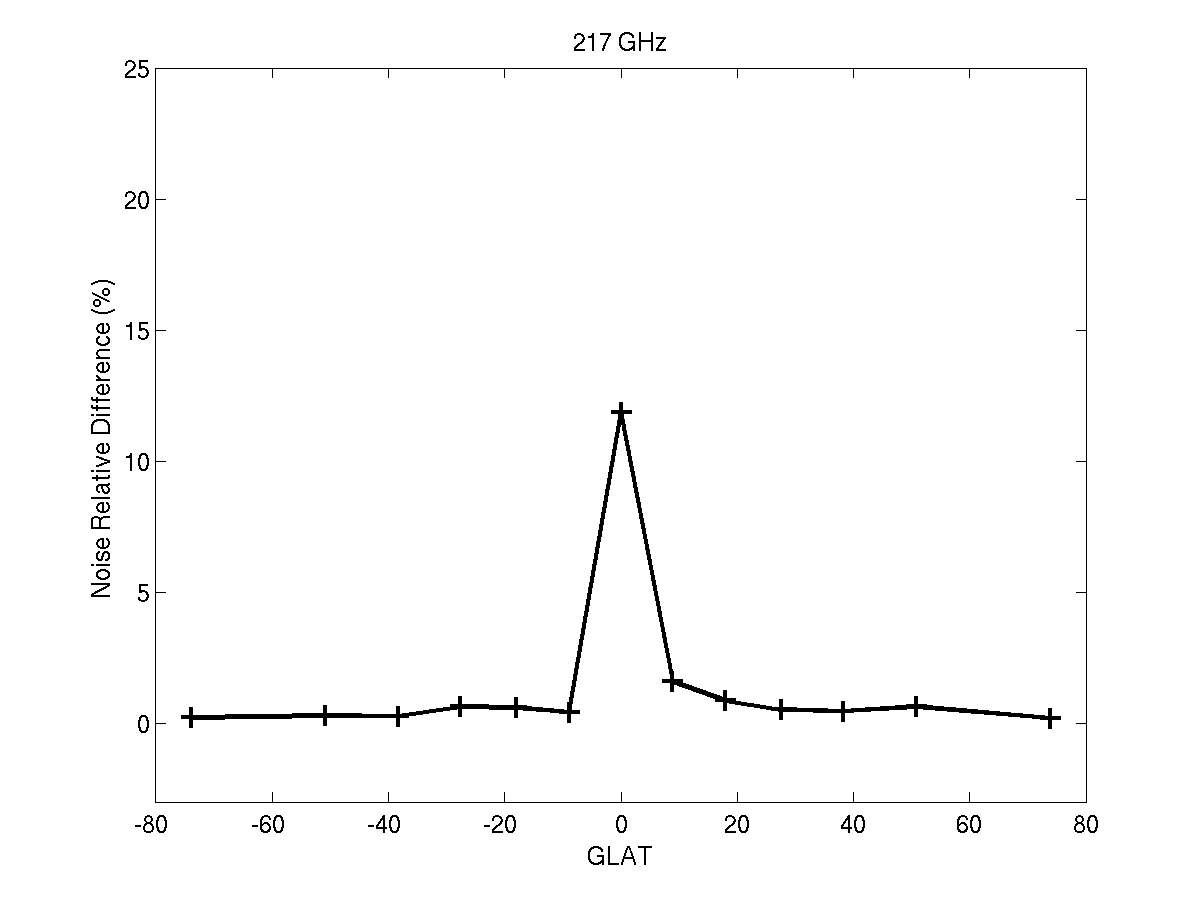}
\includegraphics[width=0.32\textwidth]{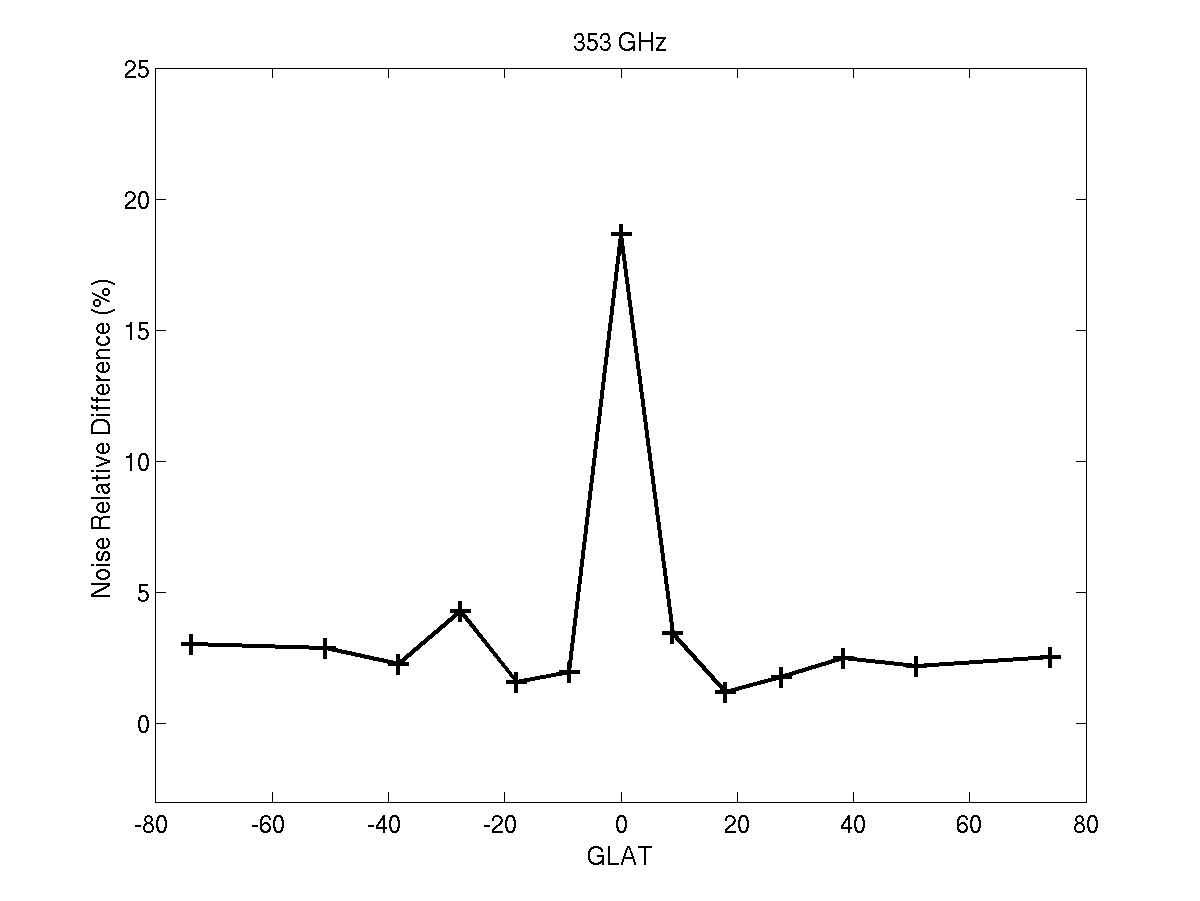}
\includegraphics[width=0.32\textwidth]{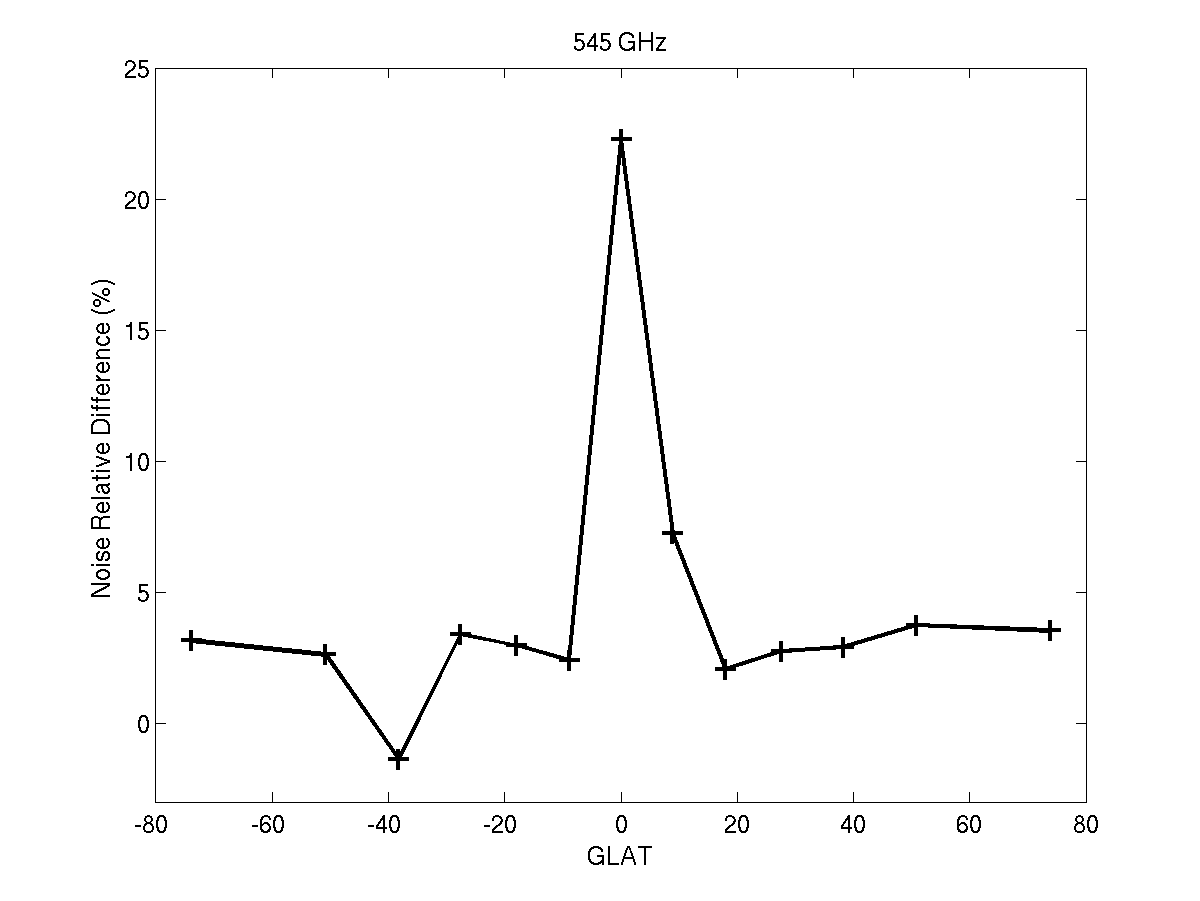}
\includegraphics[width=0.32\textwidth]{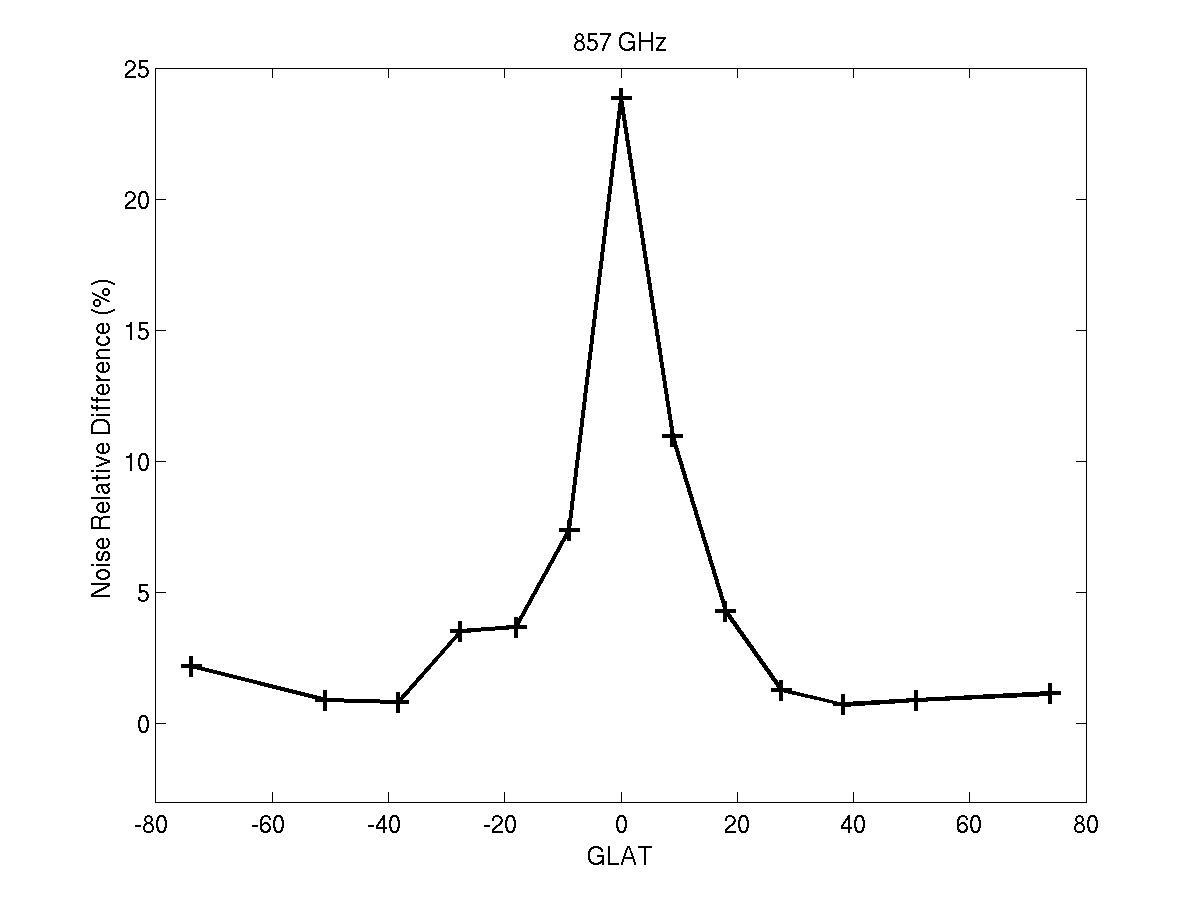}
 \caption{The average relative improvement (of BAF with respect to MHW2) in $\sigma_\omega$ is given. The BAF is able to improve the estimation of the noise up to $25\%$ in particularly complex regions, as in those areas dominated by galactic emission.}
 \label{fig:mean_noise_0}
 \end{center}
 \end{figure*}

 First, in Figure \ref{fig:mean_noise_0} we represent the relative
 difference between the noise level estimated for one method and the
 other as a function of galactic latitude and frequency. This allows us
 to see the overall behaviour of the noise estimated with the new
 filter compared with the MHW2. In the upper panels one can see that
 the estimation of the noise in the patches filtered with the BAF is a
 few percent smaller than that of the MHW2. When we increase the
 frequency one starts to see a change in the behaviour, as expected,
 and between 143 and 857 GHz the estimation of the noise of the BAF is
 up to a $25\%$ smaller than that of the MHW2, but only in the
 vicinity of the galactic plane. Note that even though the BAF, by
 definition, includes the MHW2 and its performance should always be
 equal or better than that of the MHW2, in the 545 GHz panel of Figure
 \ref{fig:mean_noise_0} we see that for one of the bins, the relative
 difference is negative. This is not a problem of the filter but a
 rare artefact of the implementation of the algorithm that occurs due
 to the fact that we divide the sky into overlapping patches and,
 sometimes, the same source is detected with the same technique in two
 adjacent patches but with different SNR. Since we keep the ones with
 highest SNR, it could happen that a source is detected with a higher
 SNR with the BAF in one patch and with the MHW2 in the adjacent
 one. In this case we cannot guarantee that the noise estimation of
 the BAF is equal or higher than the MHW2 because they are in fact
 looking at slightly different regions of the sky.

\begin{figure*}
 \begin{center}
\includegraphics[width=0.32\textwidth]{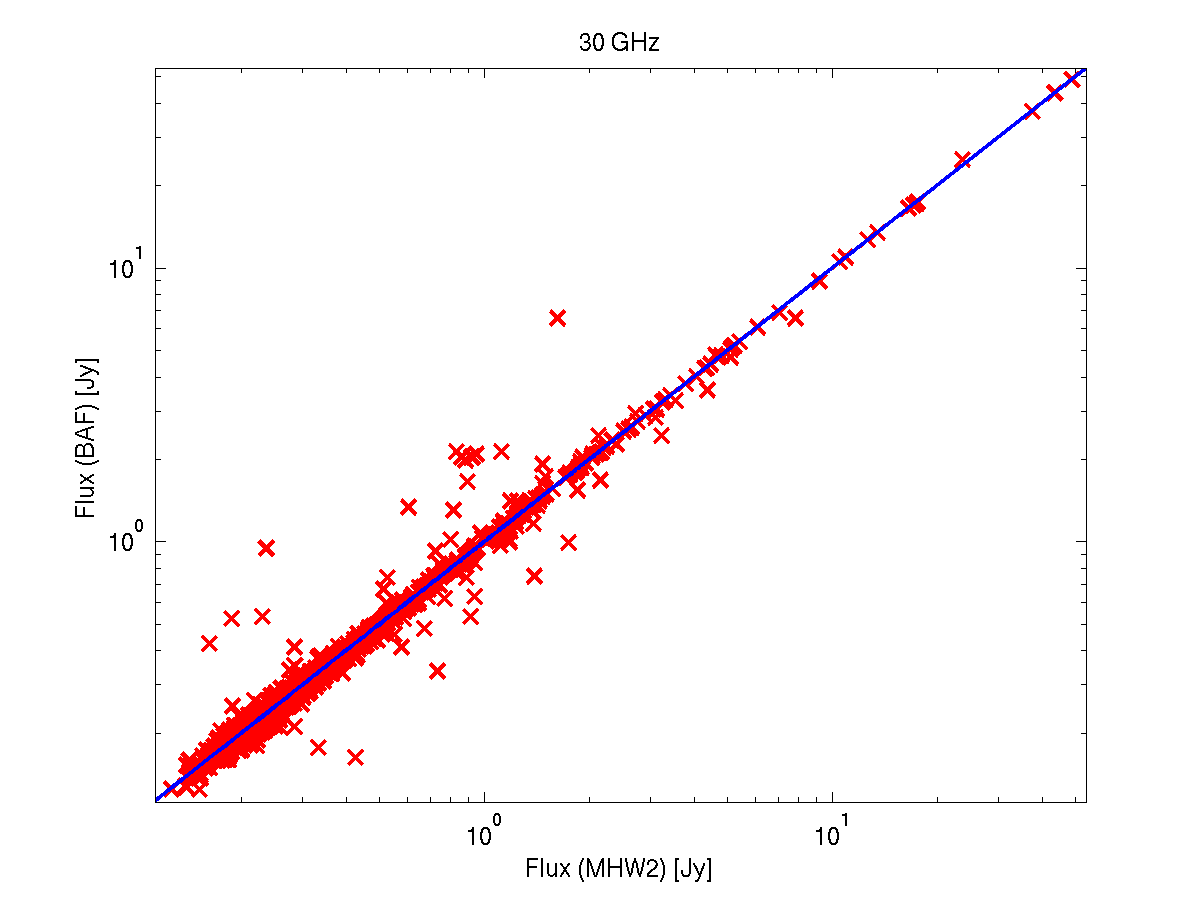}
\includegraphics[width=0.32\textwidth]{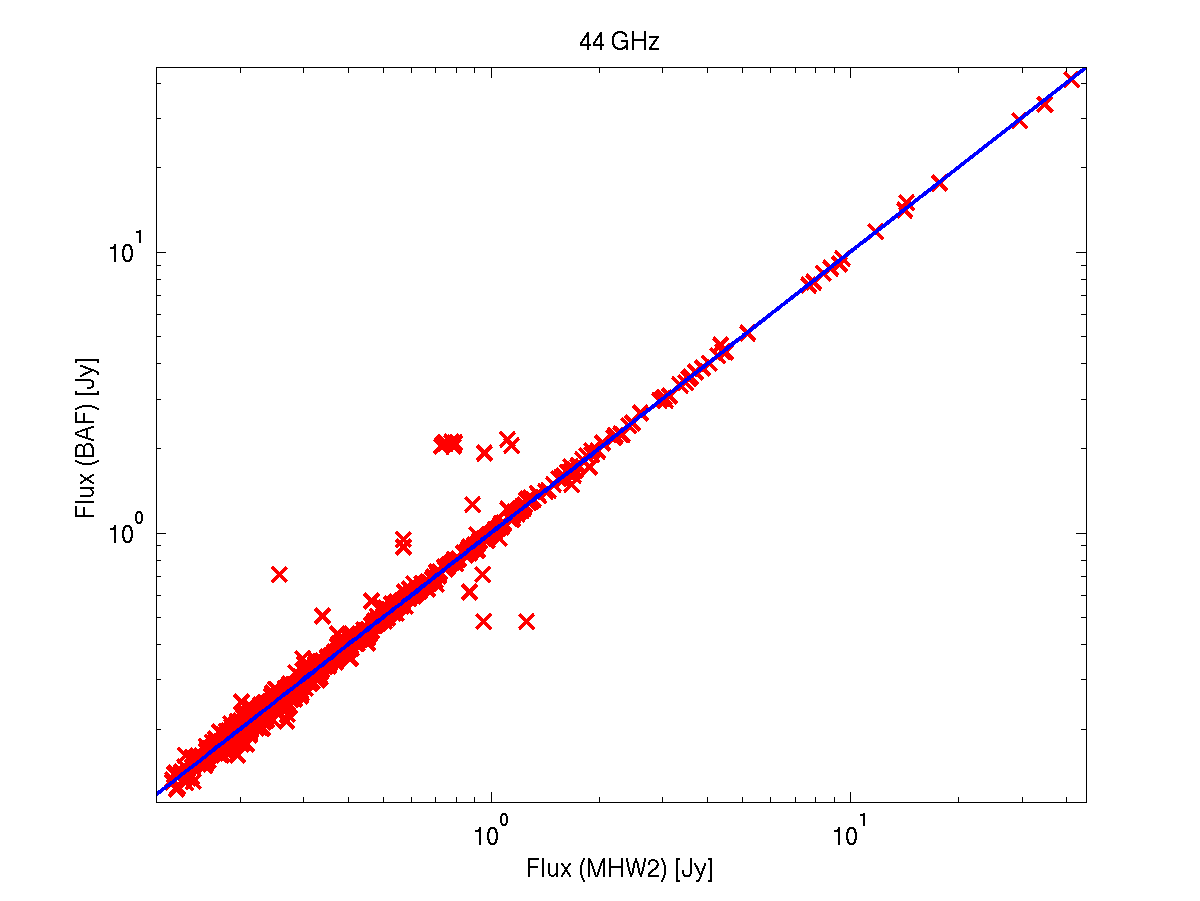}
\includegraphics[width=0.32\textwidth]{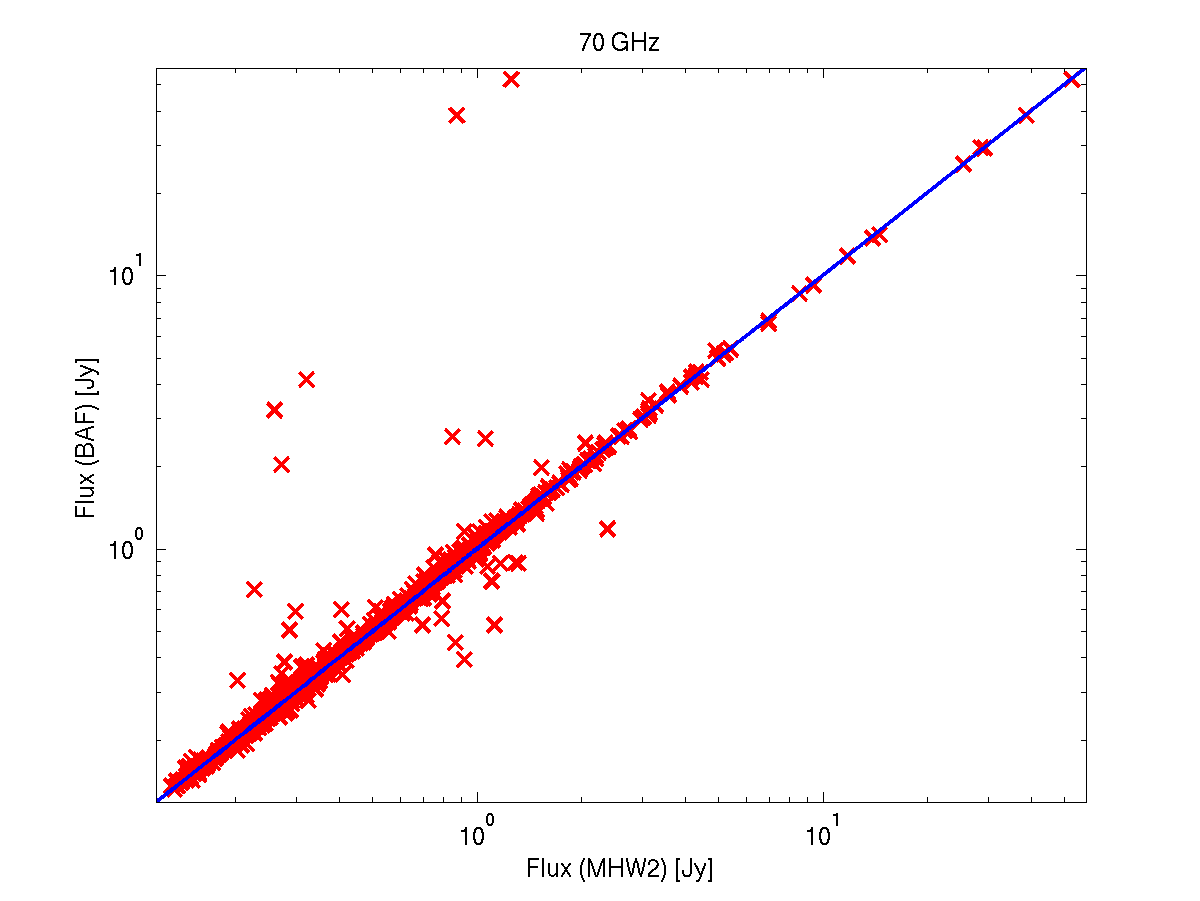}
\includegraphics[width=0.32\textwidth]{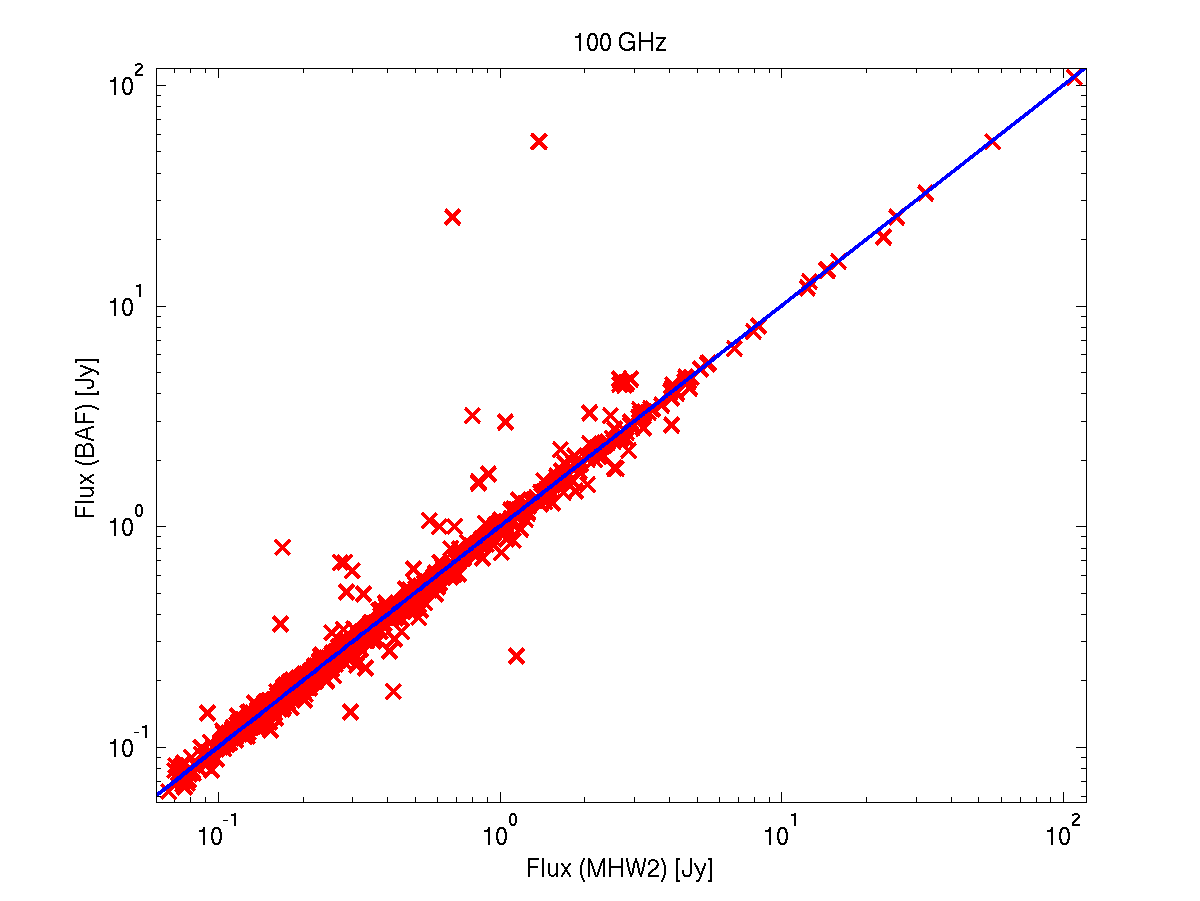}
\includegraphics[width=0.32\textwidth]{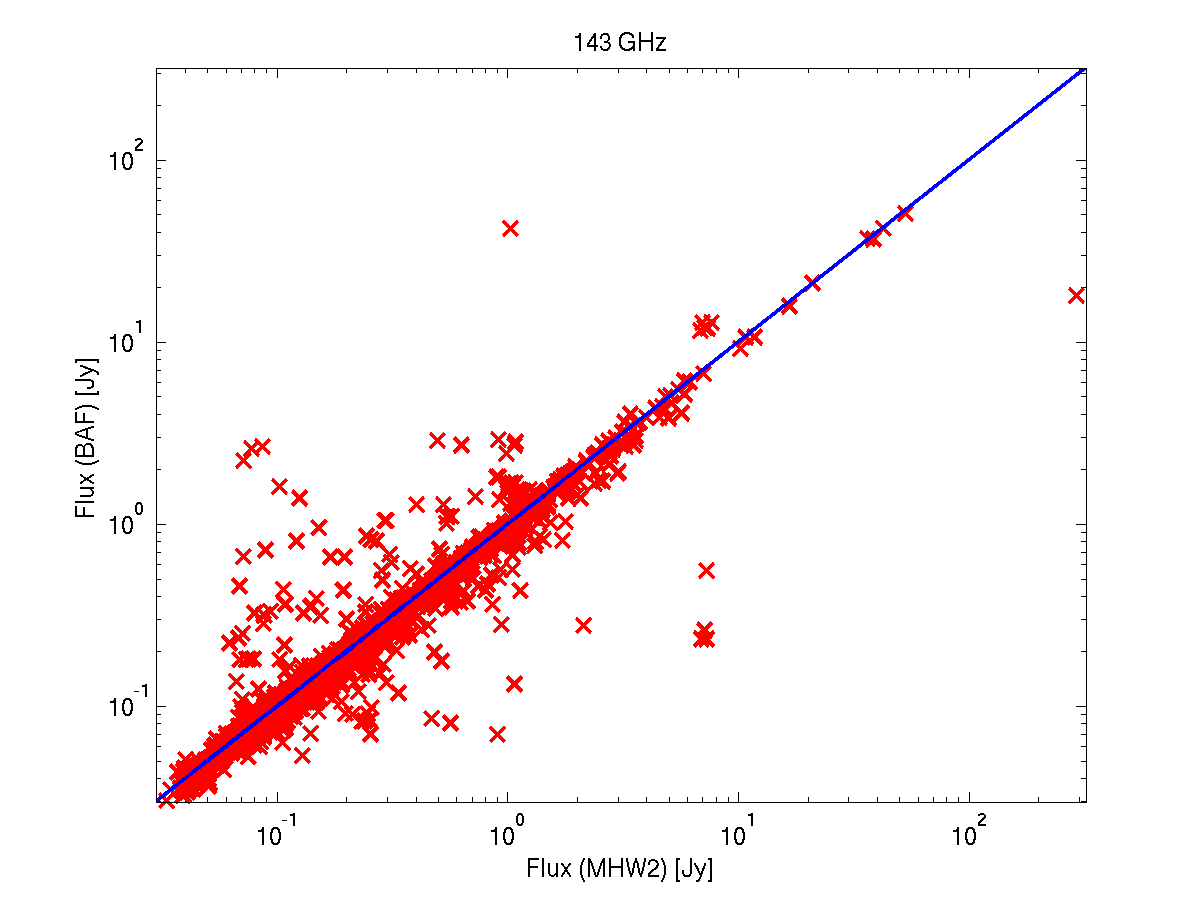}
\includegraphics[width=0.32\textwidth]{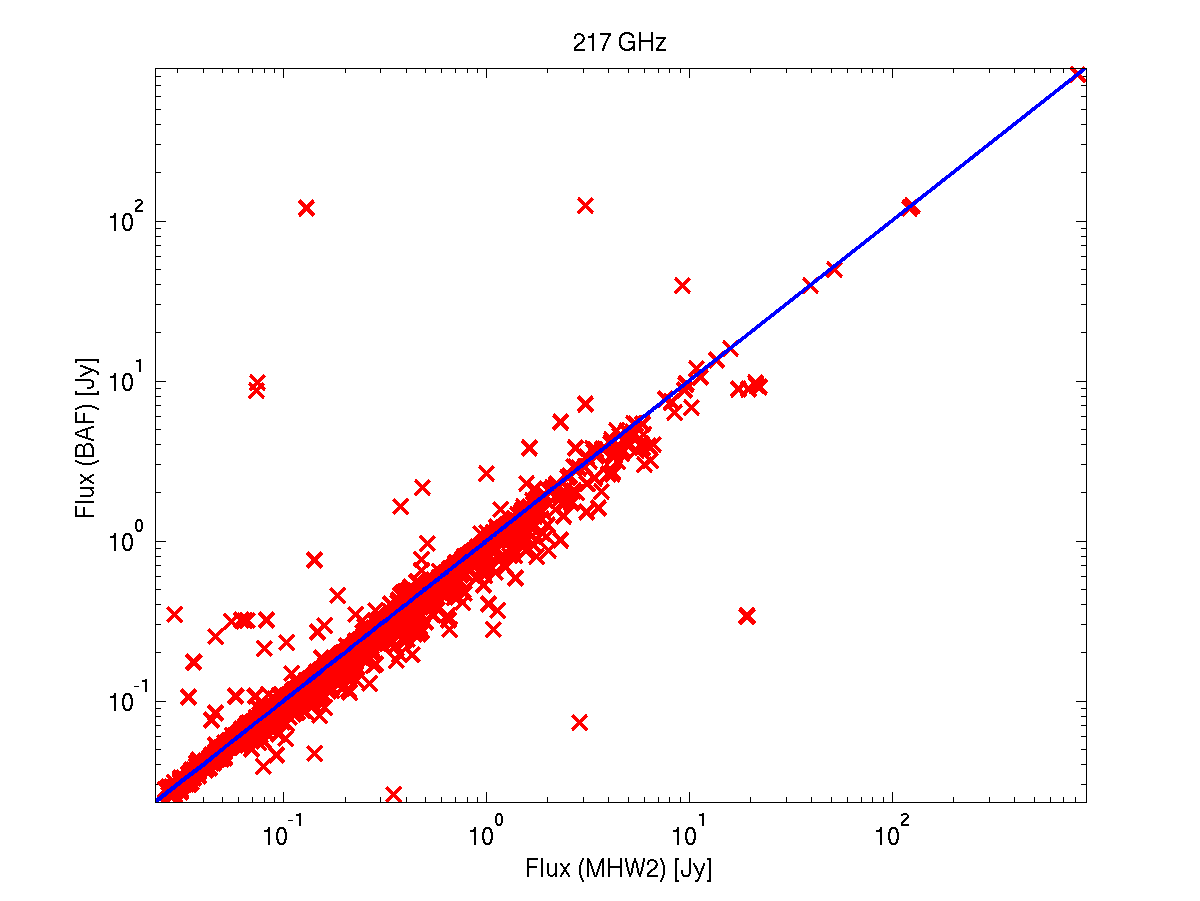}
 \caption{Comparison of the estimated flux density of the common sources detected by both methods for channels 30 to 217 GHz.}
 \label{fig:flux_0}
 \end{center}
 \end{figure*}

\begin{figure*}
\begin{center}
\includegraphics[width=0.32\textwidth]{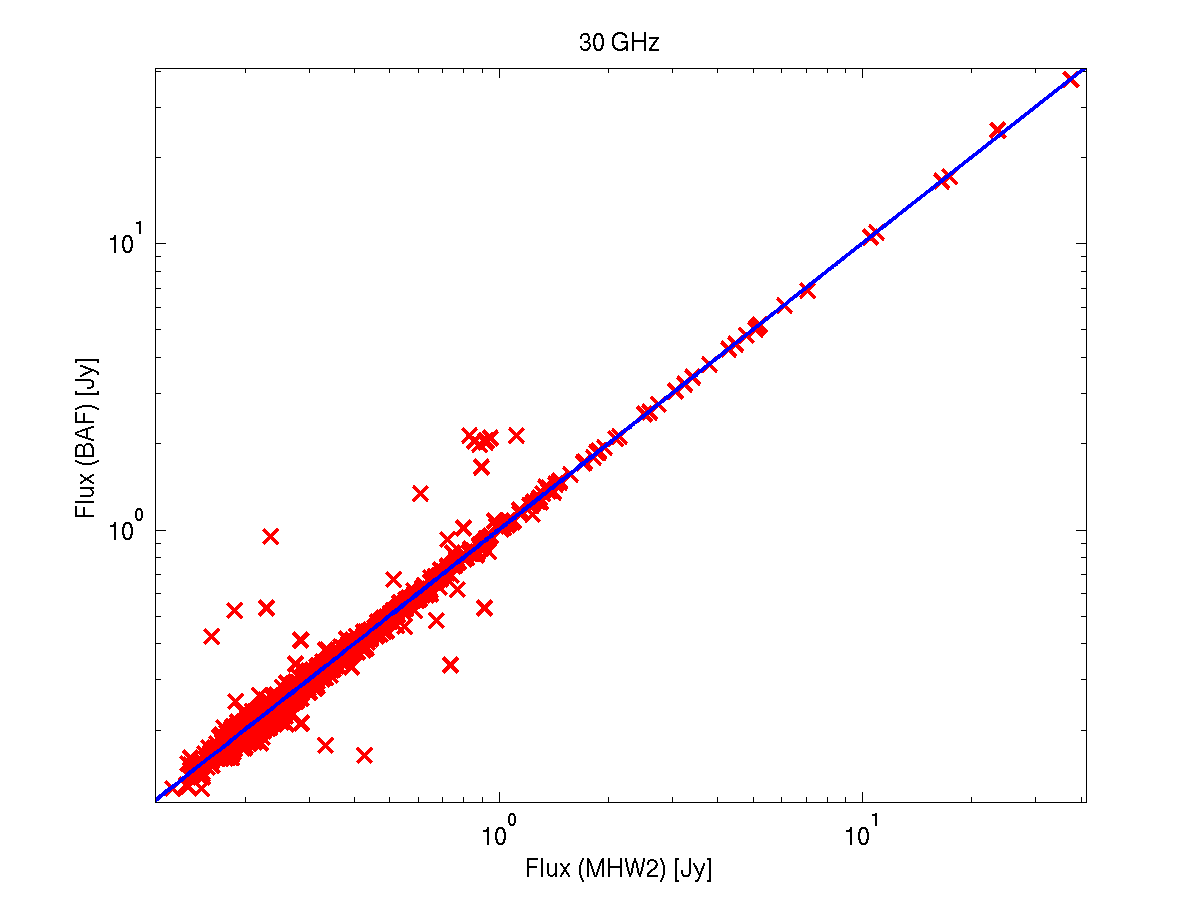}
\includegraphics[width=0.32\textwidth]{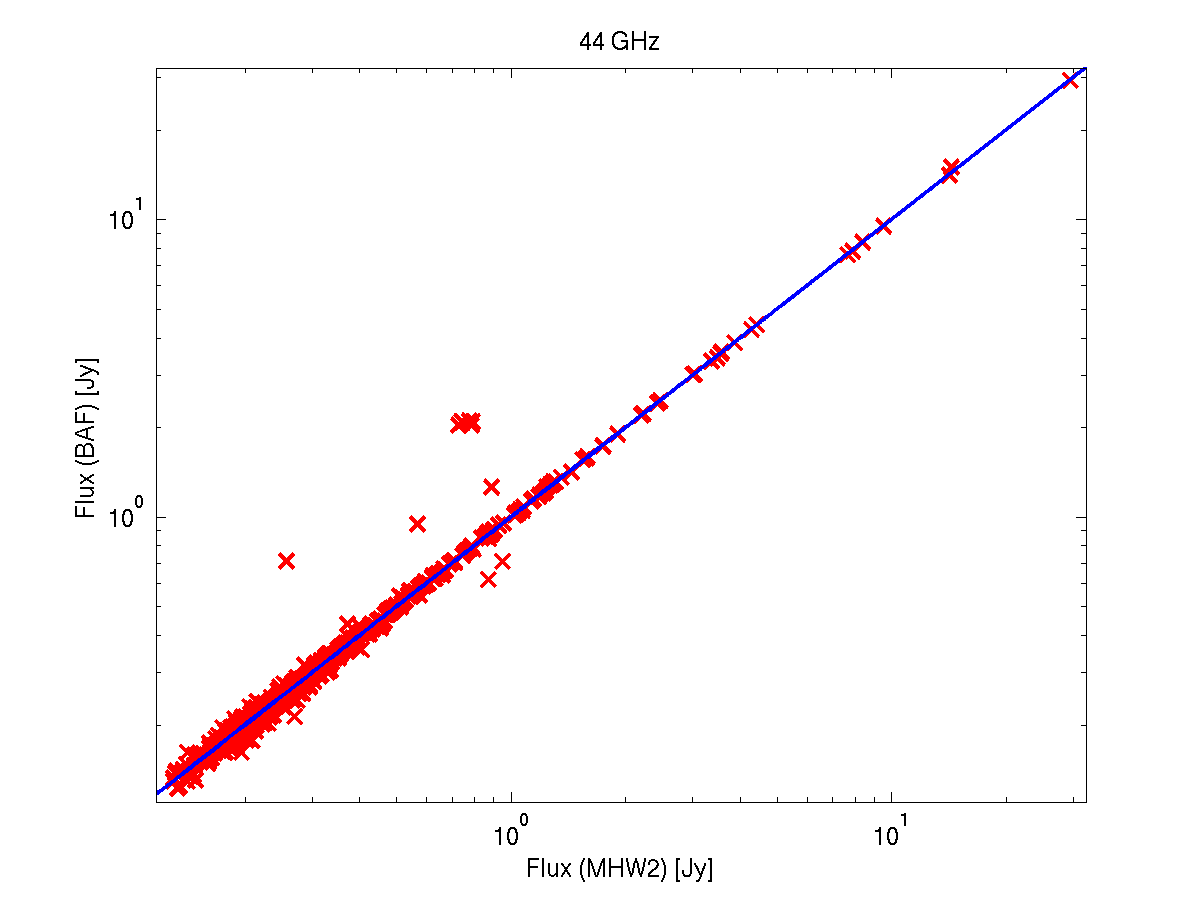}
\includegraphics[width=0.32\textwidth]{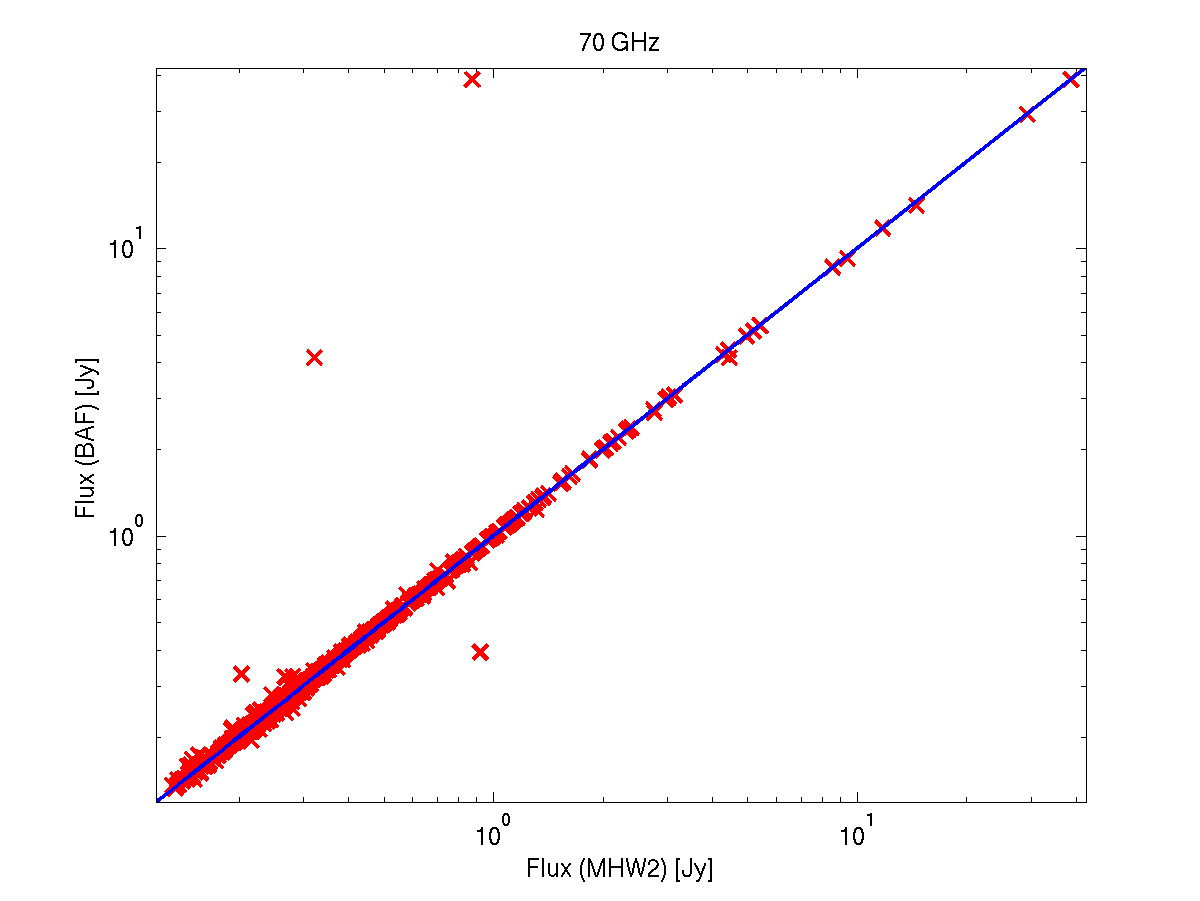}
\includegraphics[width=0.32\textwidth]{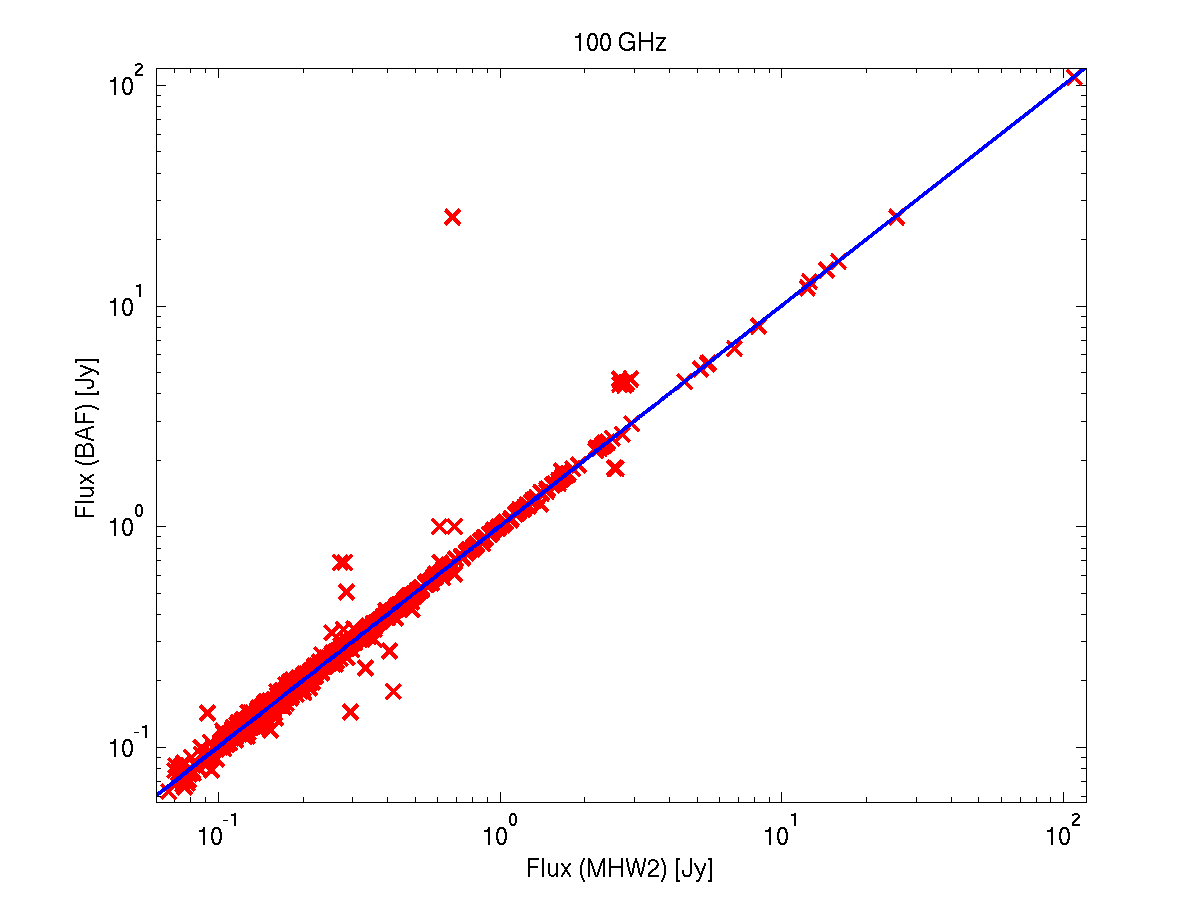}
\includegraphics[width=0.32\textwidth]{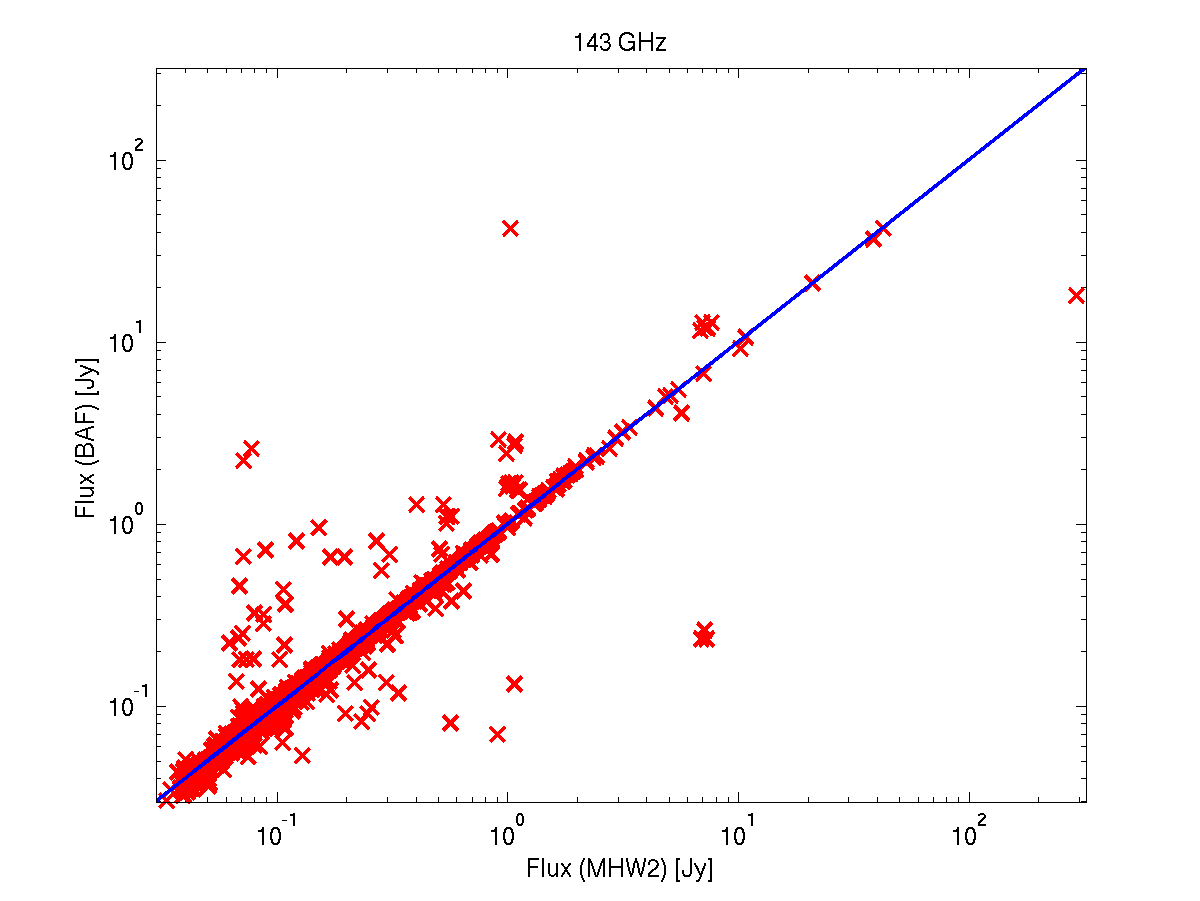}
\includegraphics[width=0.32\textwidth]{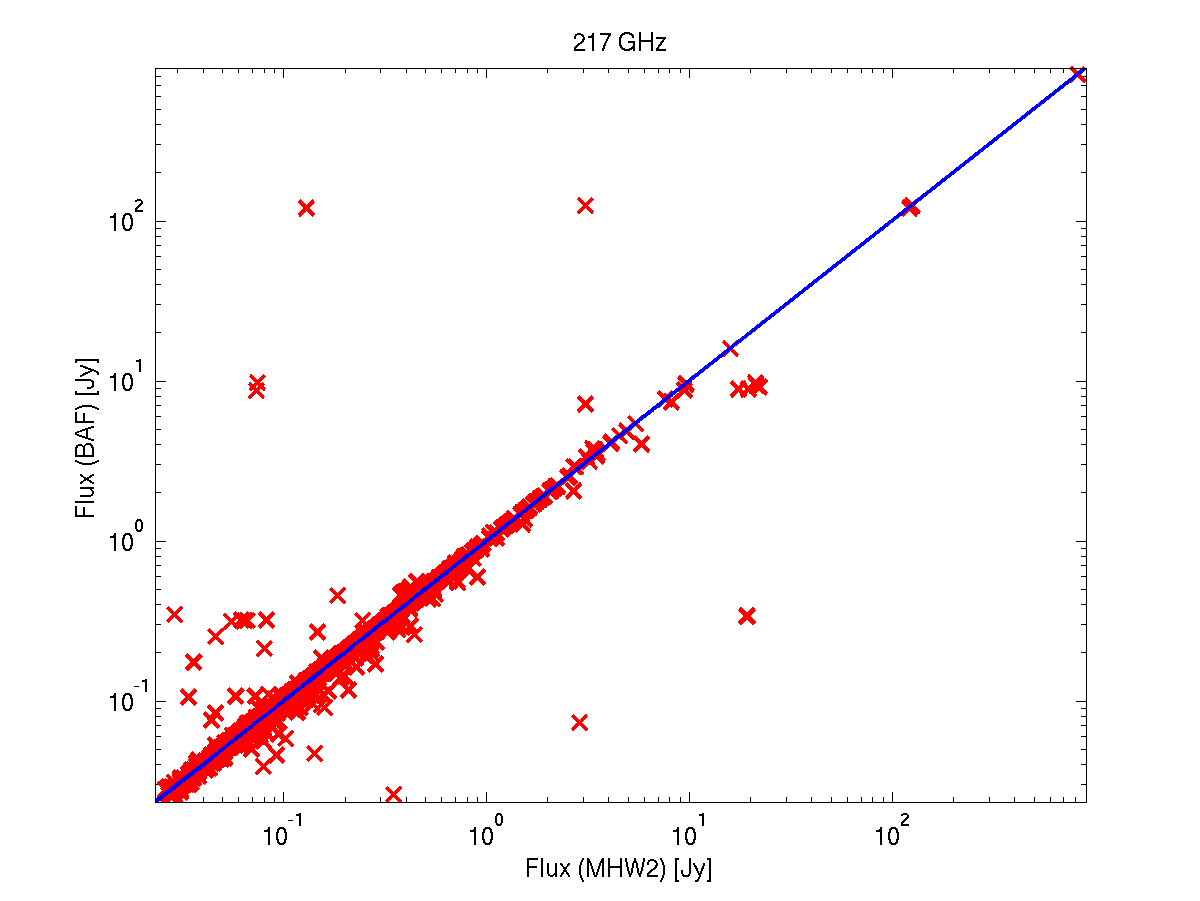}
 \caption{As in the previous figure, but applying a $\pm3$ degrees three galactic cut to show that the small bias found in the 143 and 217 GHz panels rapidly disappears when we exclude from the figure sources very close to the galactic plane}
 \label{fig:flux_3}
 \end{center}
 \end{figure*}

 Second, with respect to the flux density estimation of the sources
 obtained with the BAF or the MHW2 techniques, in Figure
 \ref{fig:flux_0} one can see that between 30 and 217 GHz the fluxes
 obtained with both methods follow a clear one-to-one line. At 143 GHz
 one starts to see an increase in the dispersion and a small bias that
 is more obvious at higher frequencies. These differences mitigate
 when all the objects that lie within a $\pm3$ degrees cut in latitude
 are excluded. These plots can be seen in Figure \ref{fig:flux_3},
 where most of the scatter and bias has now disappeared.

\begin{figure*}
\begin{center}
\includegraphics[width=0.32\textwidth]{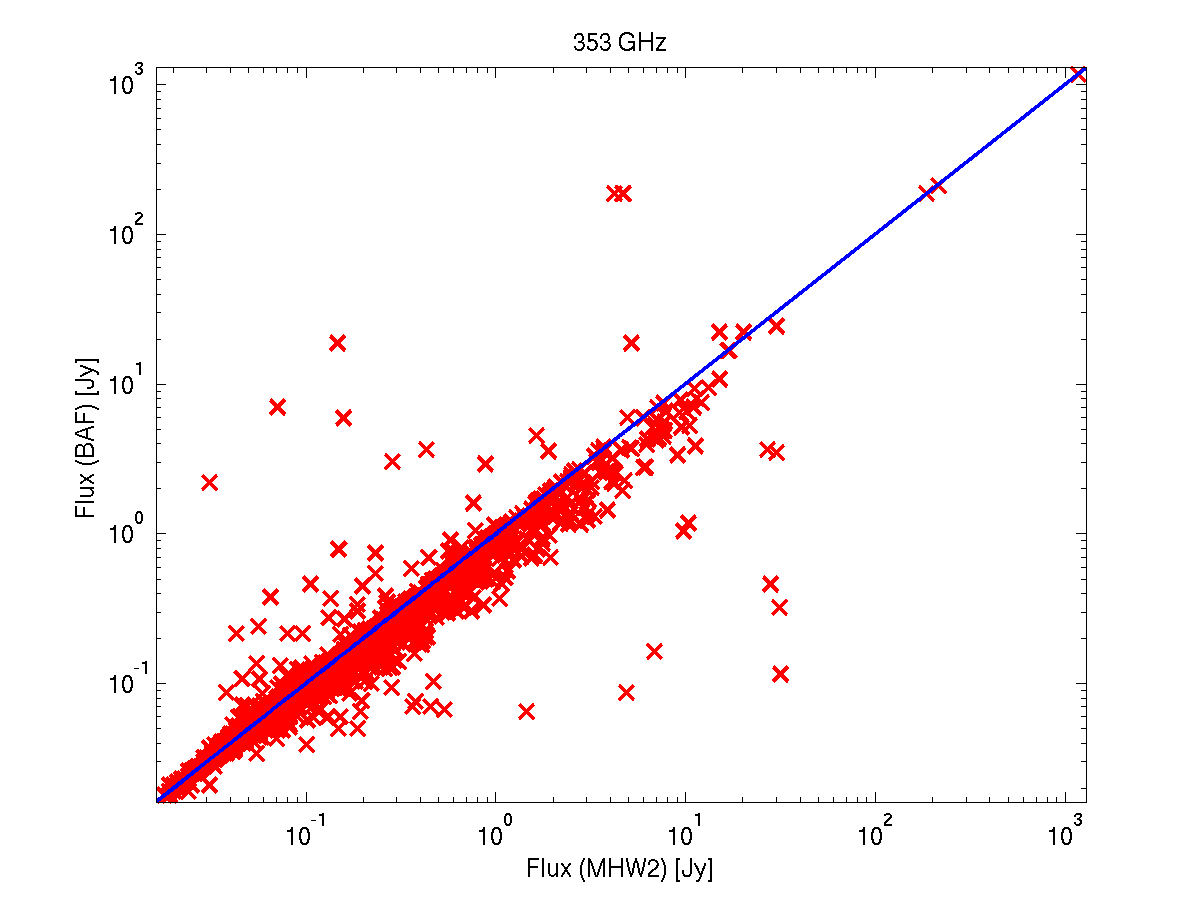}
\includegraphics[width=0.32\textwidth]{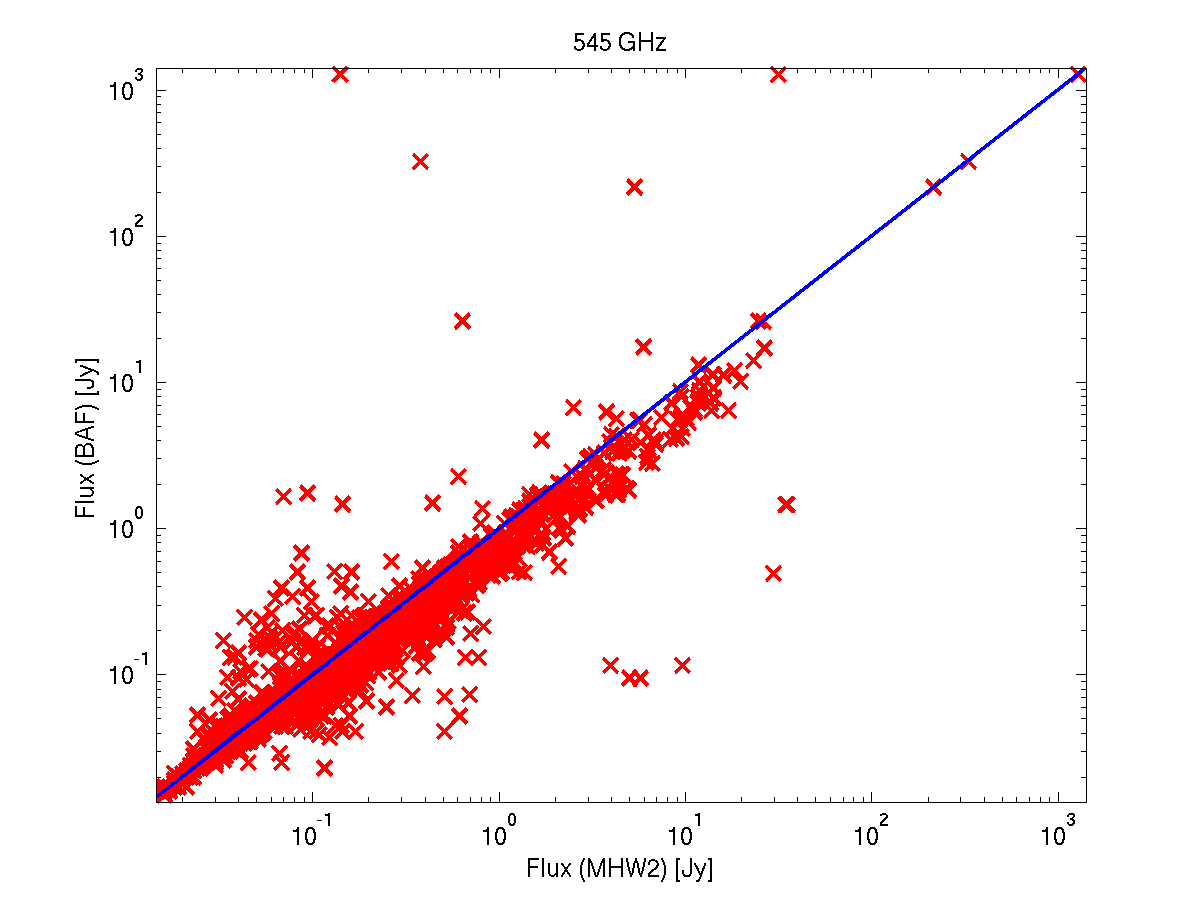}
\includegraphics[width=0.32\textwidth]{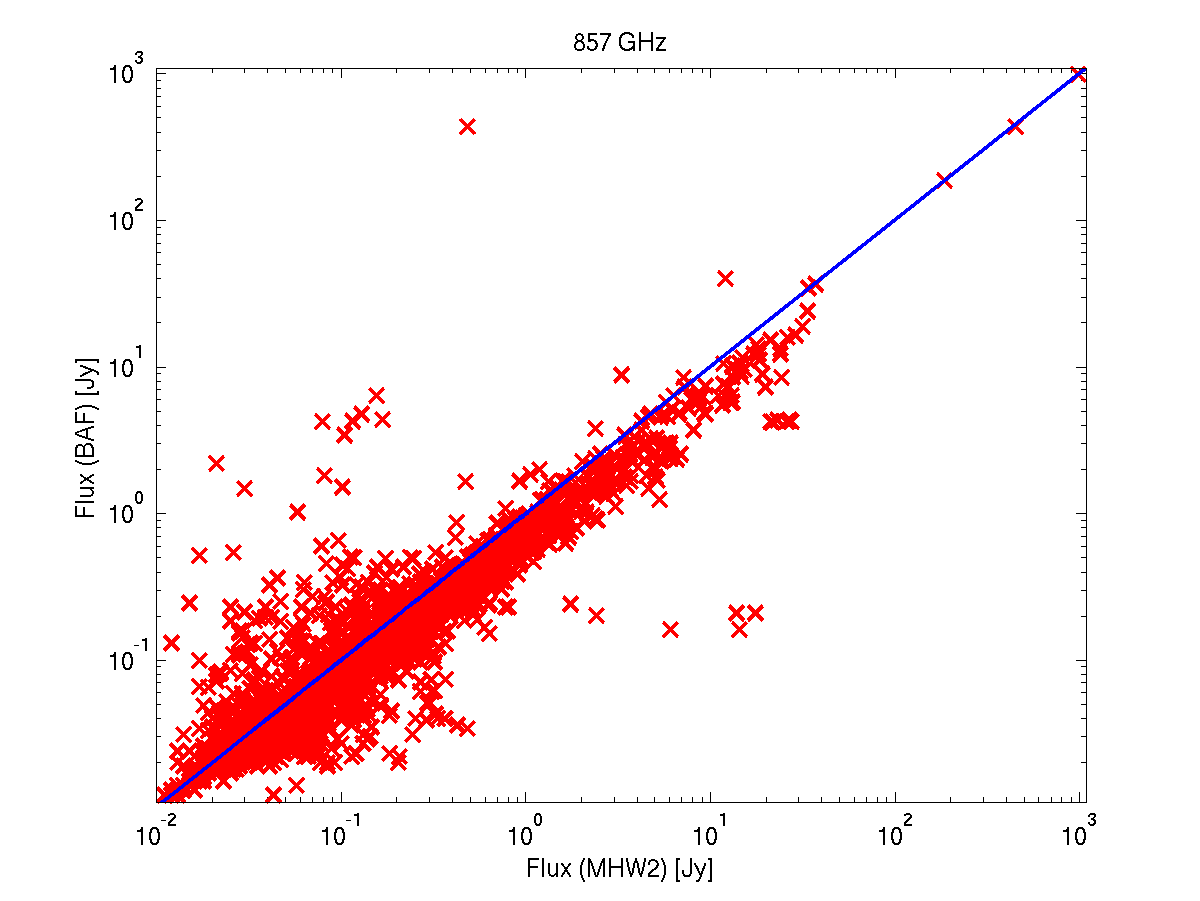}
\includegraphics[width=0.32\textwidth]{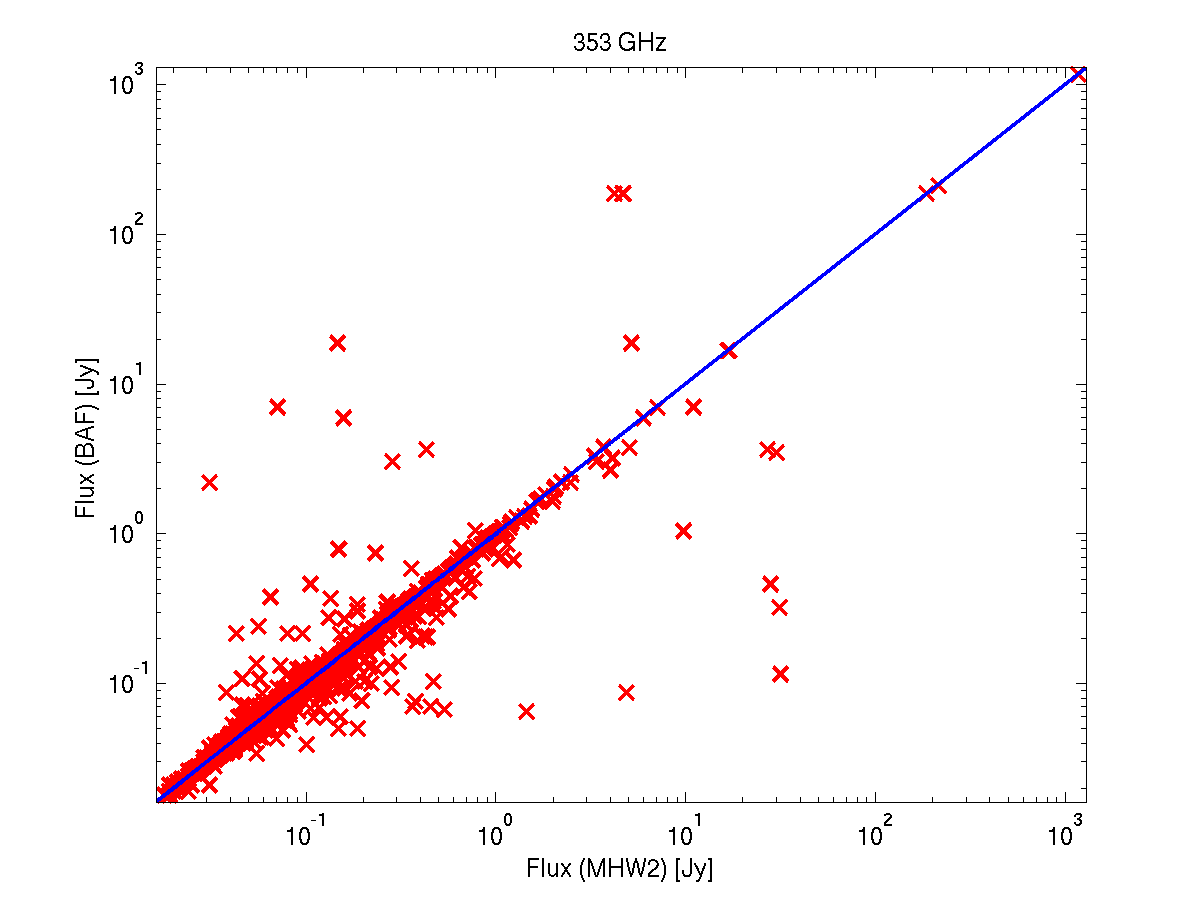}
\includegraphics[width=0.32\textwidth]{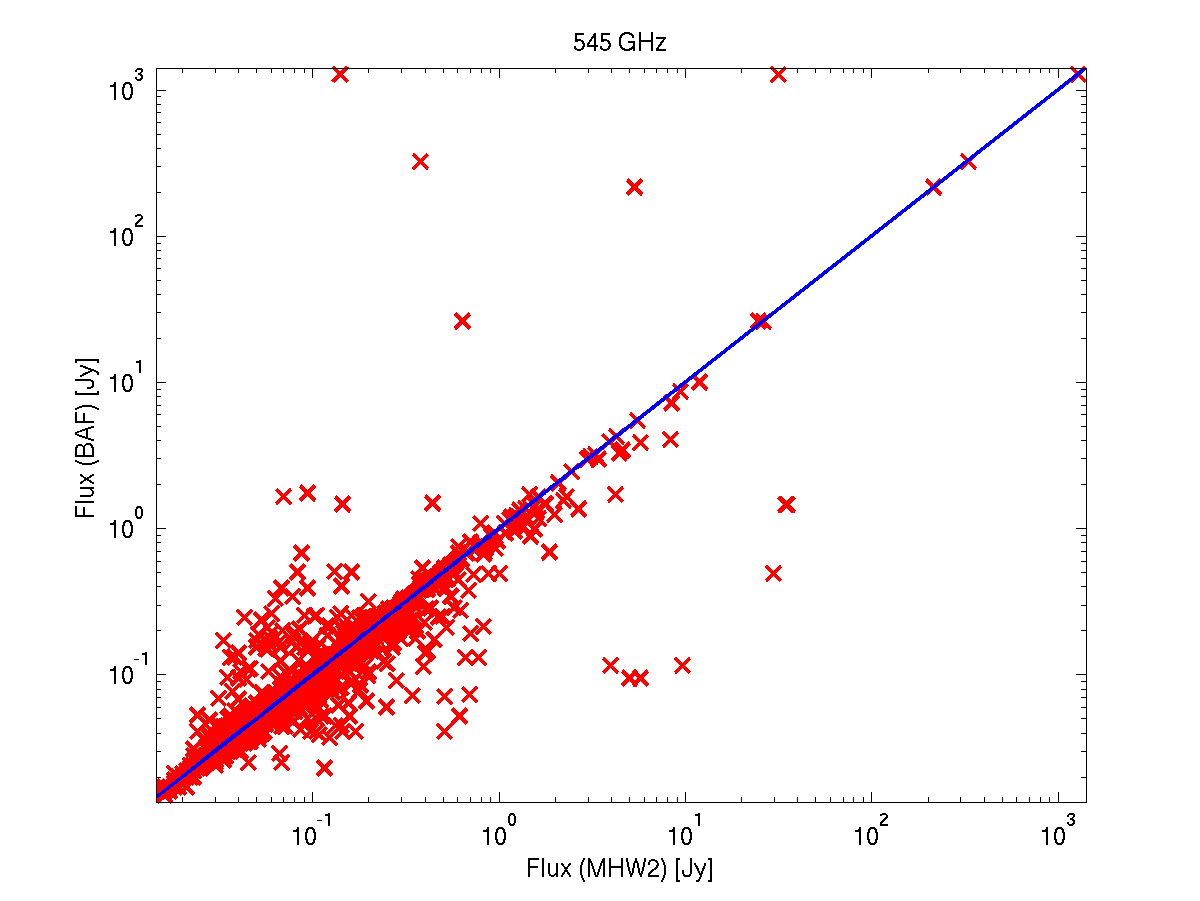}
\includegraphics[width=0.32\textwidth]{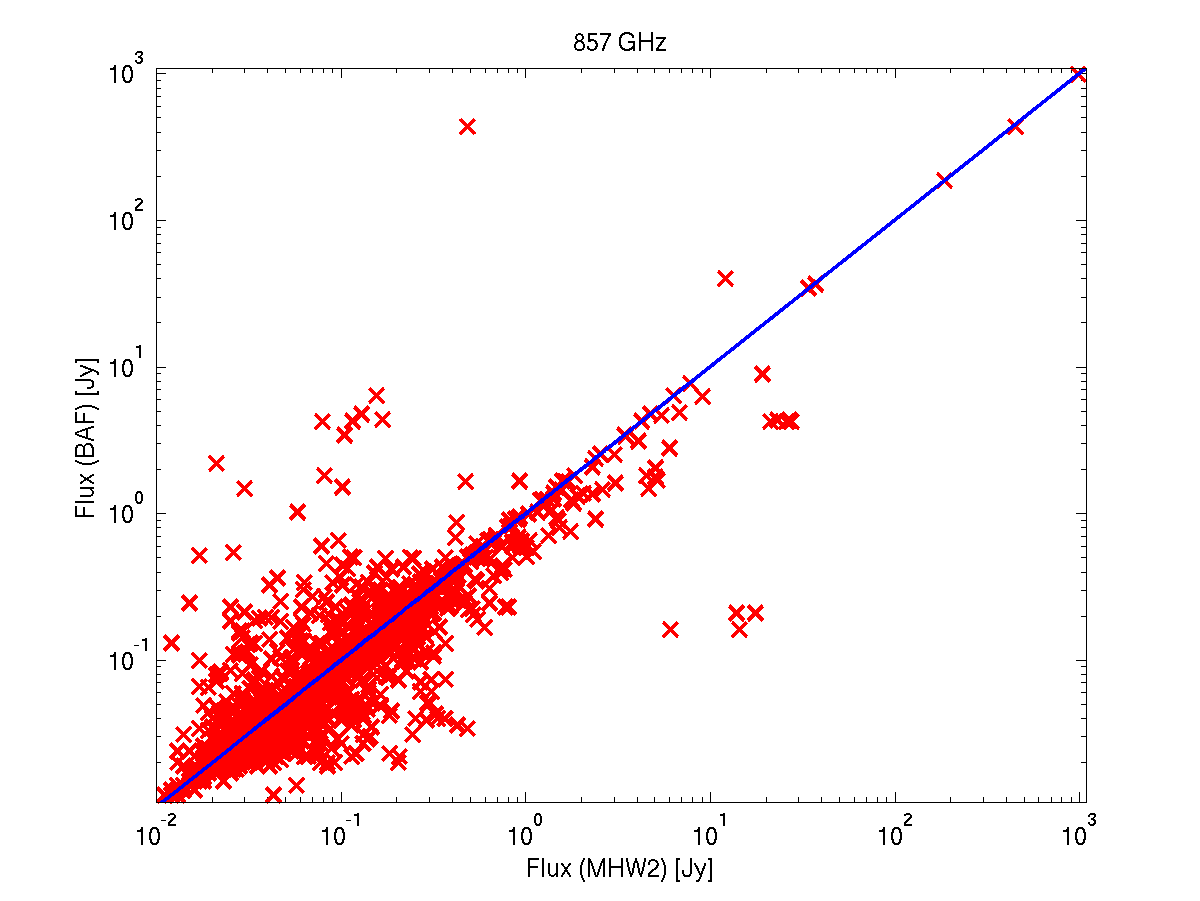}
\includegraphics[width=0.32\textwidth]{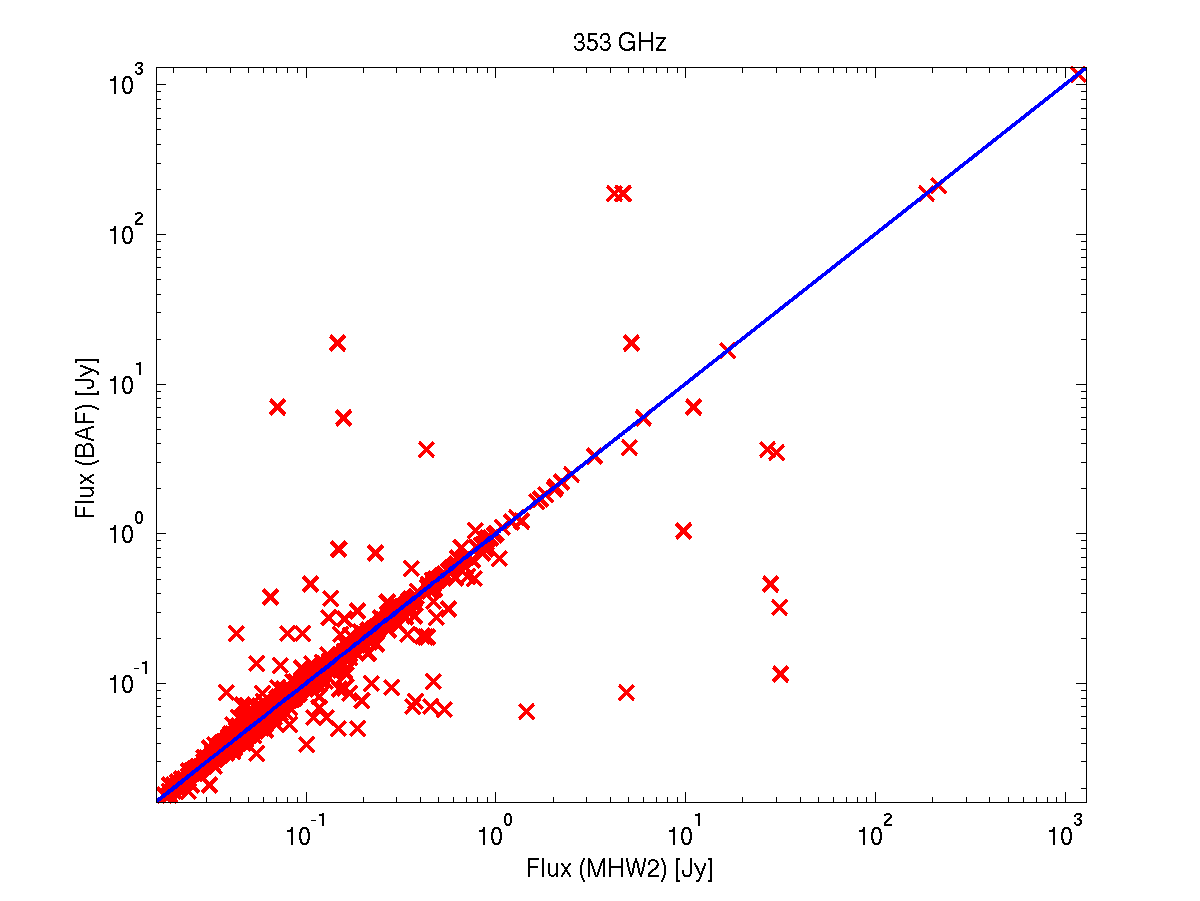}
\includegraphics[width=0.32\textwidth]{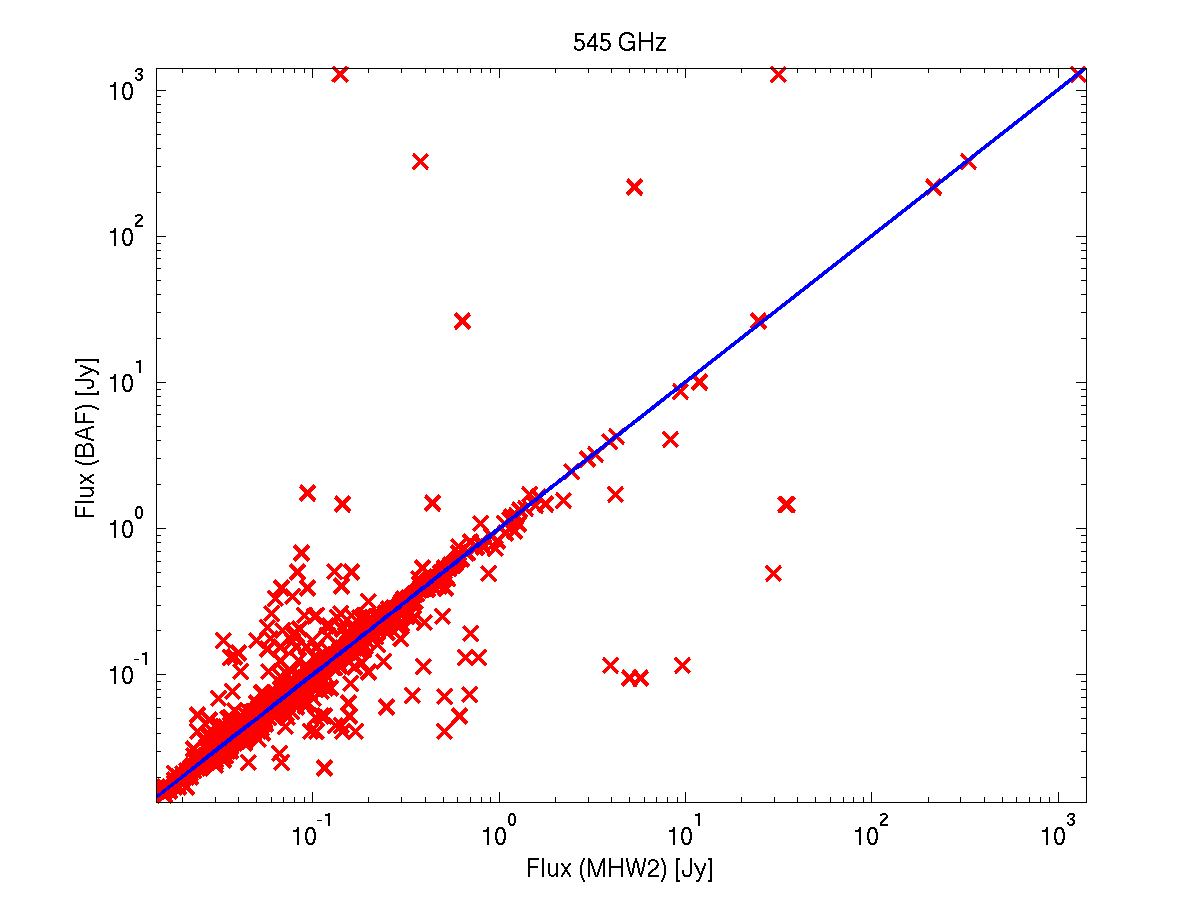}
\includegraphics[width=0.32\textwidth]{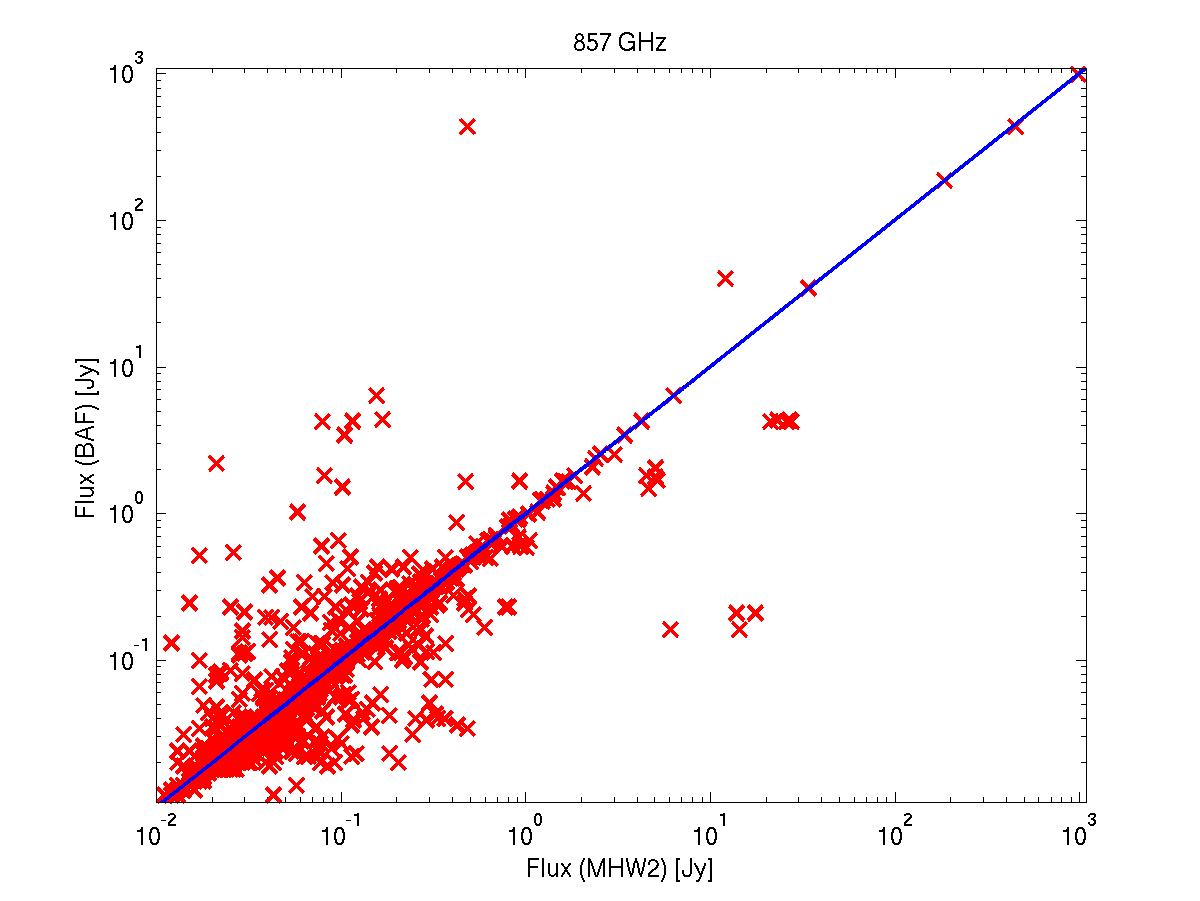}
 \caption{Comparison of the estimated flux density of the common sources detected by both filters for channels 353, 545 and 857 (left, center and right panels) and applying three galactic cuts (none, $\pm3$ degrees and $\pm15$ degrees) to show that the bias found in the figures rapidly disappears when we exclude from the figure sources very close to the galactic plane}
 \label{fig:flux_hfi}
 \end{center}
 \end{figure*}

 To further investigate this correlation between biased fluxes and
 extended galactic regions at high frequencies, we have plotted the
 flux of the MHW2 vs. that of the BAF for 353, 545 and 857 GHz using
 an additional galactic cut of $\pm15$ degrees. In Figure
 \ref{fig:flux_hfi} we show the fluxes without applying any galactic
 cut (upper panels), applying a $\pm3$ degree cut (middle panels) and
 applying a $\pm 15 $ degree cut (lower panels). Again, the scatter
 and bias found in the upper panels decreases significantly when
 increasing the galactic cut, as we expected.

This result is telling us that both filters produce unbiased flux density 
estimations outside the galactic plane and is highlighting a problem in its 
vicinity. We suspect that the problem may be related to any or both of the 
following effects. First, the fact that the extremely bright and spatially 
variable background signal is contributing to the recovered flux density of the 
point sources. In other words, the filters are not able to fully remove the 
contribution from the background. Second, the filters are not able to recover 
unbiased flux density estimations of the extended sources present in the galactic 
plane, because they have not been designed for this purpose and their response to 
these objects can be different. In order to test these ideas, we have
 performed additional simulations injecting 200 point sources with
 the same flux density and spatially distributed along the galactic plane of
 the 545 GHz simulated map. Then we have attempted to estimate their
 flux densities using the BAF and MHW2 techniques and obtain that we are able 
 to recover them with differences smaller than $1\%$, i.e., much lower than the 
 ones in Figure \ref{fig:flux_hfi}. Therefore, we are confident that the reason 
 for the bias seen at 217 GHz and above is due to the uneven performance of both 
 filters when dealing with extended objects, a situation for which they have not been designed.
  
\begin{figure*}
 \begin{center}
\includegraphics[width=0.45\textwidth]{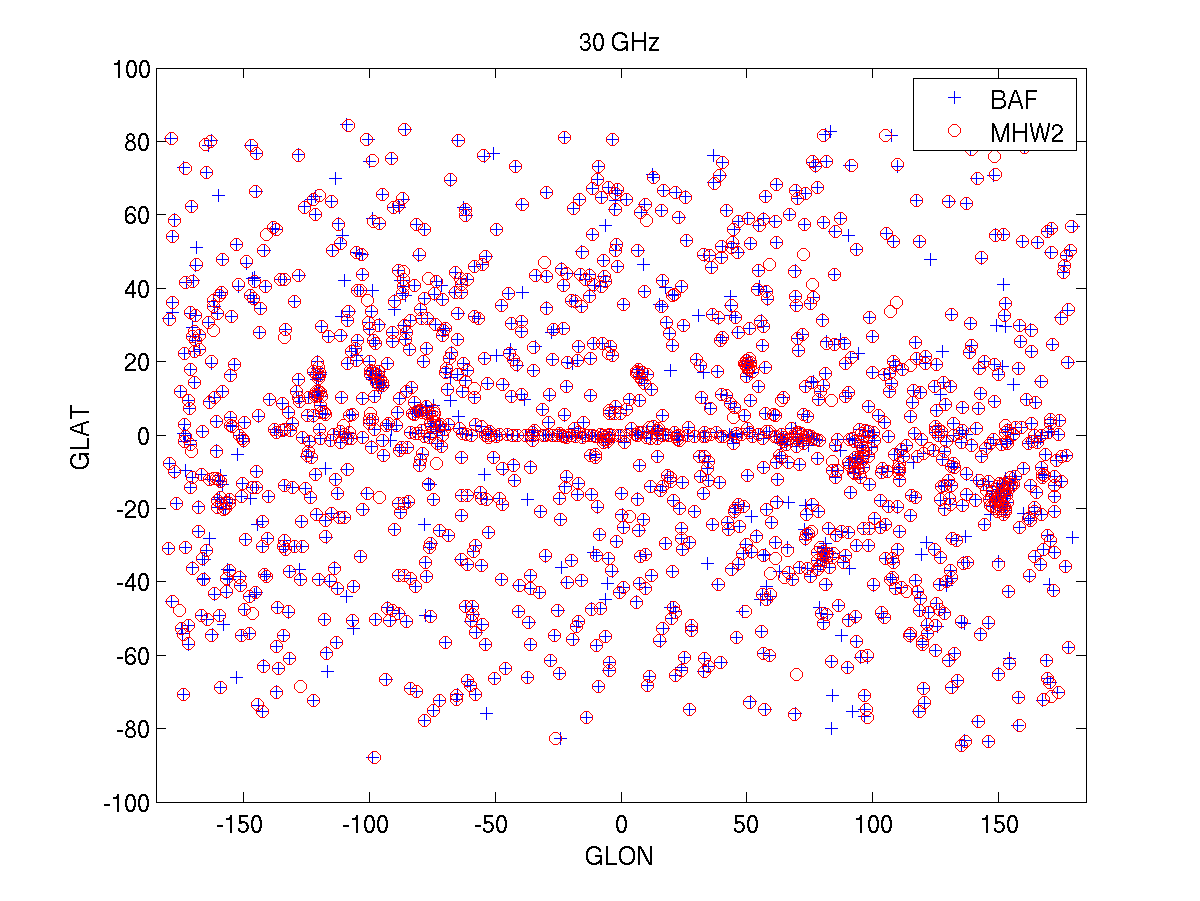}
\includegraphics[width=0.45\textwidth]{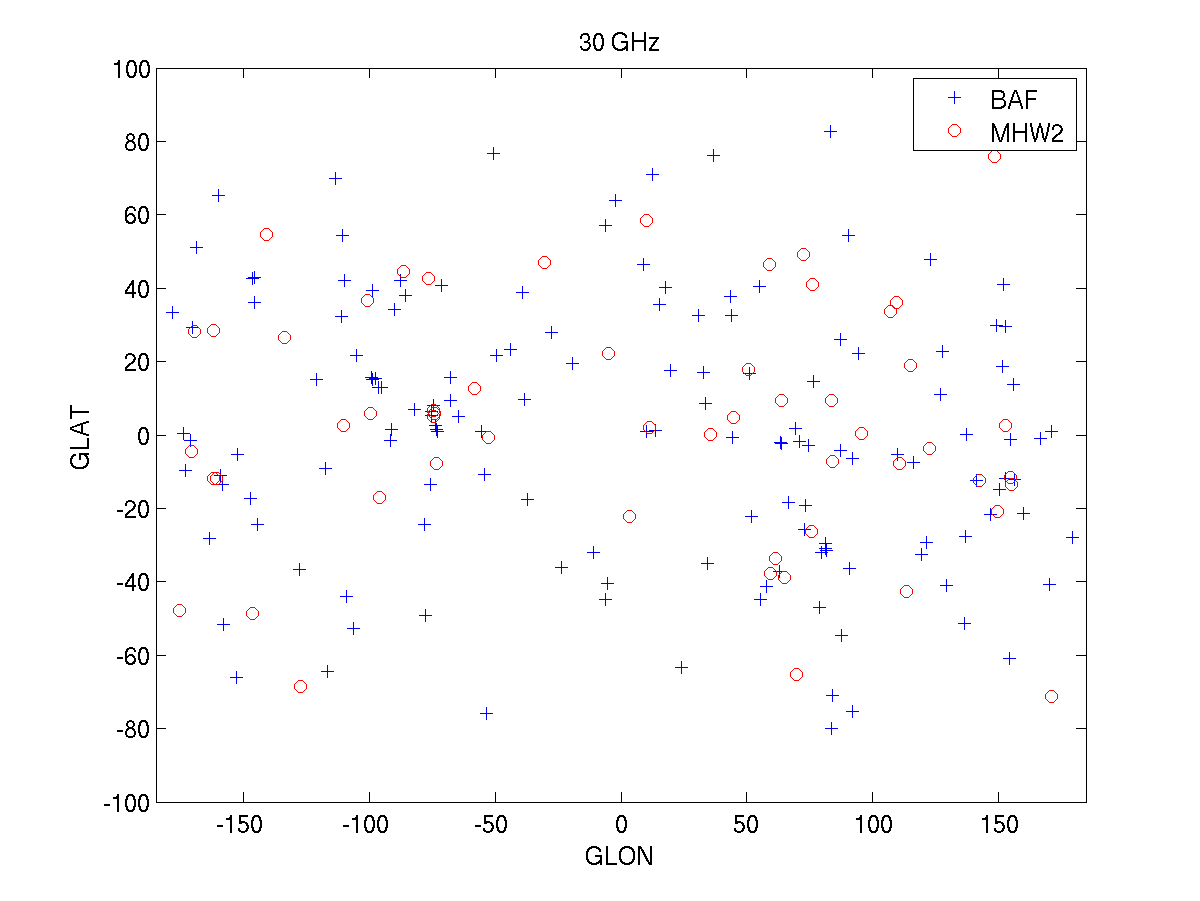}
\includegraphics[width=0.45\textwidth]{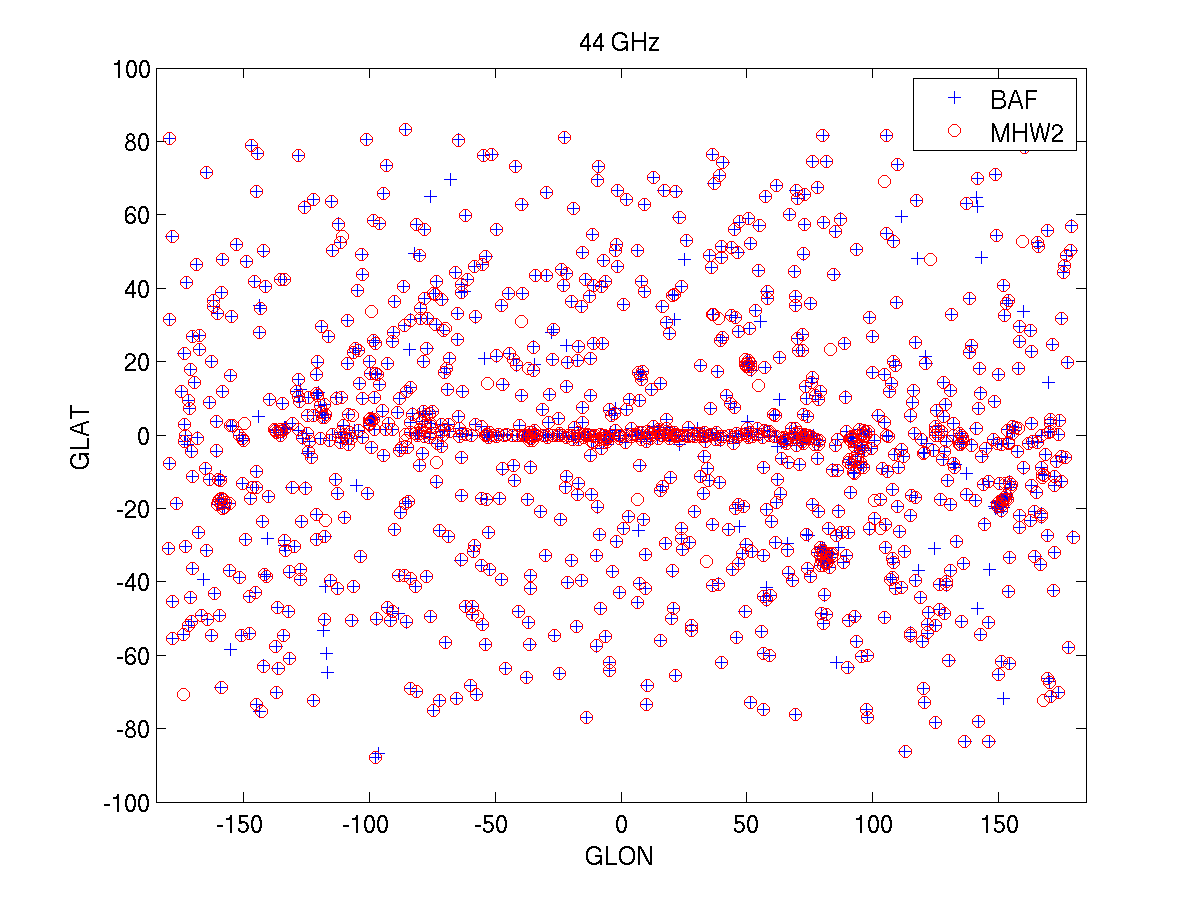}
\includegraphics[width=0.45\textwidth]{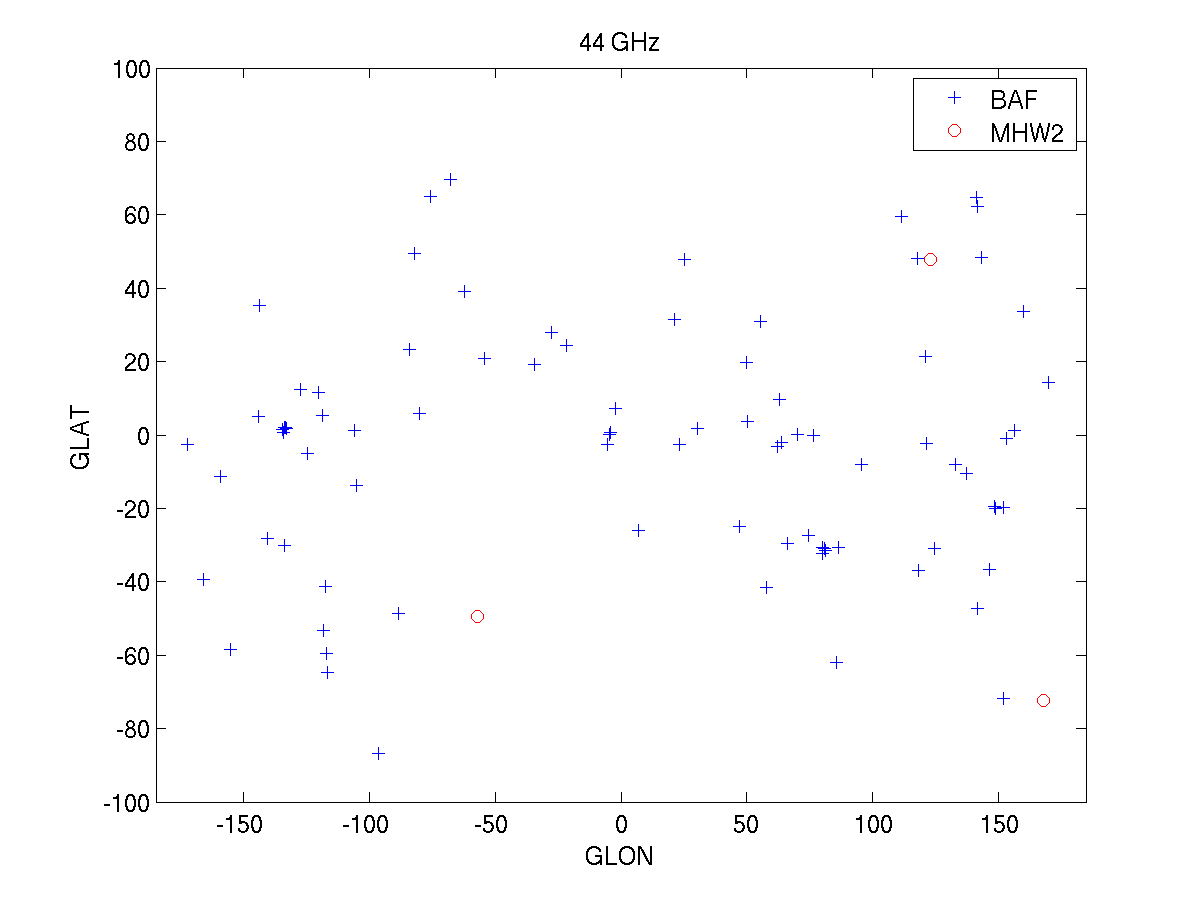}
\includegraphics[width=0.45\textwidth]{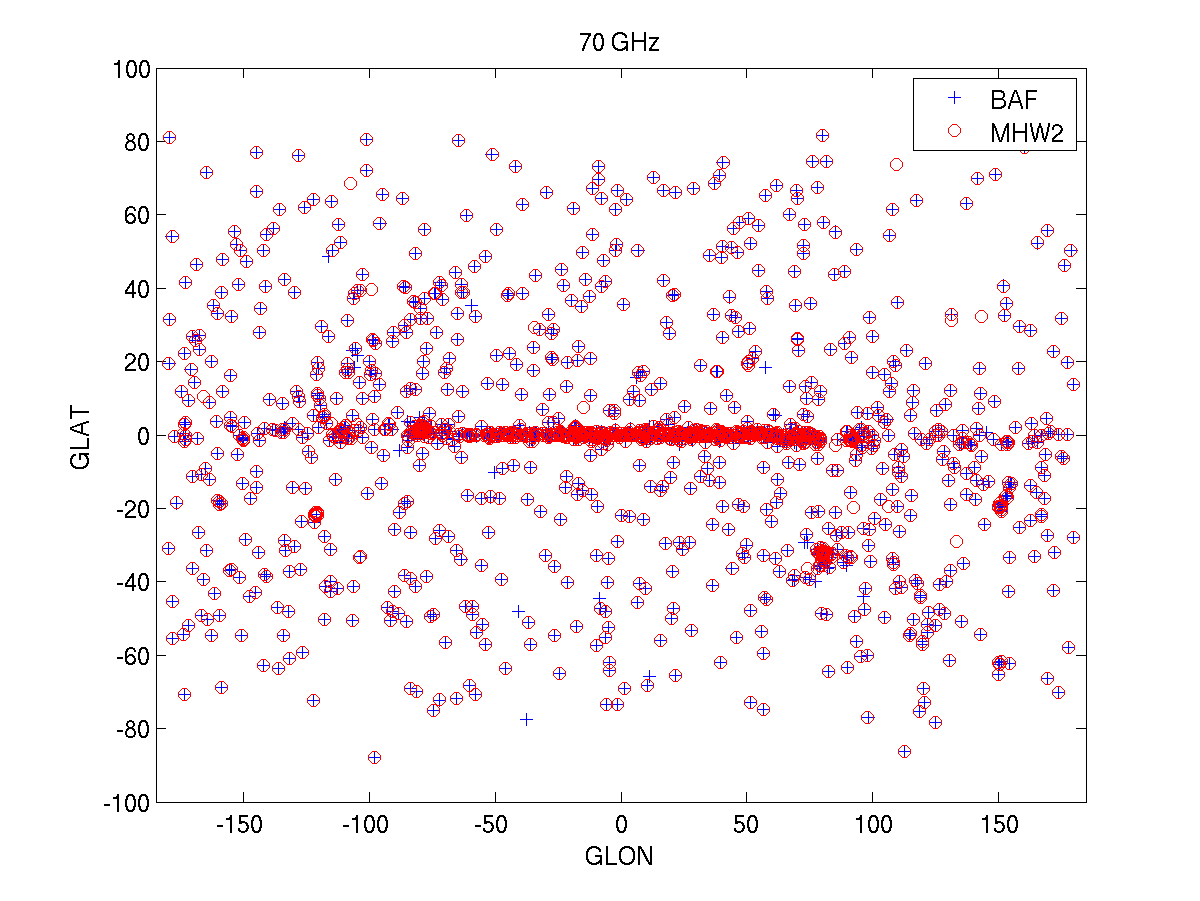}
\includegraphics[width=0.45\textwidth]{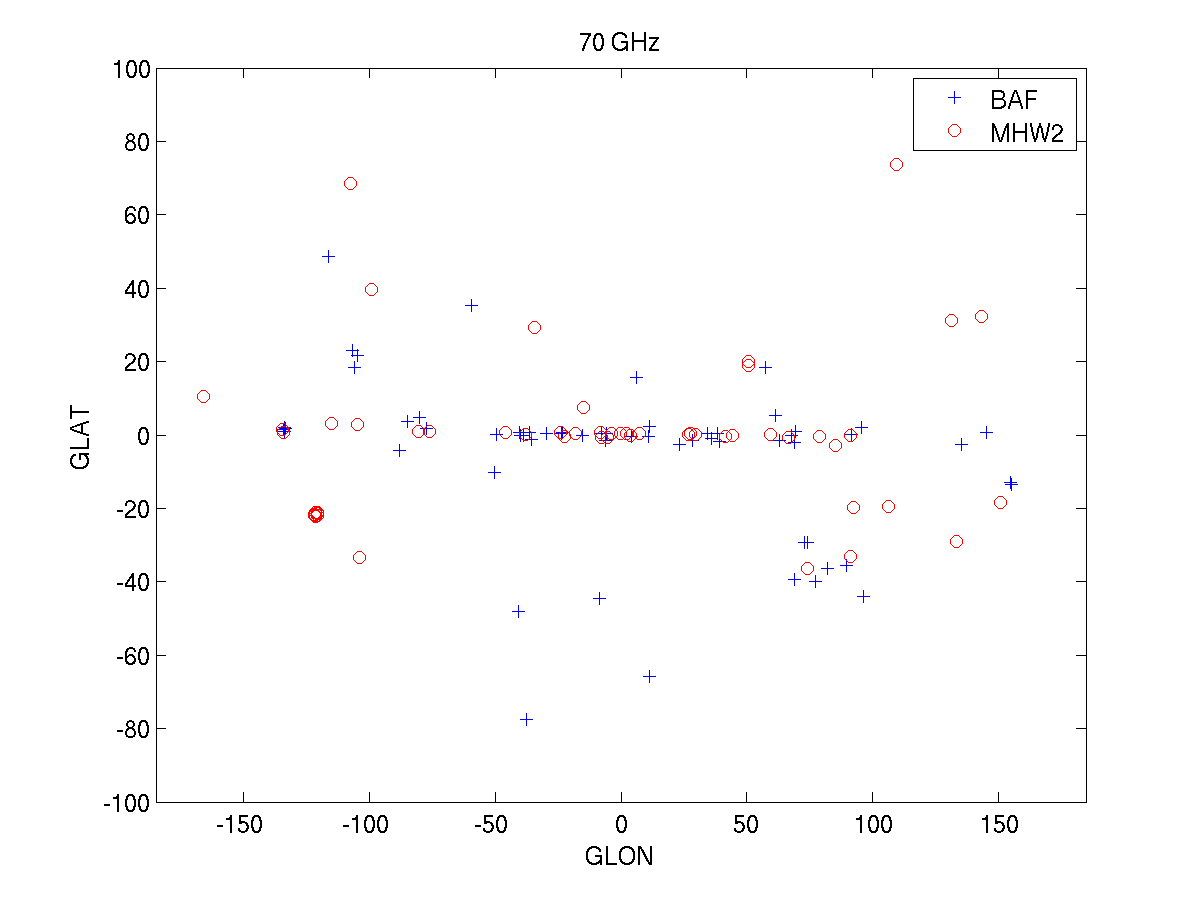}
 \caption{Position of the detected sources at 30 (upper panels), 44 (medium panels) and 70 GHz (lower panels) for the two considered techniques, MHW2 and BAF.In the left panels we show the position in the sky of all the objects detected above $SNR>5$. In the right panels we show only those objects $SNR>5$ that were detected by one method and not by the other, and viceversa.}
 \label{fig:positions1}
 \end{center}
 \end{figure*}

\begin{figure*}
 \begin{center}
\includegraphics[width=0.45\textwidth]{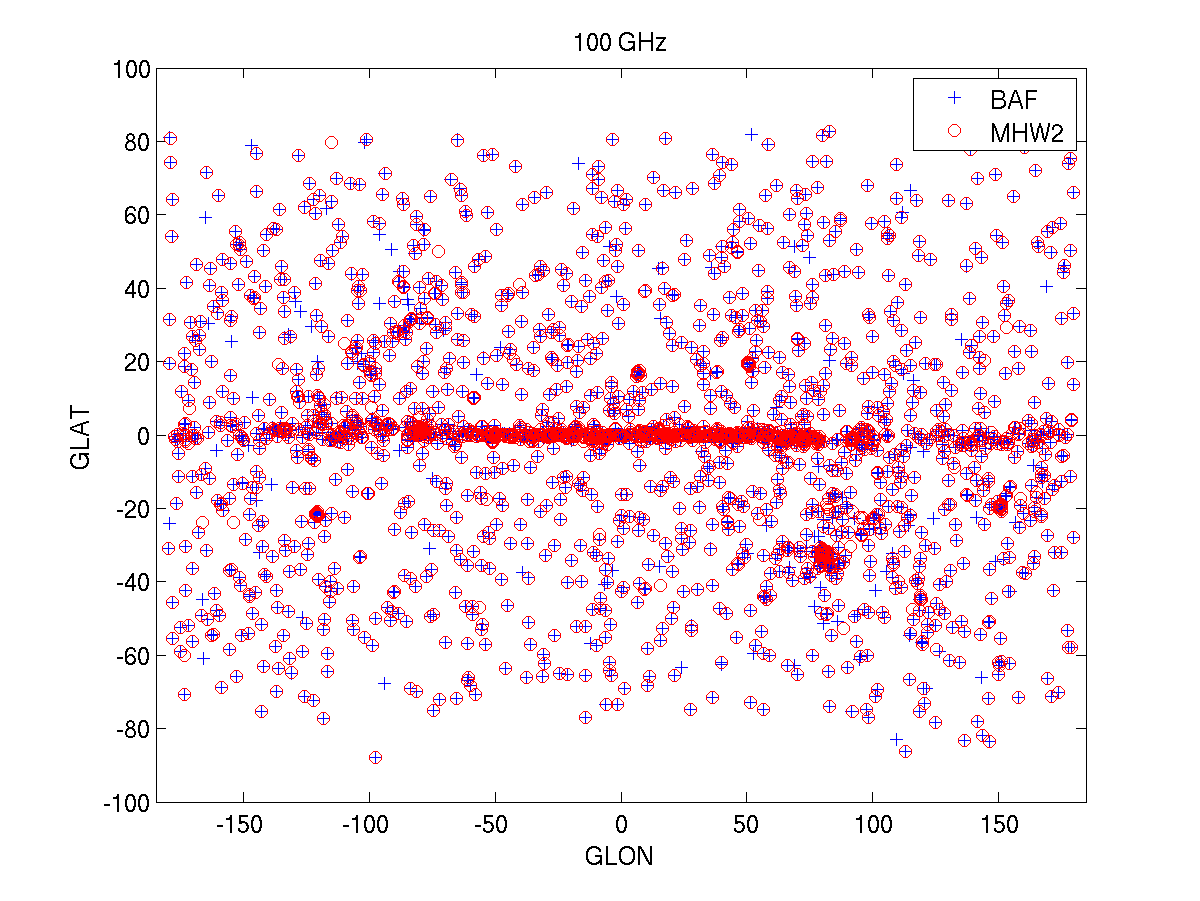}
\includegraphics[width=0.45\textwidth]{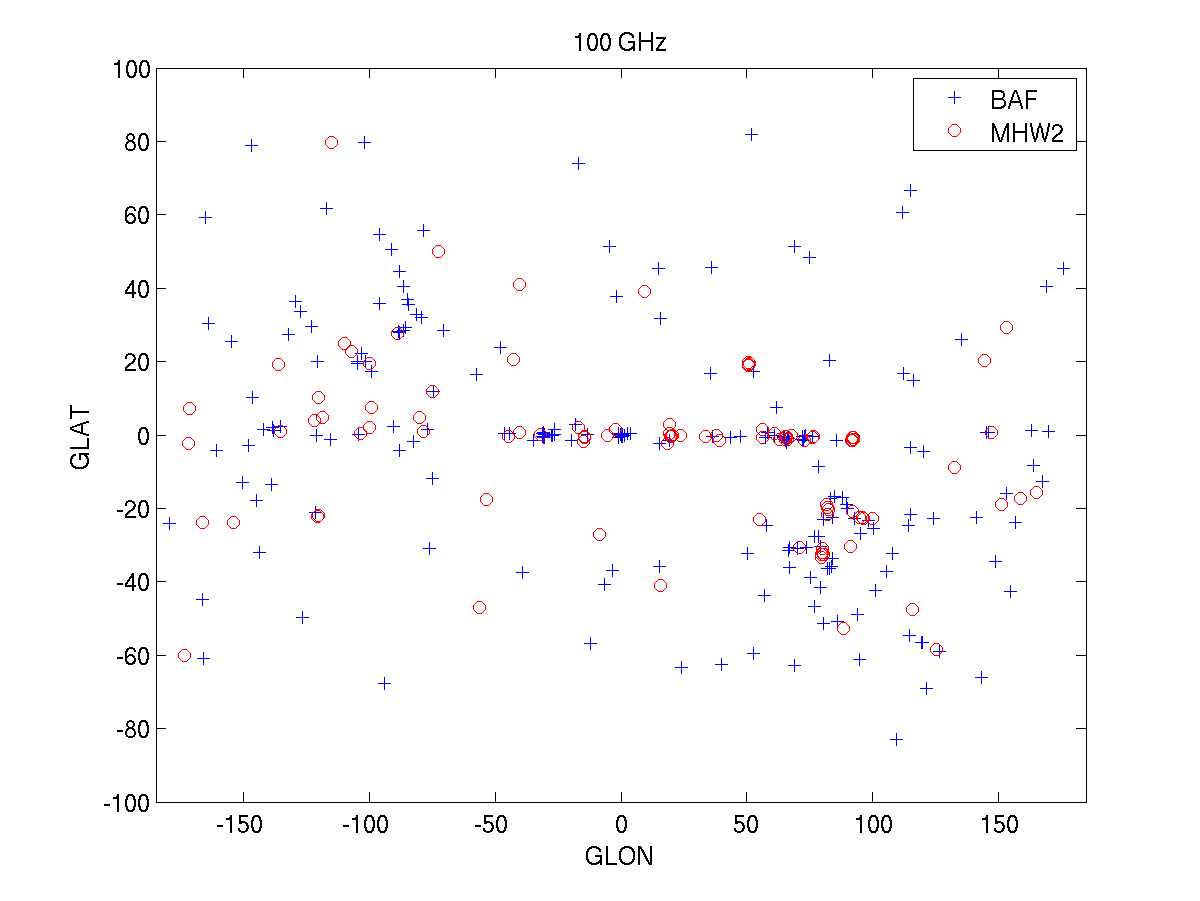}
\includegraphics[width=0.45\textwidth]{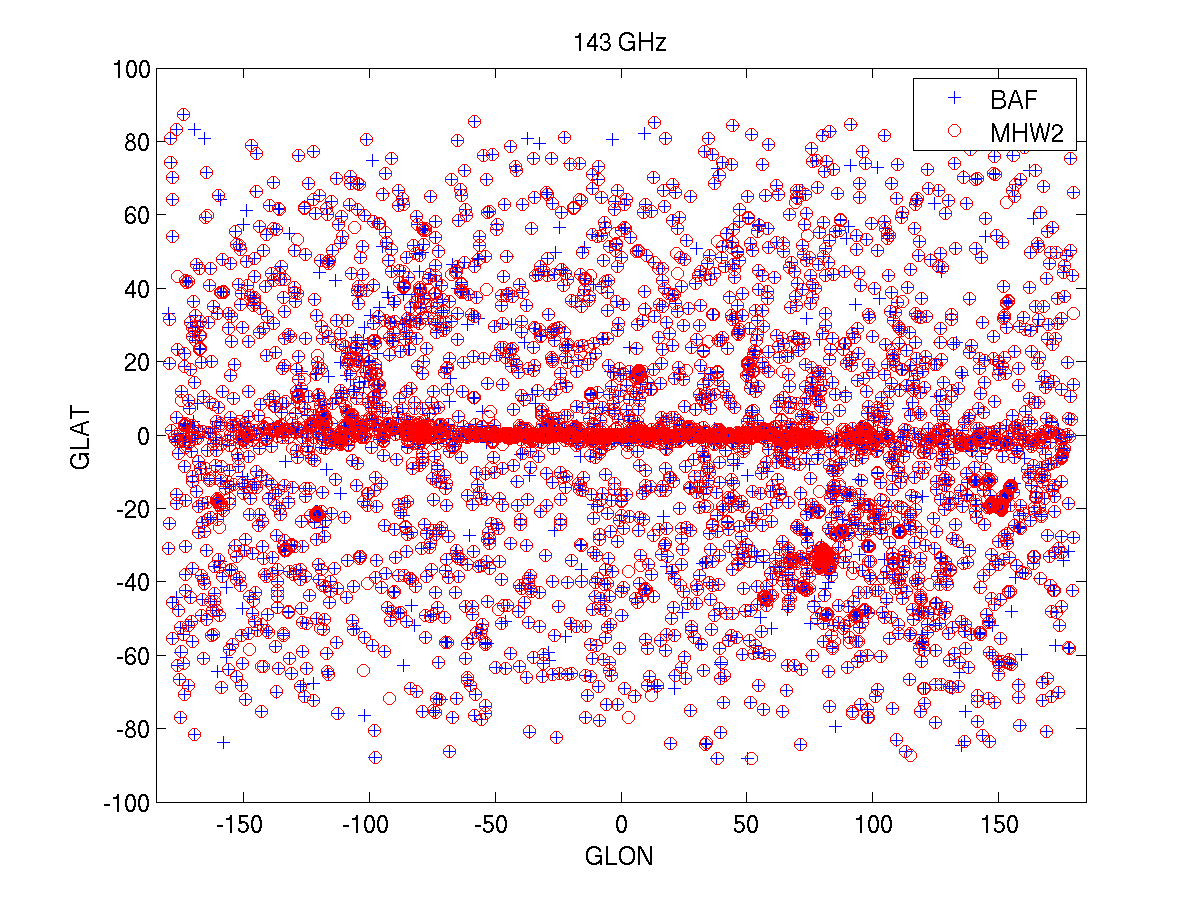}
\includegraphics[width=0.45\textwidth]{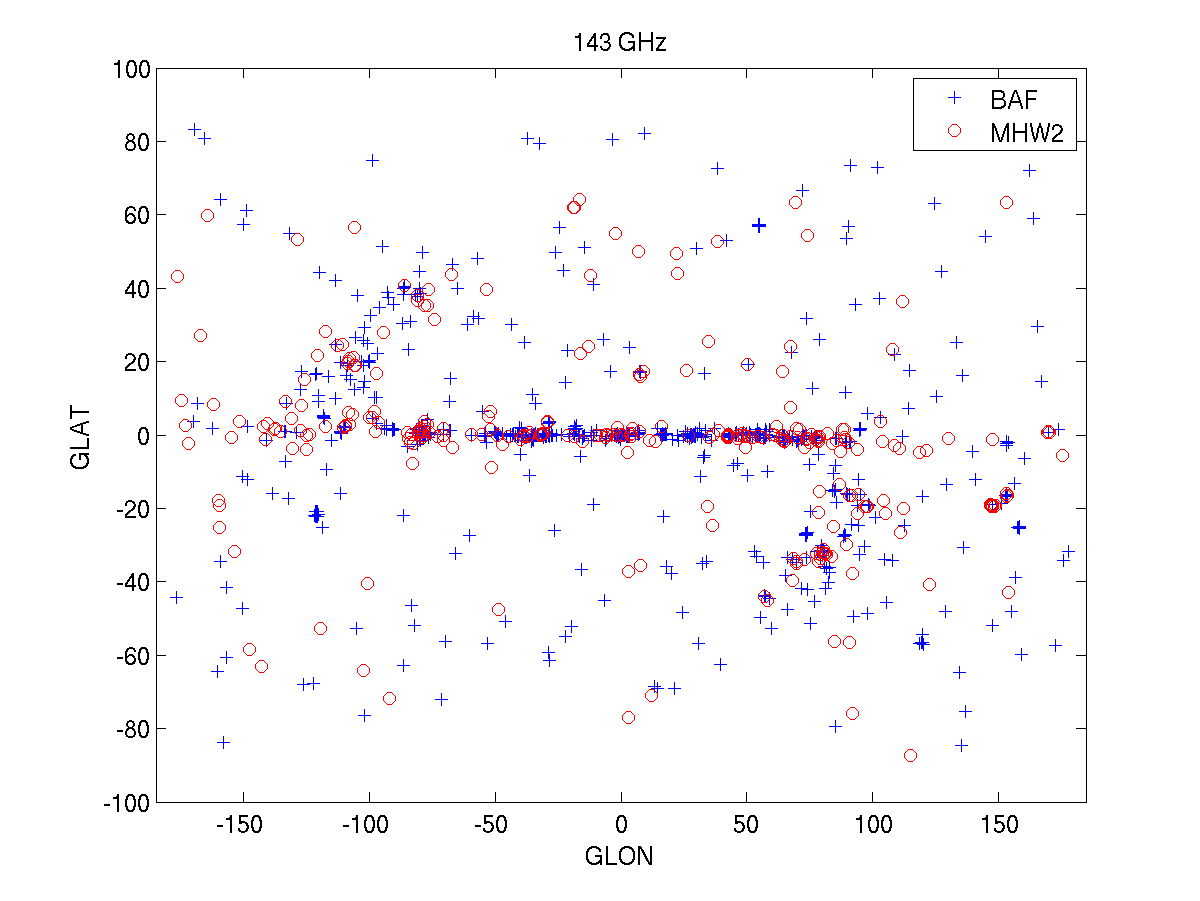}
\includegraphics[width=0.45\textwidth]{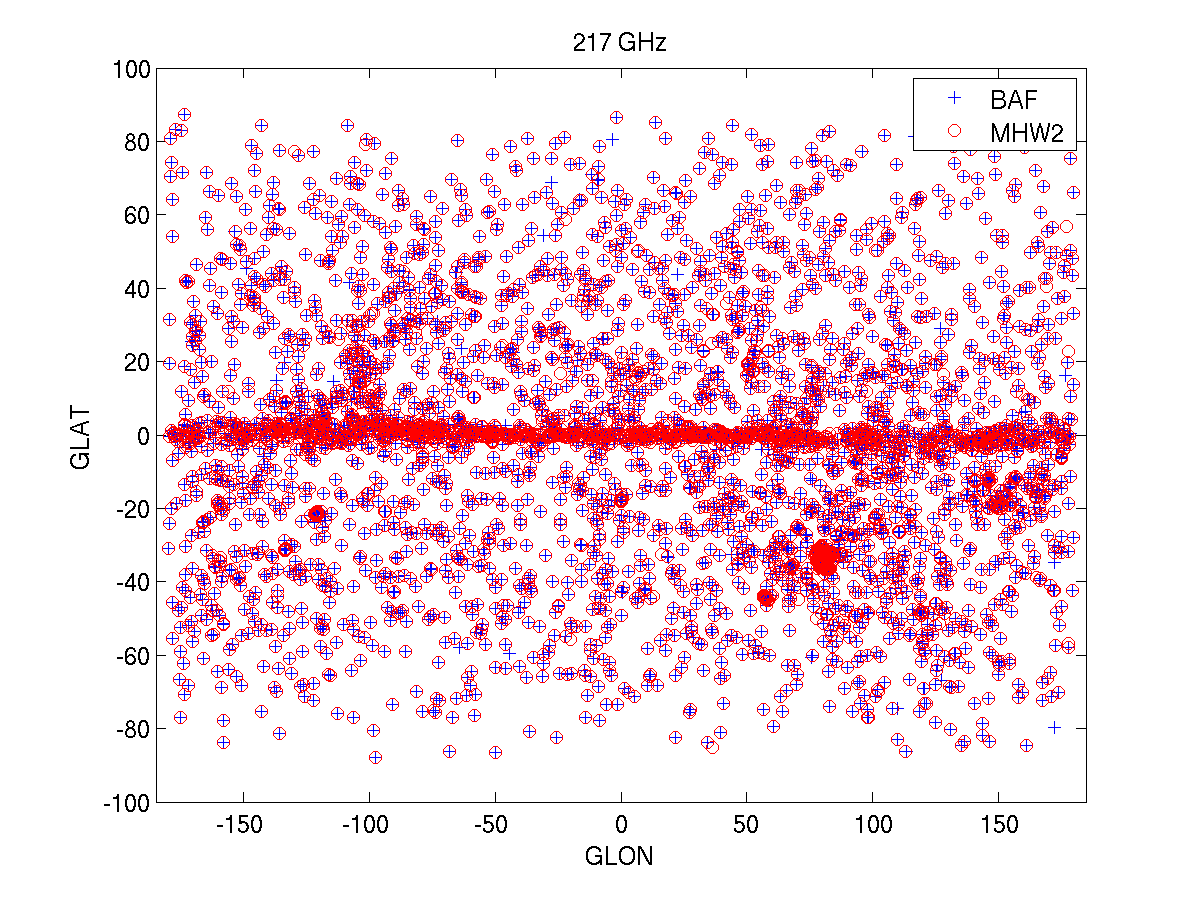}
\includegraphics[width=0.45\textwidth]{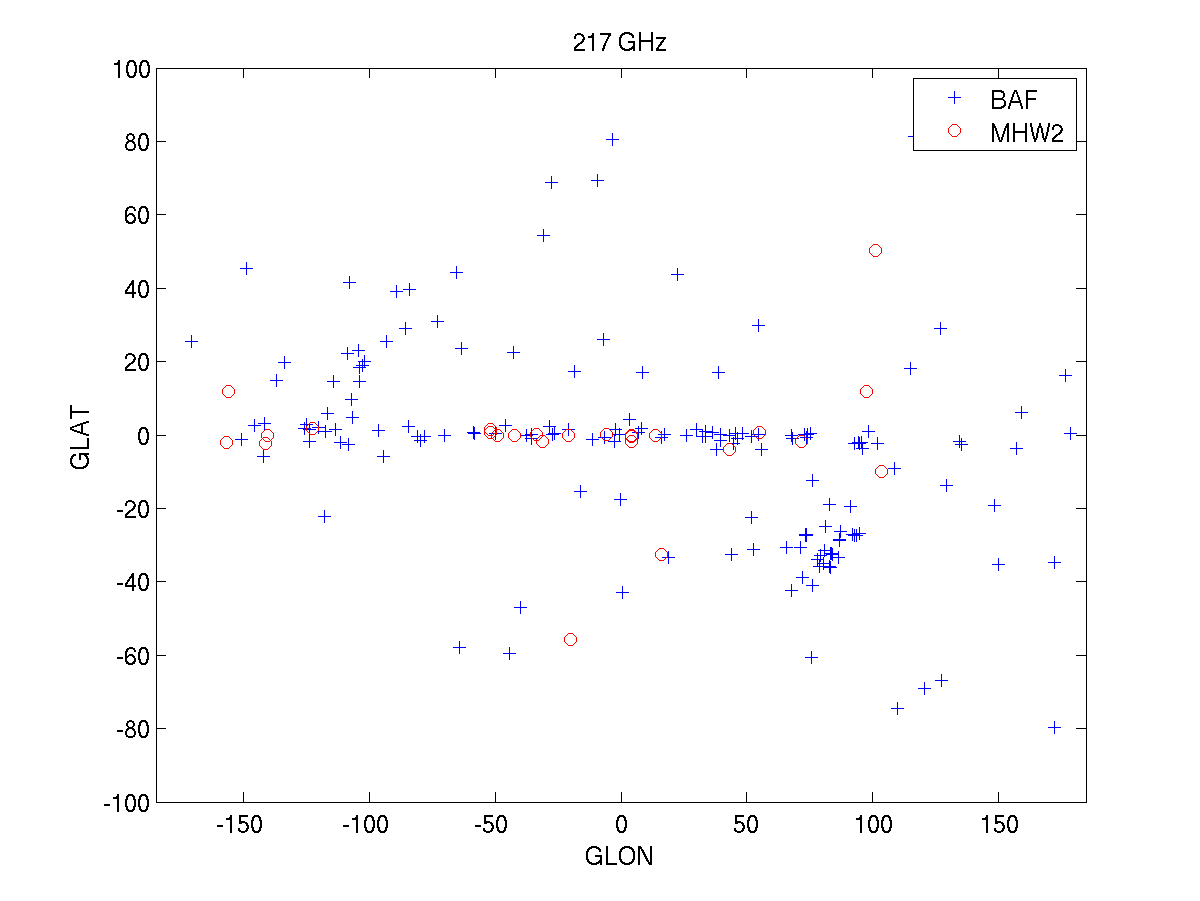}
 \caption{Position of the detected sources between 100 and 217 GHz for the two considered techniques, MHW2 and BAF.In the left panels we show the position in the sky of all the objects detected above $SNR>5$. In the right panels we show only those objects $SNR>5$ that were detected by one method and not by the other, and viceversa.}
 \label{fig:positions2}
 \end{center}
 \end{figure*}

\begin{figure*}
 \begin{center}
\includegraphics[width=0.45\textwidth]{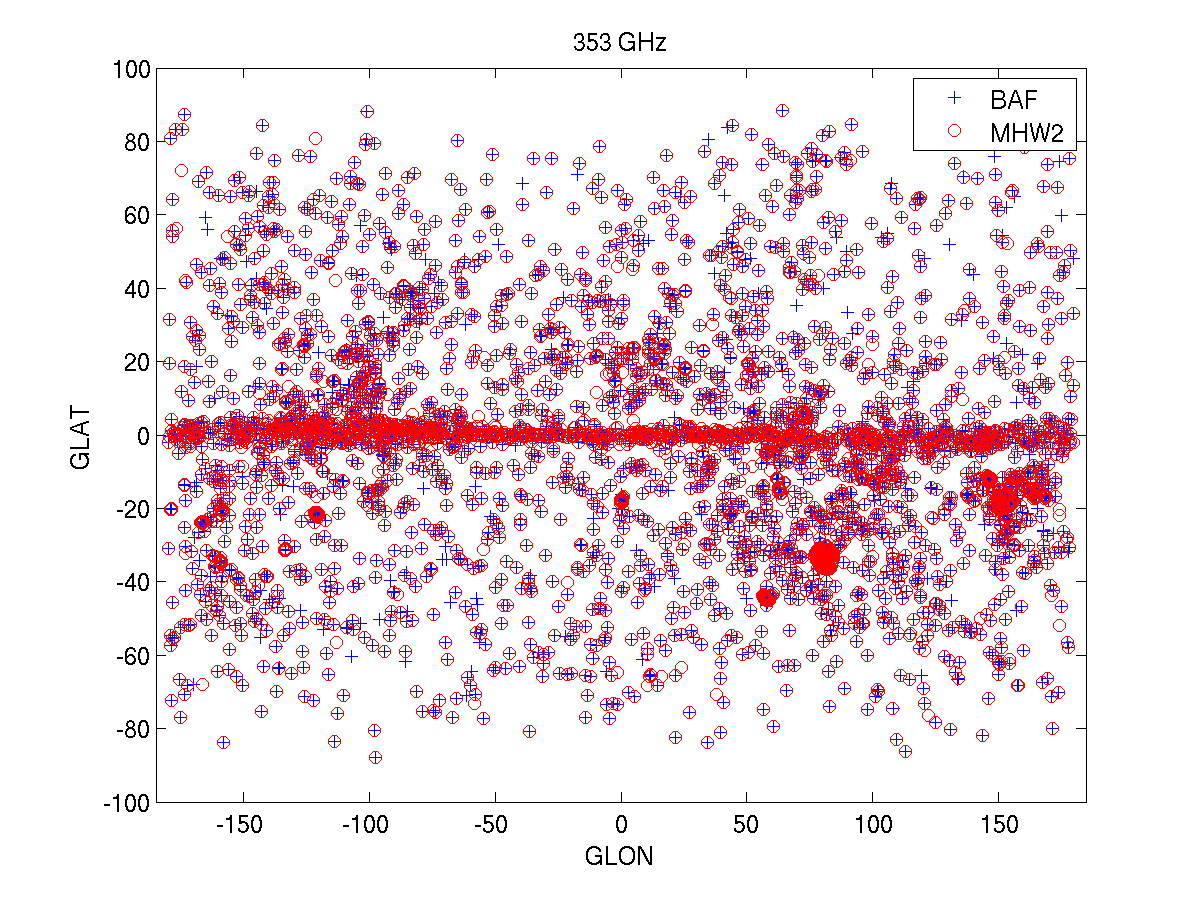}
\includegraphics[width=0.45\textwidth]{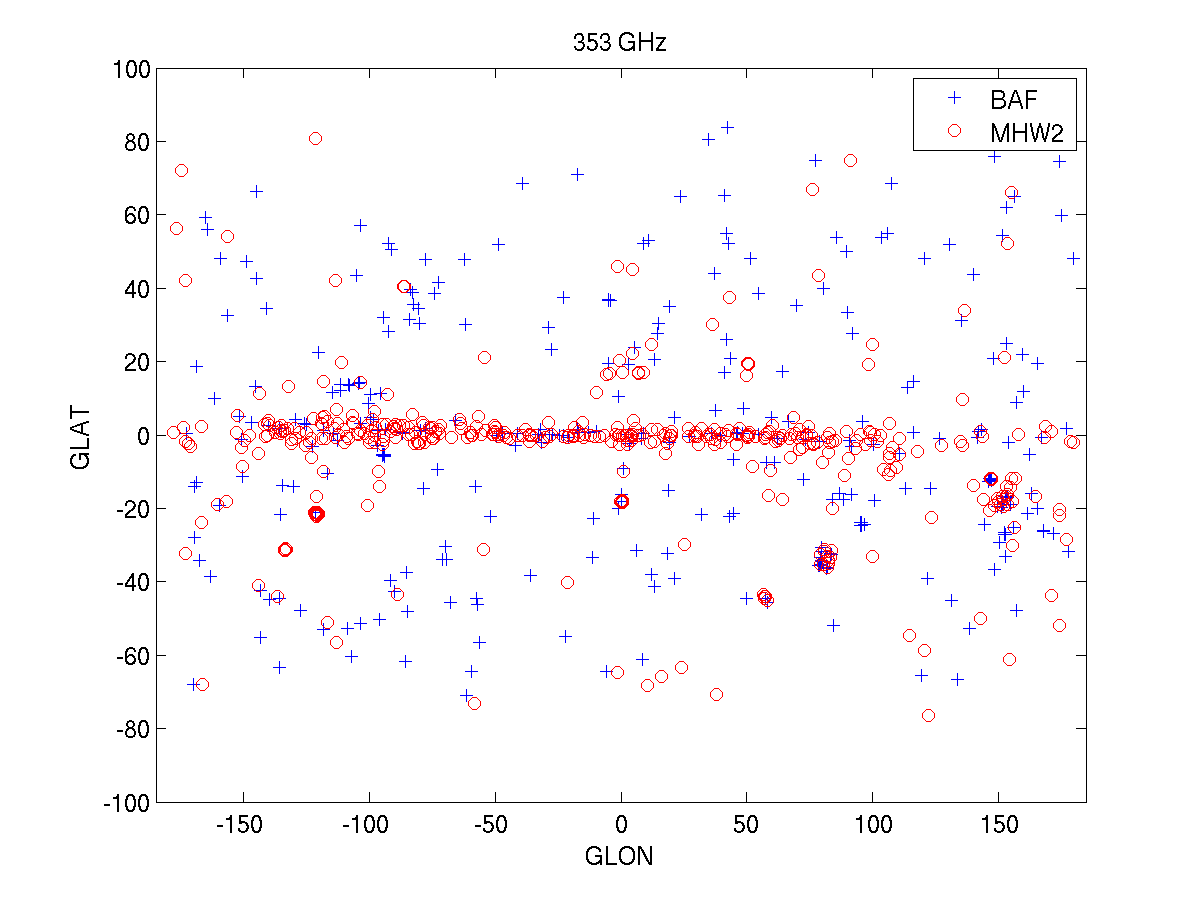}
\includegraphics[width=0.45\textwidth]{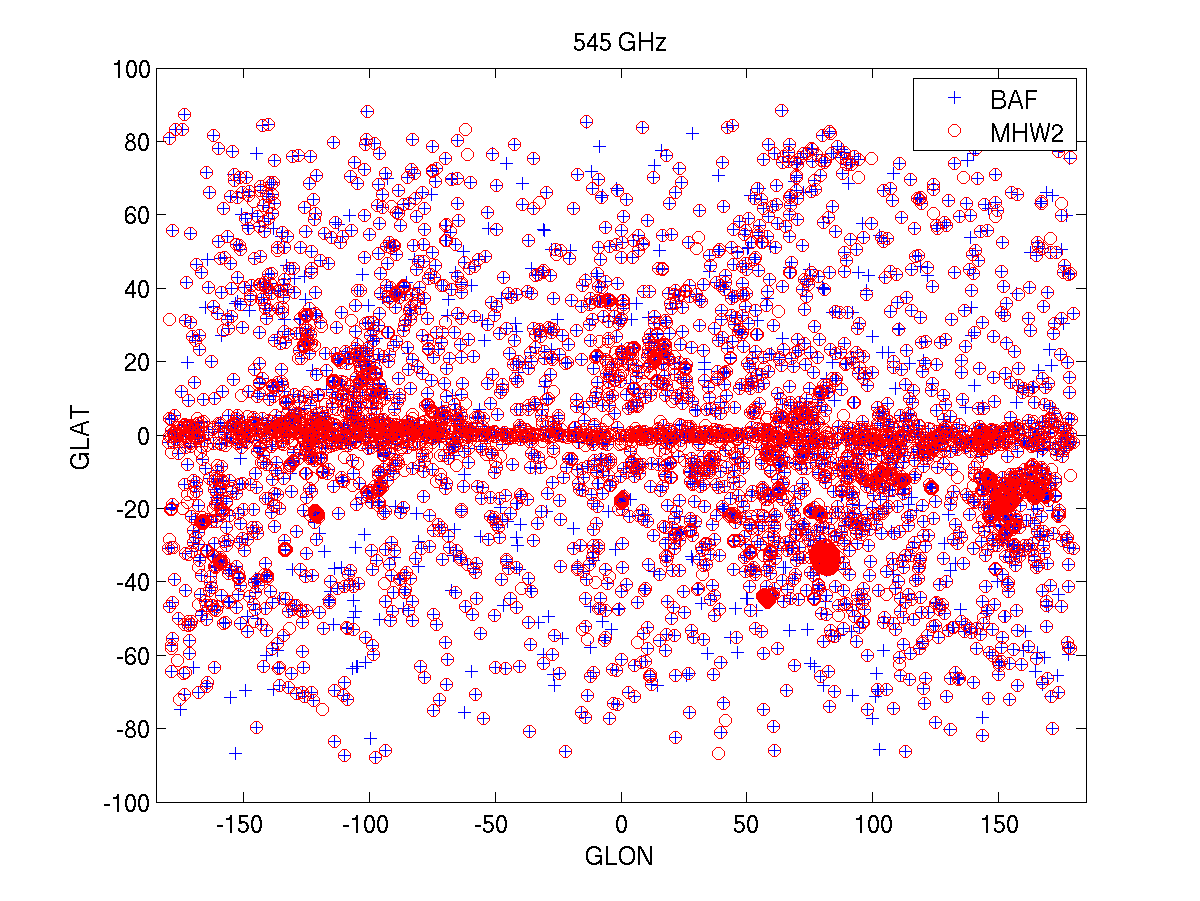}
\includegraphics[width=0.45\textwidth]{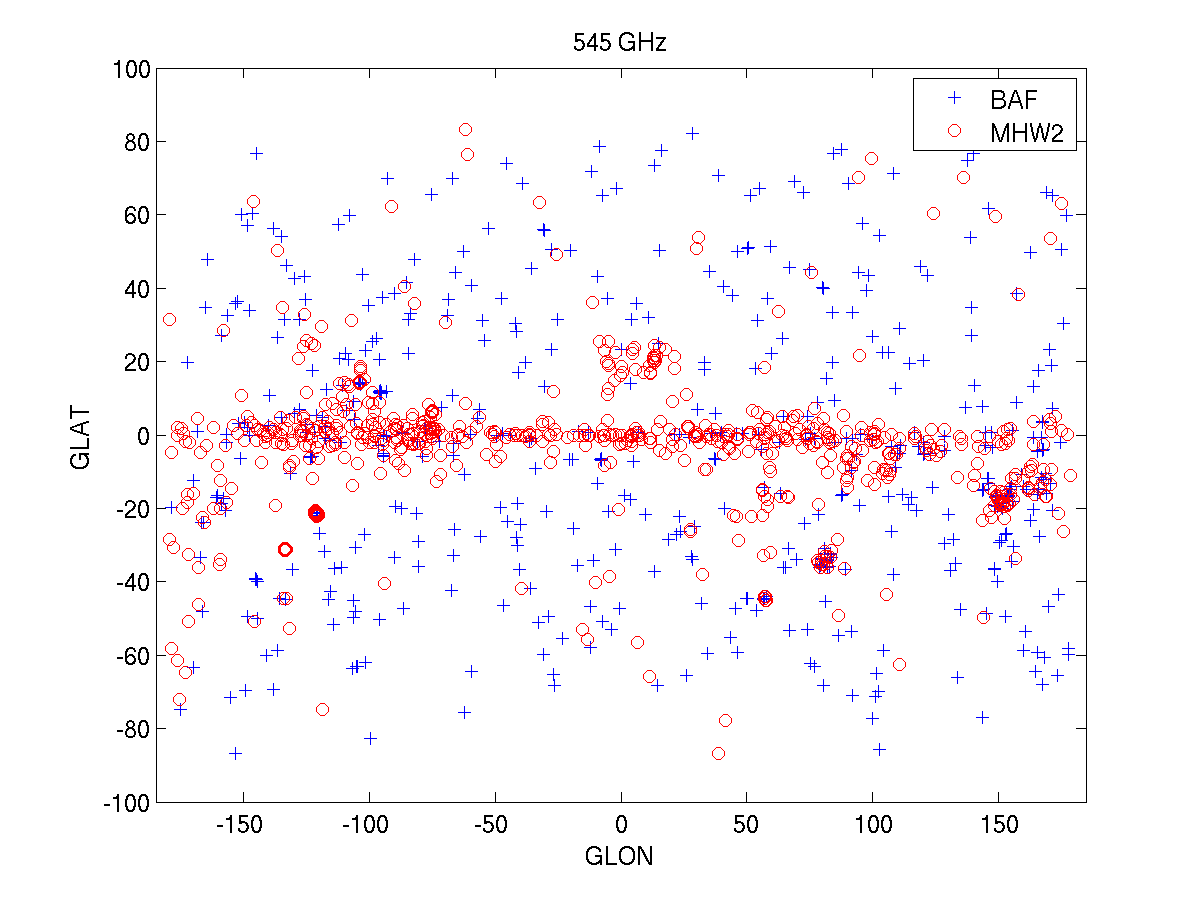}
\includegraphics[width=0.45\textwidth]{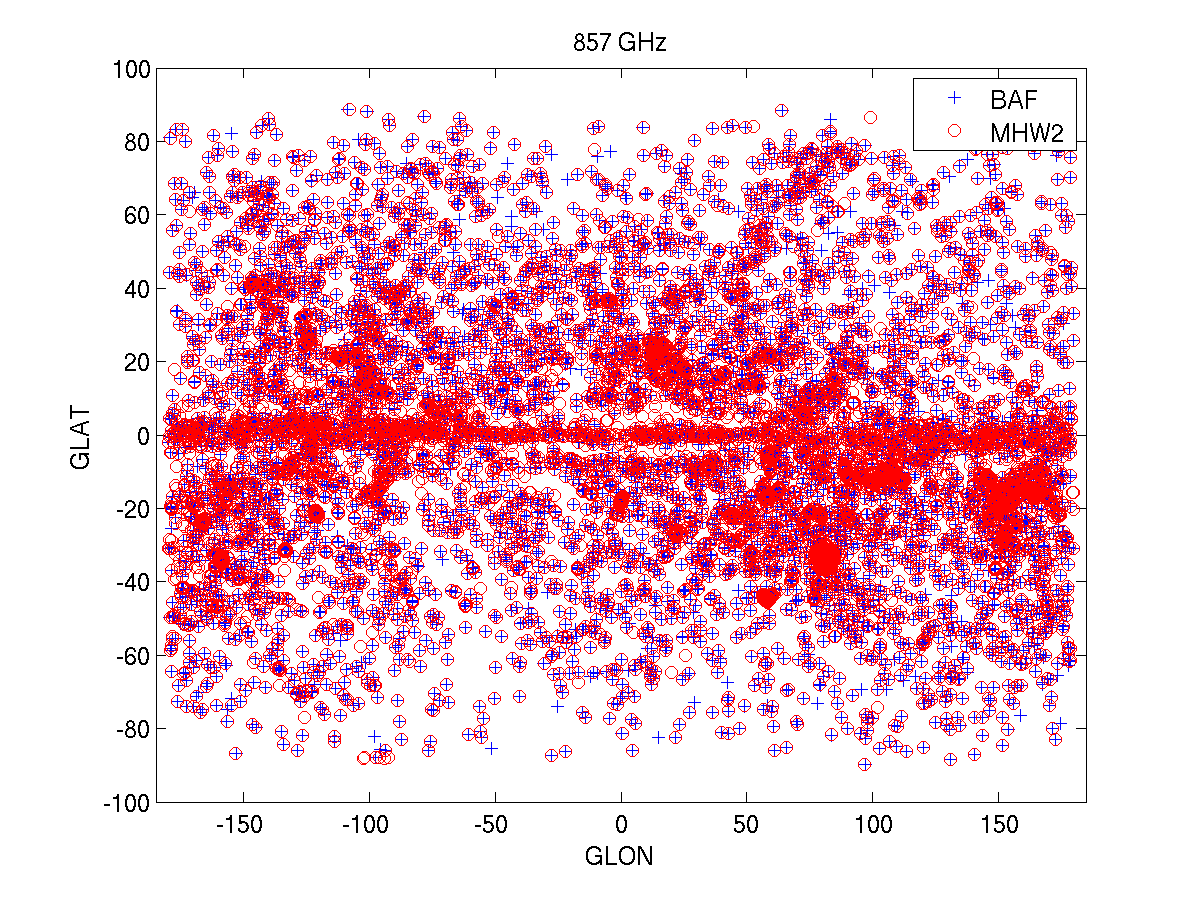}
\includegraphics[width=0.45\textwidth]{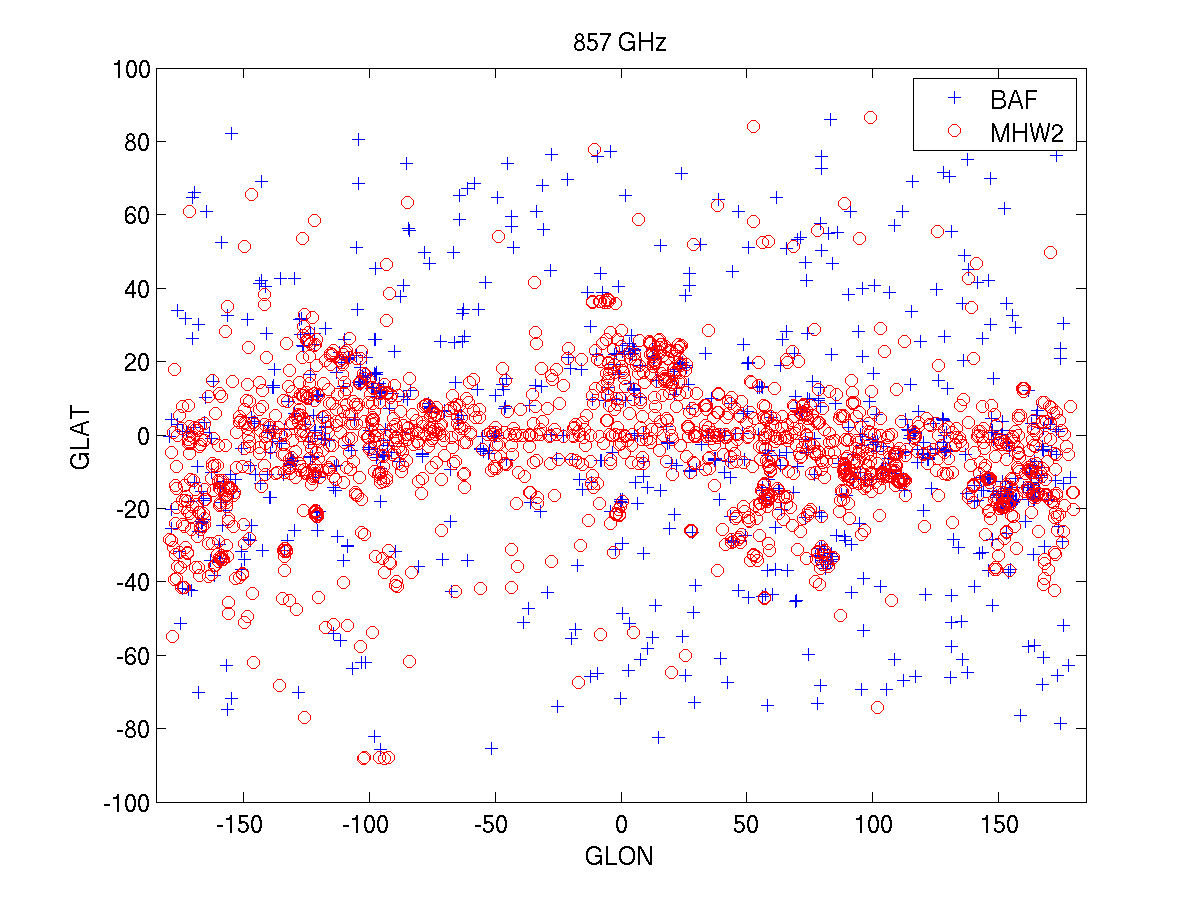}
 \caption{Position of the detected sources between 353 and 857 GHz for the two considered techniques, MHW2 and BAF.In the left panels we show the position in the sky of all the objects detected above $SNR>5$. In the right panels we show only those objects $SNR>5$ that were detected by one method and not by the other, and viceversa.}
 \label{fig:positions3}
 \end{center}
 \end{figure*}

 Moreover, to illustrate the performance of the new filter we will do a simple
 comparison with the MHW2 in terms of the number of detections above
 $SNR>5$ obtained by each method when applied to the nine simulated
 maps. In the left and right panels of Figures \ref{fig:positions1},
 \ref{fig:positions2} and \ref{fig:positions3} we show the positions
 in the sky of the sources detected by the BAF (left panels) and those detected by the BAF or the
 MHW2 only (right panels). In these figures one can see that the number of
 objects detected in the vicinity of the galactic plane, as well as
 other complex regions such as the LMC, Orion, etc., is large and
 experience tell us that a fraction of those detections are not
 true sources but bright compact galactic structure that looks like
 sources. Starting at 217 GHz, one can notice that the number of
 detections increases rapidly due to the change in the dominant
 population of sources, from non-thermal radio sources to thermal
 infra-red sources. In addition, one can see that the BAF seems to
 detect more new sources all over the sky as opposed to the MHW2 that
 concentrates most of its new detections in the vicinity of the
 galactic plane, many of which tend to be spurious detections caused by filaments and
 extended structures in the galaxy. This is an important result
 because it shows the potential of the BAF, a filter that better
 adapts itself to the local properties of the background, removing
 part of the noise and large scale emission more effectively. This
 will imply a reduction in the number of detections in complex regions
 (that are likely to be spurious) while increasing the number of real
 detections in cleaner regions of the sky due to the improved
 estimation of the noise.

 In Table \ref{tab:tabla_comparacion}, we show the number of detections with a $SNR>5$ for both
 techniques when applied to the same regions of the sky, in one case
 covering the whole sky and in the other case applying a large
 galactic cut of $\pm 30$ degrees. One can see that the BAF 
 detects, in general, more sources, specially between 30 to
 143 GHz. In some channels the improvement is almost inexistent, for
 example at 70 GHz. Here, due to the properties of the background, the
 optimal value of $g$ is essentially $4$, which corresponds to the
 MHW2. On the contrary, in the upper frequency channels (353, 545 and
 857 GHz) the number of $SNR>5$ sources produced by the MHW2 is a bit
 larger than those of the BAF. Note that these results apply when
 considering the detection above $SNR>5$ in the whole sky, including
 the galactic plane where we know that the MHW2 is detecting more
 sources likely to be galactic emission rather than extragalactic
 compact sources. If one looks at the second case, where a galactic
 cut has been applied, the BAF detects more sources than the MHW2 in
 all cases except for one, 217 GHz, where the difference is very
 small. Note that these numbers are in agreement to what it was
 mentioned above. If we inspect the right panels of Figures
 \ref{fig:positions1}, \ref{fig:positions2} and \ref{fig:positions3}
 one can see that the MHW2 tends to detect more objects in complex
 regions than the new filter, many of which are most likely not true
 sources. This is telling us that the new filter is removing the
 background more efficiently which implies not only an improved
 estimation of the SNR but also a decrease in the number of spurious
 detections. 
 
\begin{table}
\begin{center}
\begin{tabular}{|c|c|c|c|c|}
\hline $\nu$ & $BAF$ & $MHW2$ & $BAF_{30}$ & $MHW2_{30}$ \\ 
\hline  30 & 1400 & 1298 & 598 & 545   \\ 
\hline  44 & 1082 & 1037 & 434 & 412   \\ 
\hline  70 & 1175 & 1172 & 401 & 398   \\ 
\hline 100 & 1975 & 1889 & 752 & 695   \\ 
\hline 143 & 4001 & 3608 & 1531 & 1440   \\ 
\hline 217 & 3753 & 3856 & 1557 & 1570   \\ 
\hline 353 & 3495 & 3563 & 1313 & 1247   \\ 
\hline 545 & 4325 & 4461 & 1582 & 1440   \\ 
\hline 857 & 8460 & 9137 & 3429 & 3382   \\ 
\hline 
\end{tabular} 
\label{tab:tabla_comparacion}
 \caption{This table shows the number of detections above $SNR>5$ that we have found applying the MHW2 and the BAF techniques at the nine simulated maps. In addition we show the number of detections above $SNR>5$ and galactic cut of $\pm30$ degrees to have an idea of the effect of the galactic emissions when doing this kind of simple comparisons.}
\end{center}
\end{table}

 In addition, we have compared the catalogs of detections obtained with the BAF and the MHW2, searching for those objects detected only by one method and not by the other. We find that the BAF provides a higher number of unique detections in the maps from 30 to 217 GHz, whereas the MHW2 detects more objects at the highest frequencies (see Table \ref{tab:tabla_comp_mask}). A deeper comparison is made when distinguishing between unique detections in highly contaminated regions (e.g., in the vicinity of the Galactic plane) and cleaner areas in the sky. Obviously, as mentioned above, the former will correspond (at high probability) with spurious detections associated to extended objects, whereas the latter will correspond most probably with true point sources. In order to identify the contaminated regions, we generate two types of masks. The first one includes the $15\%$ of the brightest pixels in each map. The second one, more conservative than the other, includes the $25\%$ of brightest pixels in each map. We use the less conservative $15\%$ masks between 30 and 217 GHz and the most conservative one between 353 and 857 GHz, where the galactic dust emission extends to higher latitudes. In practice, these masks identify very well the galactic emission and a few complex regions across the sky (e.g., Magellanic clouds, Orion, Ophiucos). In Table \ref{tab:tabla_comp_mask} one can also see that BAF detects more unique objects both inside and outside the mask between 30 and 217 GHz, whereas in the most contaminated channels at 353 GHz and above, it also detects more unique objects outside the mask while the MHW2 detects up to three times more unique objects inside the mask, many of which are most likely galactic extended emission rather than extragalactic point sources.
 
\begin{table}
\begin{center}
\begin{tabular}{|c|c|c|c|c|c|c|}
\hline 
$\nu$ &  & $MHW2$ & & & $BAF$ &  \\ 
\hline [GHz] & all-sky & out & in & all-sky & out & in \\
\hline  30 &  55  &  33  & 22  & 146 & 111 & 35  \\ 
\hline  44 &   3  &  3   & 0   & 84  &  49 & 35  \\ 
\hline  70 &  54  &  23  & 31  & 60  &  18 & 42   \\ 
\hline 100 &  99  &  45  & 54  & 201 & 125 & 76   \\ 
\hline 143 &  287 &  104 & 183 & 731 & 279 & 452   \\ 
\hline 217 &  25  &   6  & 19  & 167 &  88 & 79   \\ 
\hline 353 &  448 &  92 & 356 & 312 & 175 & 137   \\ 
\hline 545 &  720 &  138 & 582 & 530 & 323 & 207   \\ 
\hline 857 &  1614 & 353 & 1261 & 882 & 391 & 491  \\ 
\hline 
\end{tabular} 
\label{tab:tabla_comp_mask}
 \caption{In this table we present the total number of detections above $SNR>5$ obtained by one method and not by the other, as well as the number of detections inside and outside a a mask that we have defined for each frequency and that includes the $15\%$ of the brightest pixels in the maps between 30 and 217 GHz and a more conservative $25\%$ of the brightest pixels between 353 and 857 GHz. We use these masks to define what can be considered as a complex region and what is not.}
\end{center}
\end{table}

\begin{figure*}
\begin{center}
\includegraphics[width=0.32\textwidth]{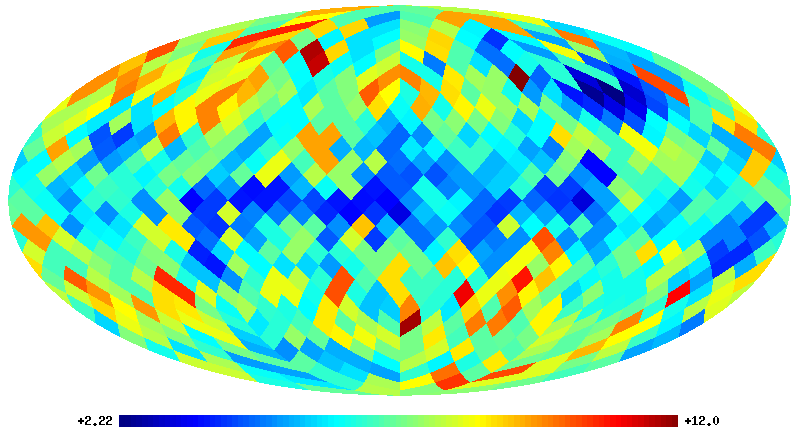}
\includegraphics[width=0.32\textwidth]{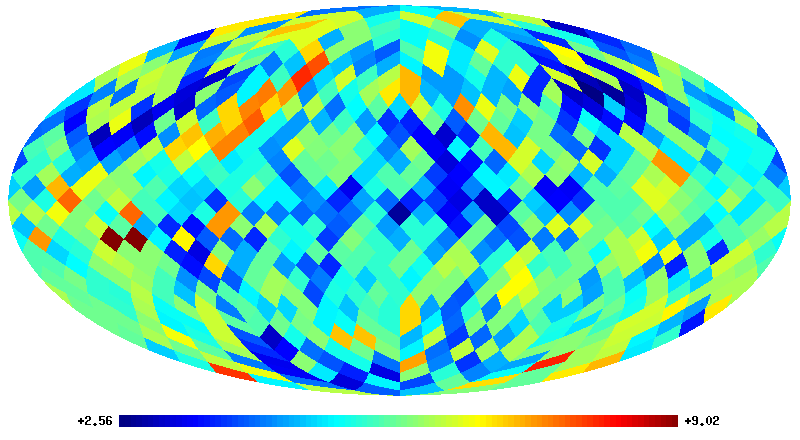}
\includegraphics[width=0.32\textwidth]{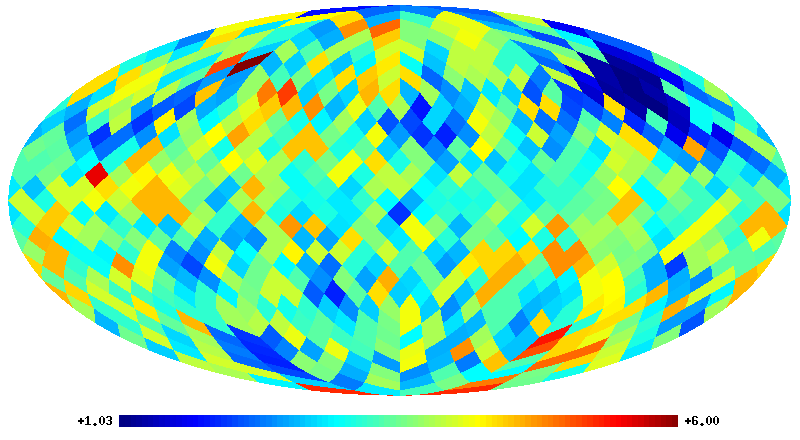}
\includegraphics[width=0.32\textwidth]{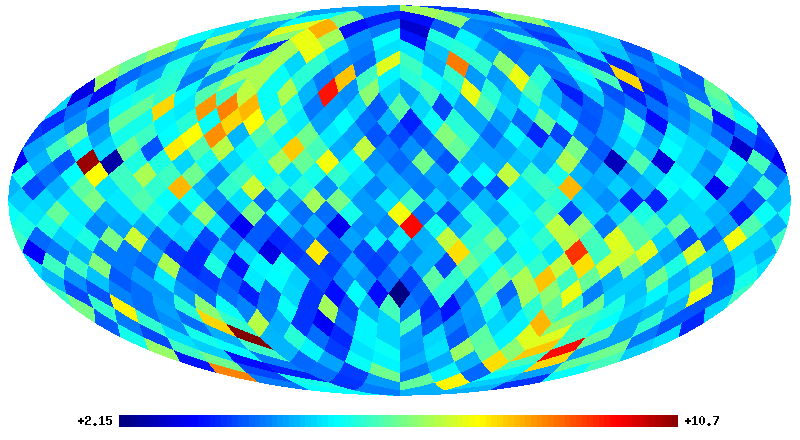}
\includegraphics[width=0.32\textwidth]{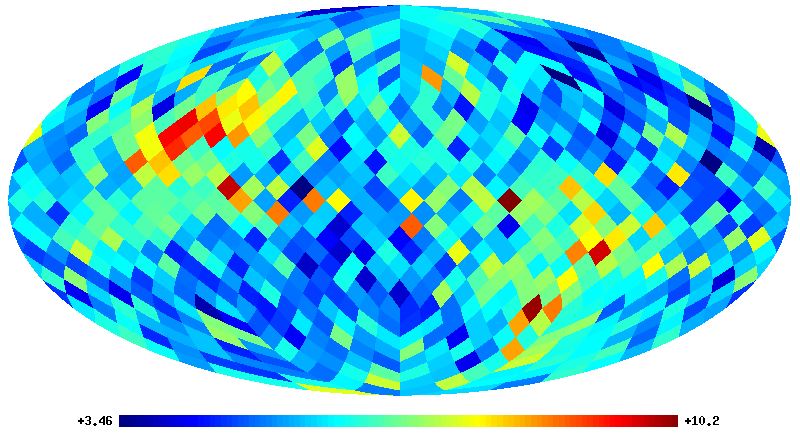}
\includegraphics[width=0.32\textwidth]{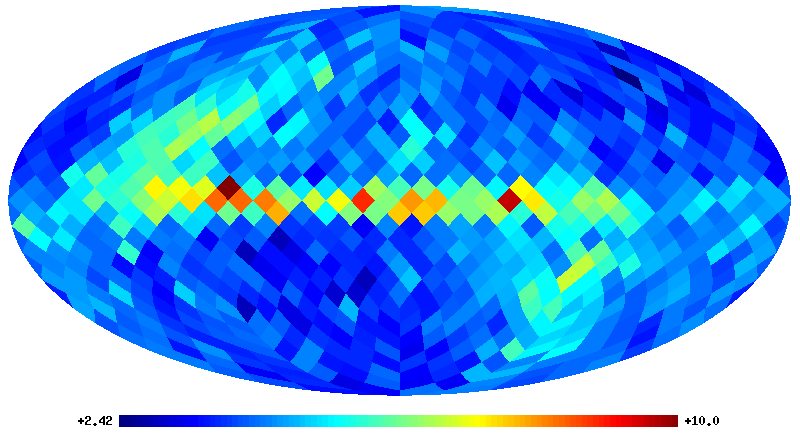}
\includegraphics[width=0.32\textwidth]{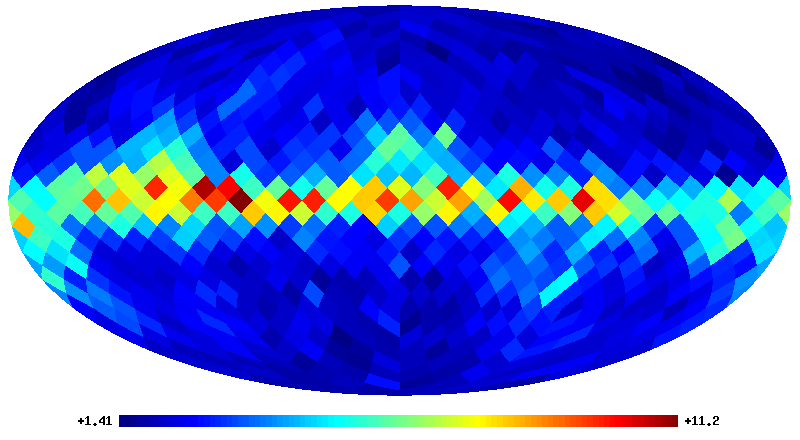}
\includegraphics[width=0.32\textwidth]{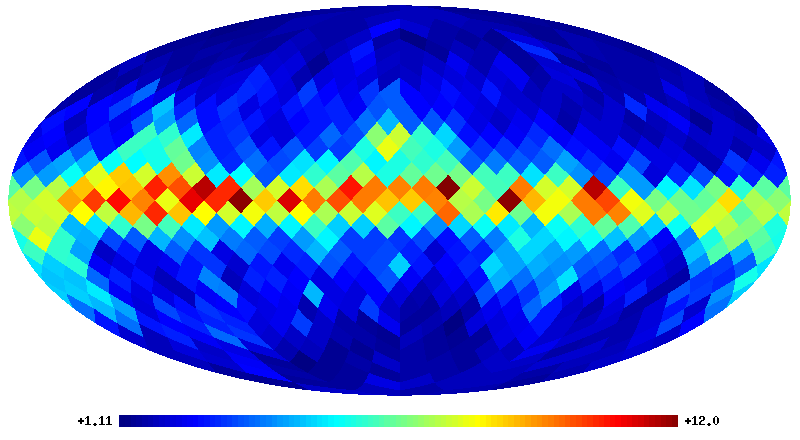}
\includegraphics[width=0.32\textwidth]{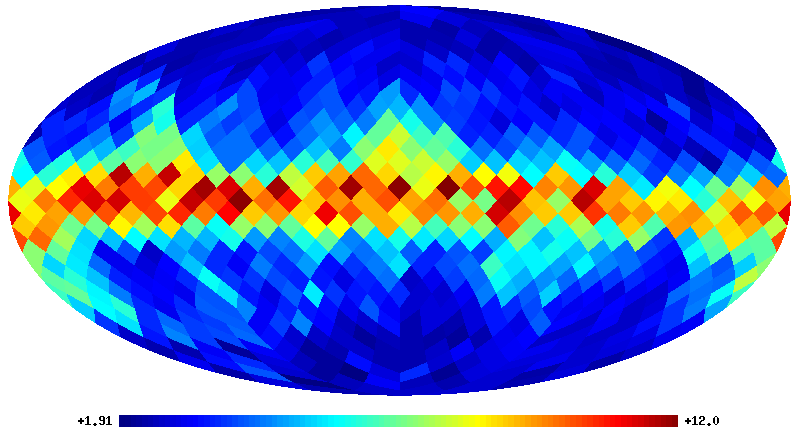}
 \caption{Spatial distribution across the sky of the values of the filter \iindex $g$ found during the analysis of nine simulations between 30 and 857 GHz. The upper panels show the 30, 44 and 70 GHz cases, the middle panels show the 100, 143 and 217 GHz cases and the lower panels show the 353, 545 ad 857 cases.}
 \label{fig:mapg}
 \end{center}
 \end{figure*}

\begin{figure*}
\begin{center}
\includegraphics[width=0.32\textwidth]{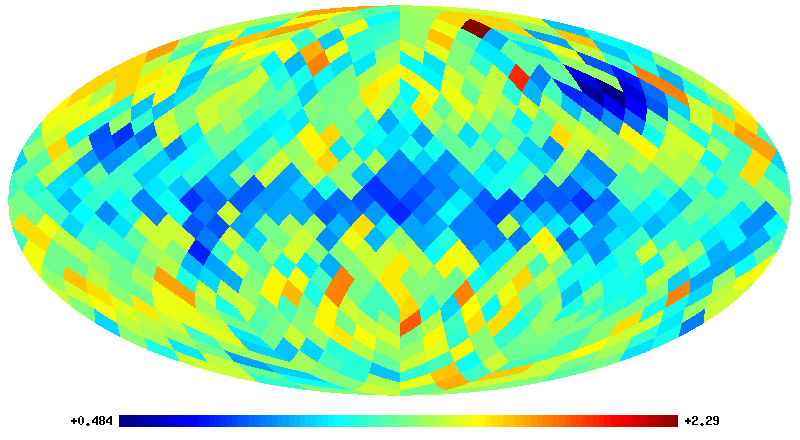}
\includegraphics[width=0.32\textwidth]{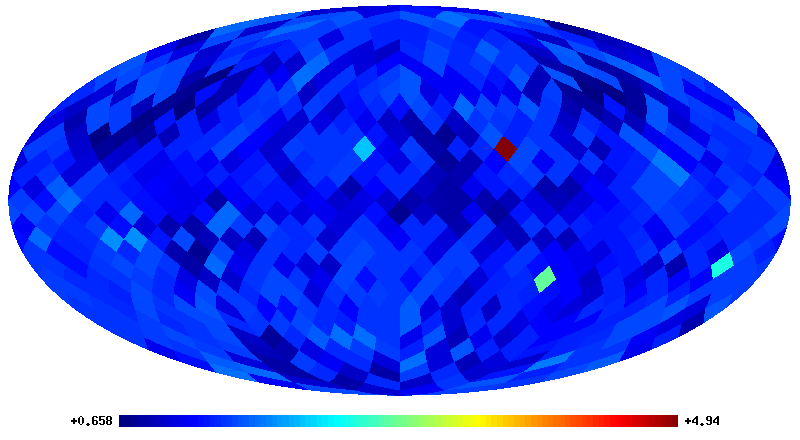}
\includegraphics[width=0.32\textwidth]{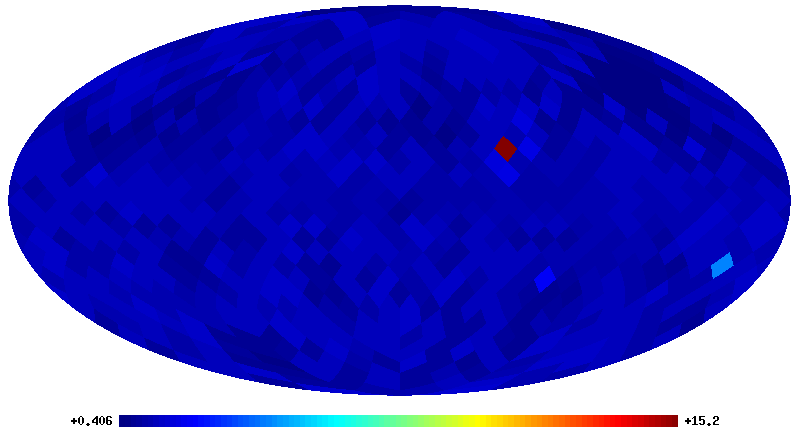}
\includegraphics[width=0.32\textwidth]{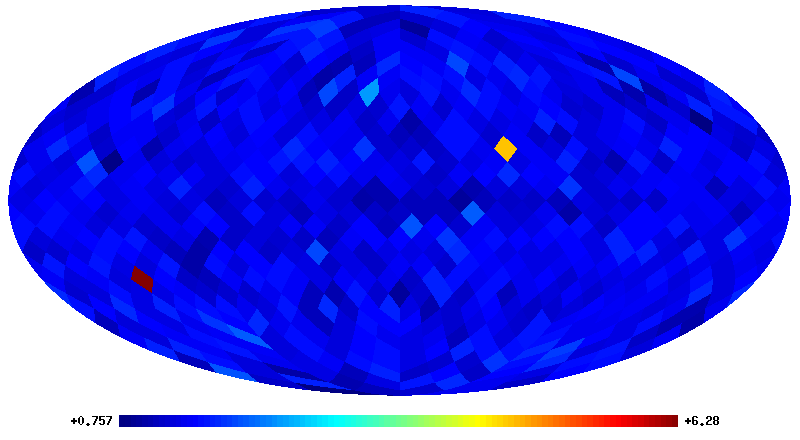}
\includegraphics[width=0.32\textwidth]{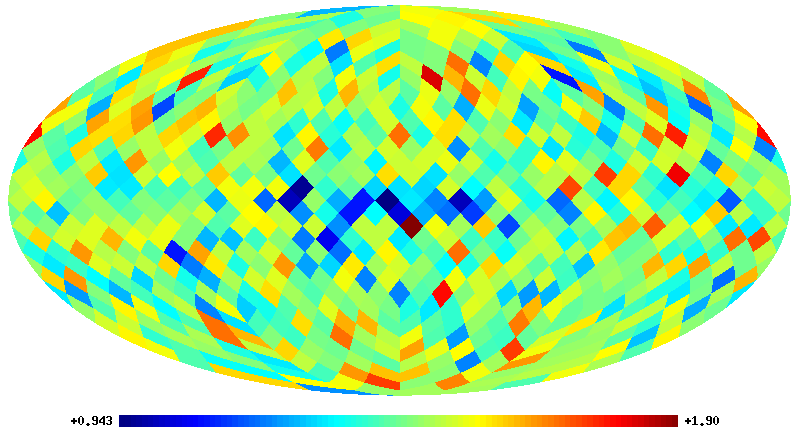}
\includegraphics[width=0.32\textwidth]{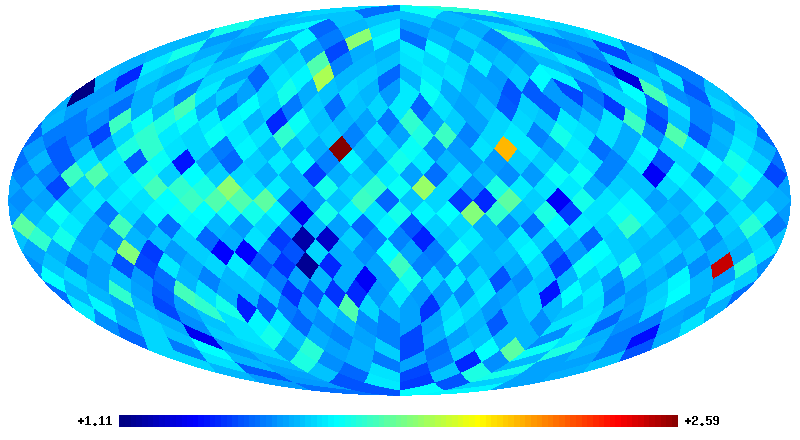}
\includegraphics[width=0.32\textwidth]{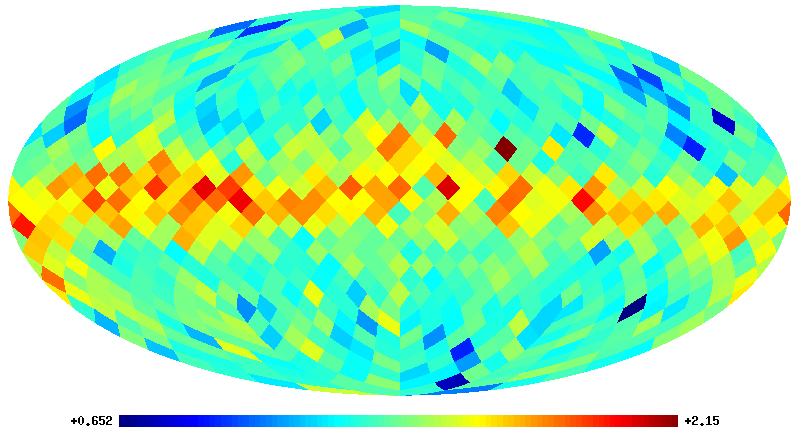}
\includegraphics[width=0.32\textwidth]{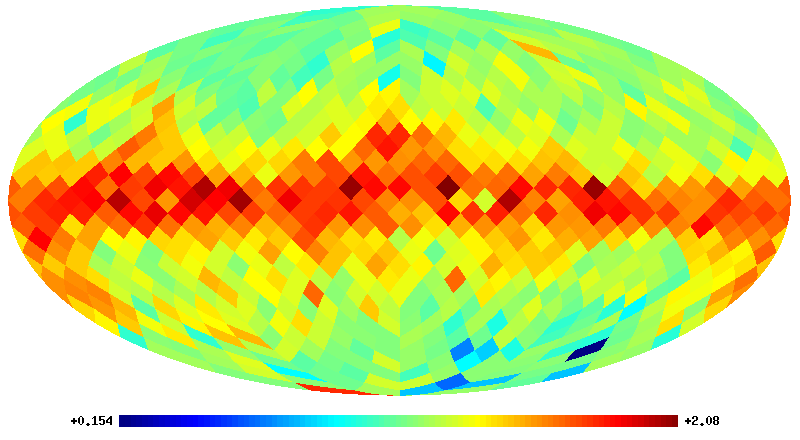}
\includegraphics[width=0.32\textwidth]{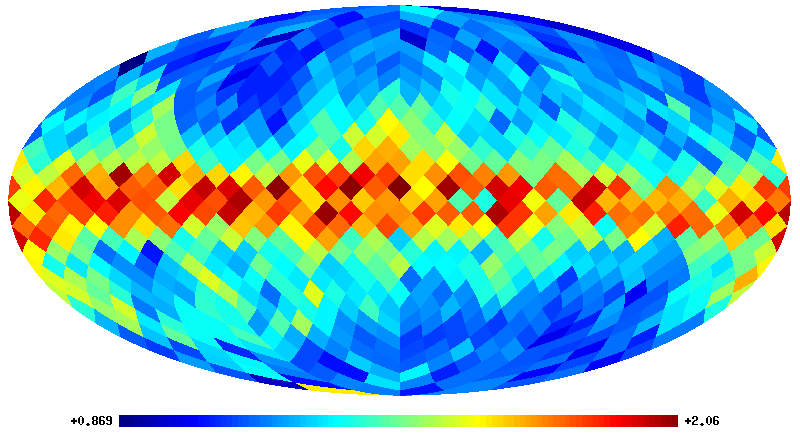}
 \caption{Spatial distribution across the sky of the values of the optimal \sscale of the filter $z=R/\sigma_b$ found during the analysis of nine simulations between 30 and 857 GHz. The upper panels show the 30, 44 and 70 GHz cases, the middle panels show the 100, 143 and 217 GHz cases and the lower panels show the 353, 545 ad 857 cases. }
 \label{fig:mapr}
 \end{center}
 \end{figure*}

\begin{figure*}
\begin{center}
\includegraphics[width=0.32\textwidth]{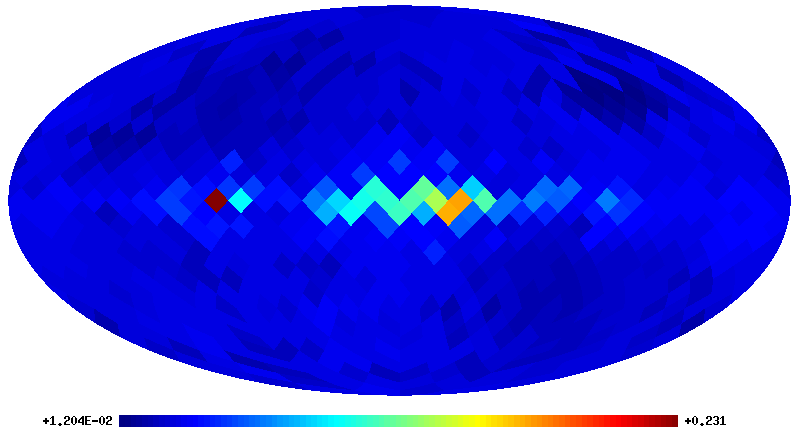}
\includegraphics[width=0.32\textwidth]{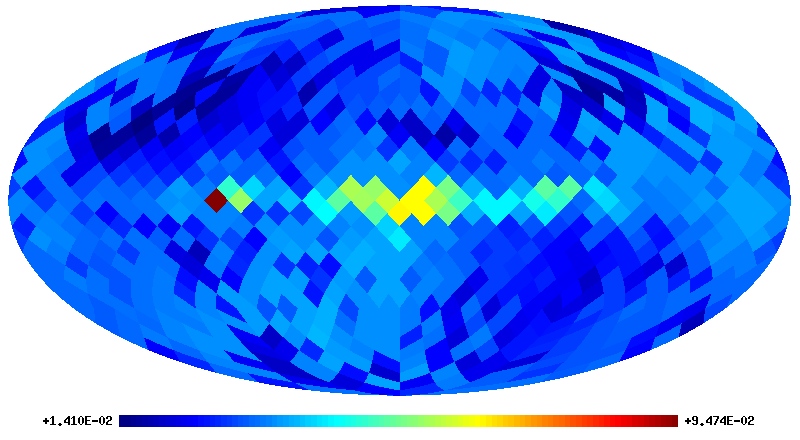}
\includegraphics[width=0.32\textwidth]{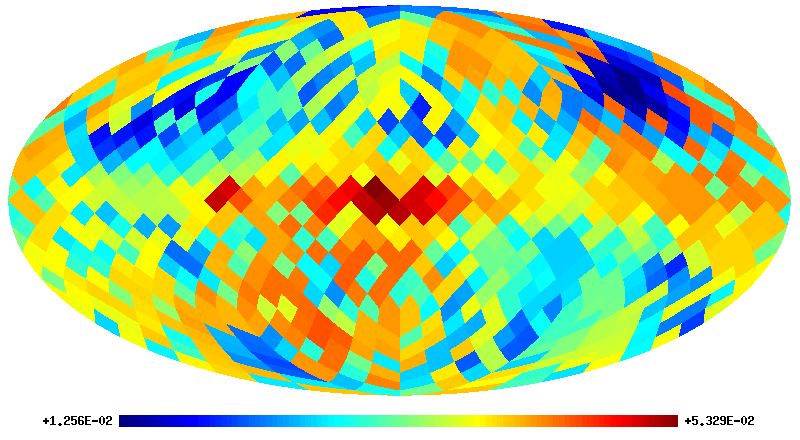}
\includegraphics[width=0.32\textwidth]{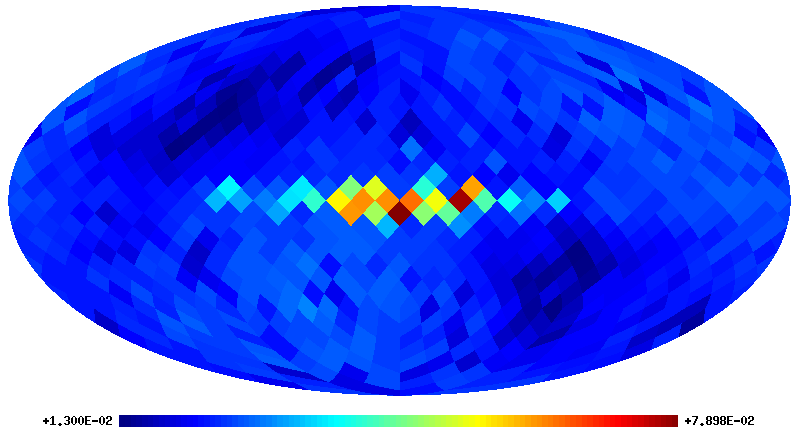}
\includegraphics[width=0.32\textwidth]{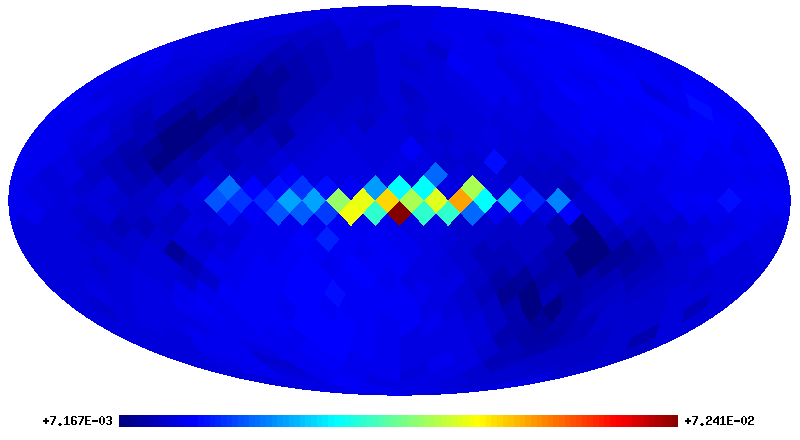}
\includegraphics[width=0.32\textwidth]{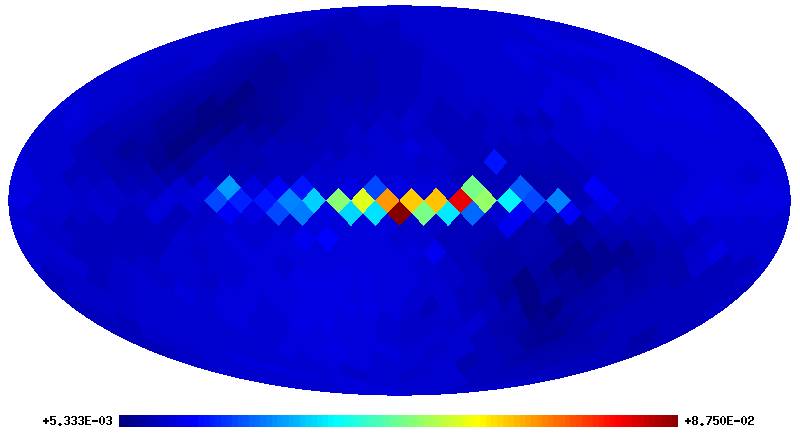}
\includegraphics[width=0.32\textwidth]{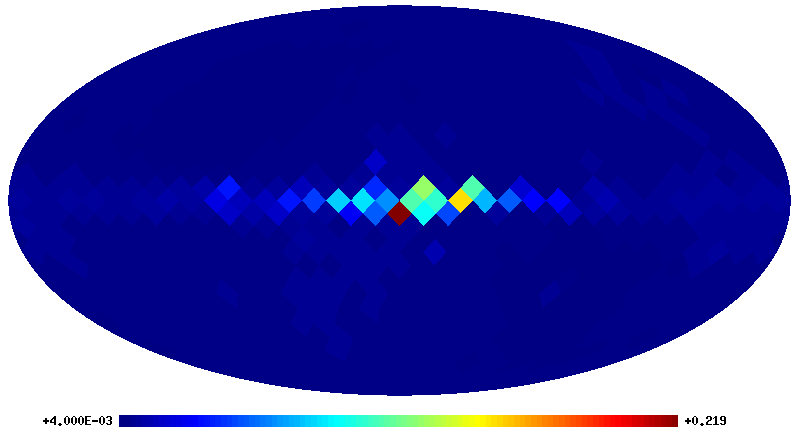}
\includegraphics[width=0.32\textwidth]{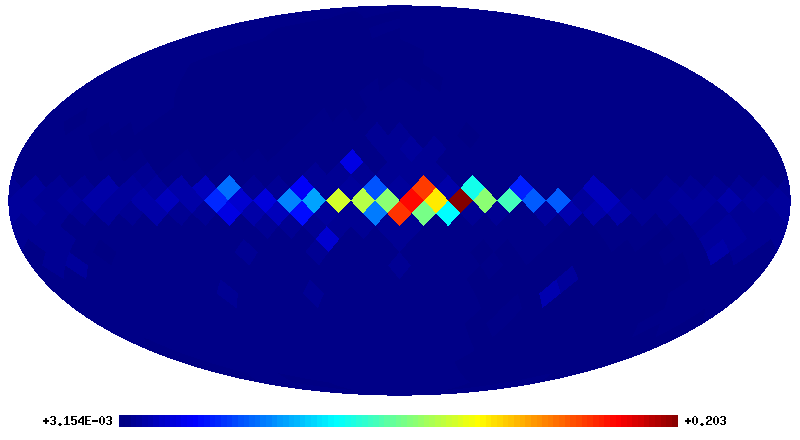}
\includegraphics[width=0.32\textwidth]{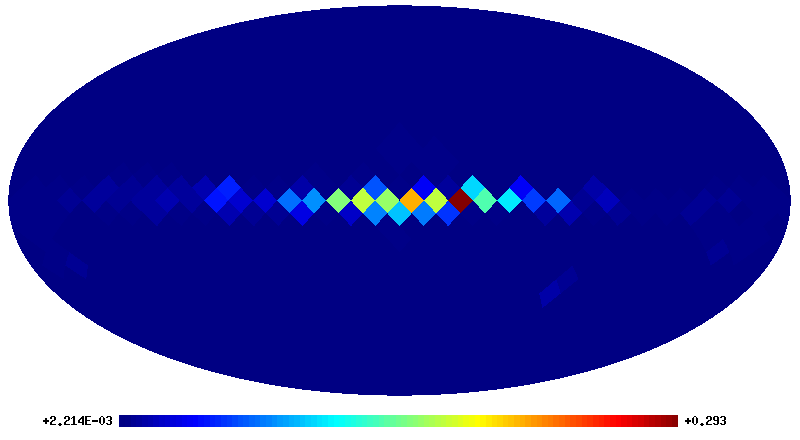}
 \caption{Spatial distribution across the sky of the estimated noise in the filtered maps $\sigma_\omega$ found during the analysis of nine simulations between 30 and 857 GHz. The upper panels show the 30, 44 and 70 GHz cases, the middle panels show the 100, 143 and 217 GHz cases and the lower panels show the 353, 545 ad 857 cases.}
 \label{fig:maps}
 \end{center}
 \end{figure*}

Finally, in order to give an idea of the spatial distribution in the
sky of the values of the filter \iindex $g$, the optimal \sscale $R$
and the noise estimation $\sigma_\omega$ of the filtered maps that we
have obtained, we have constructed a set of figures assigning at each HEALPix
pixel (NSIDE=8) the average value of $g$, $R$ and $\sigma_\omega$ of
the detected sources that fall into it (see Figures \ref{fig:mapg},
\ref{fig:mapr} and \ref{fig:maps}). It is interesting to note that, as
expected, the values of the optimal \iindex $g$ decrease a bit in the
galactic plane at 30 GHz and significantly increase in and around the
galactic plane at 217 GHz and above, while they remain fairly
homogeneous across the sky between 70 and 143 GHz. A similar behaviour
can be seen in the optimal \sscale $R$ showing that the \sscale of the
filter increases with increasing complexity of the background. In
addition, one can see that the estimated noise in the filtered maps
$\sigma_\omega$ follows the galaxy and a few other complex regions in
the sky very well.

\section{Conclusions}
In this work we have developed and studied the performance of a new
filter that maximizes the amplification of the compact sources
embedded in a complex background using two free parameters, the
\sscale $R$ and the \iindex $g$ of the filter. This new filter is
called biparametric adaptive filter (BAF). To study the capabilities
of this new filter we have used simulations of the microwave sky at
the frequencies of Planck between 30 and 857 GHz.

In a first detailed analysis of three frequencies (30, 143 and 857
GHz) we have visually selected three regions of interest with
increasing background complexity and obtained projected patches of
$7.3\times7.3$ square degrees. We have applied the BAF to these regions and
have demonstrated that we can always find a combination of $R$ and $g$
that maximizes the amplification of the sources in the filtered map
with respect to the original map. In addition, we have found that it
is important to optimize not only the \sscale of the filter but also
the \iindex $g$ that adapts to scaling properties of the background,
finding values of $g$ in the range [1.8-12] for the nine discrete
patches that we have studied.

In a second test, we have explored the performance of the BAF when doing a full-sky analysis. We divide the sky into 1344 overlapping
patches, $7.3\times7.3$ square degrees each, that effectively cover the $100\%$
of the sky and apply our new filtering technique to each one of them,
detecting sources above $SNR>5$. The results not only confirm what we
found in the preliminary study of nine interesting regions but also
show how the \iindex $g$ changes in a smooth and coherent way when we
move from low galactic latitudes to high ones in the presence of
strong galactic emission such as the synchrotron radiation at 30 GHz
or the dust emission and the Far-Infrared background starting at 217
GHz and above. The most extreme cases are 353, 545 and 857 GHz, where
the \iindex $g$ changes from very low values close to $g=2$ for the
highest galactic latitudes up to $g=12$ for the regions very close to
the galactic plane. Even more, in these channels that we could
consider cleaner from galactic emission, in the sense that the CMB is
the dominant component of the background, one can see that the
behaviour of the \iindex is fairly flat at all galactic latitudes and
takes values, on average, between $g=4$ and $6$. In particular, at
70 GHz, the \iindex is always very close to $g=4$, the \iindex of the
MHW2. In addition we have qualitatively compared the performance of the
filter with that of the MHW2, in terms of the number of detections above
$SNR>5$ in two cases, in the whole sky and above a galactic cut of
$\pm30$ degrees. We find that in the first case, the BAF detects more
sources than the MHW2 below 217 GHz and the MHW2 detects more source
at 217 GHz and above, although from the inspection of the position of
the sources in the sky one can see that most of the new detections at
high frequency are very close to the galactic plane and are likely to
be spurious detections due to bright compact emission from the
galaxy. In the second case, were the galactic cut has been applied, we
find the BAF detects more sources at all frequencies. Moreover, we have looked 
at the number of unique detections obtained by one method and by the other, inside and outside a galactic mask, and concluded that, first, the BAF detects more unique objects at all bands inside and outside the mask up to 217 GHz, and, second, that at 353 GHz and above the BAF detects less objects than the MHW2 inside the mask, where most of the detections are spurious, as mentioned above.

We have demonstrated that a tool to detect compact sources like the
MHW2 is a very good compromise for the kind of backgrounds that one can
find in microwave experiments, both at low and high galactic
latitudes, at all the Planck frequencies, but its performance can be
improved if we use the BAF, a filter that explores the combination of
the \sscale $R$ and \iindex $g$ that best adapt to the profile of the
sources and to the local properties of the background maximizing the
amplification and the SNR of the detections.

\section{Acknowledgements}
The authors thank Rita Belen Barreiro for useful discussions. The
authors acknowledge partial financial support from the Spanish
Ministerio de Ciencia e Innovaci\'on projects AYA2010-21766-C03-01 and
CSD2010-00064. MLC thanks the Spanish Spanish Ministerio de Ciencia e
Innovaci\'on for a Juan de la Cierva fellowship. PV thanks the
Spanish Ministerio de Ciencia e Innovaci\'on for a Ram\'on y Cajal
fellowship. We acknowledge the use of the pre-launch Planck Sky Model simulation
package \citep{delab11}. The HEALPix package \citep{gorski05} was
used throughout the analysis made in this paper. We would also like to thank an anonymous referee for very useful comments that improved this work.


\thebibliography{}

\bibitem[\protect\citeauthoryear{Arg\"ueso et al.}{2003}]{argueso03} Arg\"ueso F., Gonz\'alez-Nuevo J., Toffolatti L., 2003, ApJ, 598, 86

\bibitem[\protect\citeauthoryear{Arg{\"u}eso et al.}{2009}]{argueso09} Arg{\"u}eso F., Sanz J.~L., Herranz D., L{\'o}pez-Caniego M., Gonz{\'a}lez-Nuevo J., 2009, MNRAS, 395, 649 

\bibitem[\protect\citeauthoryear{Arg{\"u}eso et al.}{2011}]{argueso11} Arg{\"u}eso F., Salerno E., Herranz D., Sanz J.~L., Kuruo{\v g}lu E.~E., Kayabol K., 2011, MNRAS, 414, 410 

\bibitem[\protect\citeauthoryear{Beichman et 
al.}{1988}]{beichman88} Beichman C.~A., Neugebauer G., Habing 
H.~J., Clegg P.~E., Chester T.~J., 1988, iras, 1,  

\bibitem[\protect\citeauthoryear{Bertin \& Arnouts}{1996}]{bertin96} Bertin E., Arnouts S., 1996, A\&AS, 117, 393

\bibitem[\protect\citeauthoryear{Bennett et al.}{2003}]{bennet03} Bennett, et al., 2003, ApJ, 583, 1

\bibitem[\protect\citeauthoryear{Carvalho et al.}{2009}]{carvalho09}
Carvalho P., Rocha G., Hobson M. P., 2009, MNRAS, 338, 765

\bibitem[\protect\citeauthoryear{Cay\'on et al.}{2000}]{cayon00}
Cay\'on L., Sanz J. L., Barreiro R. B., Mart{\'\i}nez-Gonz\'alez E., Vielva P., Toffolatti L., Silk J., Diego J. M., Arg\"ueso F., 2000, MNRAS, 315, 757

\bibitem[\protect\citeauthoryear{Chen 
\& Wright}{2008}]{chen08} Chen X., Wright E.~L., 2008, ApJ, 681, 747 

\bibitem[\protect\citeauthoryear{Chen 
\& Wright}{2009}]{chen09} Chen X., Wright E.~L., 2009, ApJ, 694, 222 

\bibitem[\protect\citeauthoryear{Condon et al.}{1998}]{condon98} 
Condon J.~J., Cotton W.~D., Greisen E.~W., Yin Q.~F., Perley R.~A., Taylor 
G.~B., Broderick J.~J., 1998, AJ, 115, 1693 

\bibitem[\protect\citeauthoryear{Curto et al.}{2009}]{curto09}
Curto A., Mart{\'\i}nez-Gonz\'alez E., Barreiro R. B., 2009, ApJ, 703, 399

\bibitem[\protect\citeauthoryear{Delabrouille et al.}{2012}]{delab11} Delabrouille et al, 2012, in preparation.

\bibitem[\protect\citeauthoryear{de Zotti et al.}{2005}]{dezotti05}
de Zotti G., Ricci R., Mesa D., Silva L., Mazzotta P., Toffolatti L., Gonz\'alez-Nuevo J., 2005, A\&A, 431, 893

\bibitem[\protect\citeauthoryear{Dickinson et al.}{2003}]{Dickinson}
Dickinson, C., Davies, R. D., \& Davis, R. J. 2003, MNRAS, 341, 369

\bibitem[\protect\citeauthoryear{Finkbeiner et al.}{1999}]{Finkbeiner} 
Finkbeiner, D. P., Davis, M., \& Schlegel, D. J. 1999, ApJ, 524, 867

\bibitem[\protect\citeauthoryear{Gonz\'alez-Nuevo et al.}{2006}]{gonzaleznuevo06}
Gonz\'alez-Nuevo J., Arg\"ueso F., L\'opez-Caniego M., Toffolatti L., Sanz J. L., Vielva P., Herranz D., 2006, MNRAS, 369, 1603

\bibitem[\protect\citeauthoryear{Gonz\'alez-Nuevo et al.}{2008}]{gonzaleznuevo08}
Gonz\'alez-Nuevo J., Massardi M., Arg\"ueso F., Herranz D., Toffolatti L., Sanz J. L., L\'opez-Caniego M., de Zotti G., 2008, MNRAS, 384, 711

\bibitem[G{\'o}rski et al.(2005)]{gorski05} G{\'o}rski, K.~M., Hivon, E., Banday, A.~J., et al, 2005, Apj, 622, 759 

\bibitem[\protect\citeauthoryear{Gregory et 
al.}{1996}]{gregory96} Gregory P.~C., Scott W.~K., Douglas K., 
Condon J.~J., 1996, ApJS, 103, 427 


\bibitem[\protect\citeauthoryear{Griffith et 
al.}{1995}]{griffith95} Griffith M.~R., Wright A.~E., Burke B.~F., 
Ekers R.~D., 1995, ApJS, 97, 347 

\bibitem[\protect\citeauthoryear{Haslam et al.}{1982}]{Haslam} Haslam, C. G. T., Salter, C. J., Stoffel, H., \& Wilson, W. E. 1982, A\&A, 47, 1

\bibitem[\protect\citeauthoryear{Herranz \& Vielva}{2010}]{herranz10} Herranz D., Vielva P, 2010, IEEE Signal Processing Magazine, 27, 67

\bibitem[\protect\citeauthoryear{Hinshaw et al.}{2007}]{hinshaw07} Hinshaw et al., 2007, ApJS, 170, 228

\bibitem[\protect\citeauthoryear{H\"ogbom}{1974}]{hogbom74} H\"ogbom J. A., 1974, A\&AS, 15, 417

\bibitem[\protect\citeauthoryear{Komatsu et al.}{2003}]{komatsu03} Komatsu et al., 2003, ApJS, 148, 119

\bibitem[\protect\citeauthoryear{Leach et al.}{2008}]{leach} Leach et al., 2008, A\&A, 491,597L

\bibitem[\protect\citeauthoryear{L{\'o}pez-Caniego et al.}{2005}]{lopezcaniego05} L{\'o}pez-Caniego M., Herranz D., Barreiro 
R.~B., Sanz J.~L., 2005, MNRAS, 359, 993 

\bibitem[\protect\citeauthoryear{L\'opez-Caniego et al.}{2006}]{lopezcaniego06}
L\'opez-Caniego M., Herranz D., Gonz\'alez-Nuevo J., Sanz J. L., Barreiro R. B., Vielva P., Arg\"ueso F., Toffolatti L., 2006, MNRAS, 370, 2047

\bibitem[\protect\citeauthoryear{L\'opez-Caniego et al.}{2007}]{lopezcaniego07}
L\'opez-Caniego M., Gonz\'alez-Nuevo J., Herranz D., Massardi M., Sanz J. L., de Zotti G., Toffolatti L., Arg\"ueso F., 2007, ApJS, 170, 108

\bibitem[\protect\citeauthoryear{L{\'o}pez-Caniego et al.}{2009}]{lopezcaniego09} L{\'o}pez-Caniego M., Massardi M., Gonz{\'a}lez-Nuevo J., Lanz L., Herranz D., De Zotti G., Sanz J.~L., Arg{\"u}eso F., 2009, ApJ, 705, 868 

\bibitem[\protect\citeauthoryear{Mauch et al.}{2003}]{mauch03} 
Mauch T., Murphy T., Buttery H.~J., Curran J., Hunstead R.~W., Piestrzynski 
B., Robertson J.~G., Sadler E.~M., 2003, MNRAS, 342, 1117 

\bibitem[\protect\citeauthoryear{Sanz, Herranz, \& Mart{\'{\i}}nez-G{\'o}nzalez}{2001}]{sanz01} Sanz J.~L., Herranz D., Mart{\'{\i}}nez-G{\'o}nzalez E., 2001, ApJ, 552, 484 

\bibitem[\protect\citeauthoryear{Tauber et al.}{2010}]{tauber10} Tauber J., et al., 2010,A \& A, 520A,1T

\bibitem[\protect\citeauthoryear{Tegmark}{1997}]{tegmark97} Tegmark M., 1997, ApJ, 501, 1

\bibitem[\protect\citeauthoryear{Tegmark \& de Oliveira-Costa}{1998}]{tegmark98}
Tegmark M., de Oliveira-Costa A., 1998, ApJL, 500, 83L

\bibitem[\protect\citeauthoryear{Toffolatti et al.}{2005}]{toffolatti05}
Toffolatti L., Negrello M., Gonz\'alez-Nuevo J., de Zotti G., Silva L., Granato G. L., Arg\"ueso F., 2005, A\&A, 438, 475

\bibitem[\protect\citeauthoryear{Tucci et al.}{2005}]{tucci05}
Tucci M., Mart{\'\i}nez-Gonz\'alez E., Vielva P., Delabrouille J., 2005, MNRAS, 360, 935

\bibitem[\protect\citeauthoryear{Tucci et al.}{2011}]{tucci11} Tucci M., Toffolatti L., de Zotti G., Mart{\'{\i}}nez-Gonz{\'a}lez E., 2011, A\&A, 533, A57 

\bibitem[\protect\citeauthoryear{Vielva et al.}{2001}]{vielva01}
Vielva P., Mart{\'\i}nez-Gonz\'alez E., Cay\'on L., Diego J. M., Sanz J. L., Toffolatti L., 2001, MNRAS, 326, 181

\bibitem[\protect\citeauthoryear{Wright et al.}{2009}]{wright09} Wright E.~L., et al., 2009, ApJS, 180, 283
\end{document}